\documentclass[superscriptaddress,
twocolumn,
 amsmath,amssymb,
 aps,pra,]{revtex4}

\usepackage{graphicx}
\usepackage{dcolumn}
\usepackage{bm}
\usepackage{physics}
\usepackage{braket}
\usepackage{color}
\usepackage{hyperref}
\hypersetup{colorlinks=true}
\usepackage{hypcap}
\usepackage{amssymb,amsbsy,amsmath}
\usepackage[normalem]{ulem}
\usepackage[dvipsnames]{xcolor}

\newcommand\beq{\begin{equation}}
\newcommand\eeq{\end{equation}}
\newcommand\bea{\begin{eqnarray}}
\newcommand\eea{\end{eqnarray}}

\newcommand\dg{\dagger}

\newcommand\ig{\includegraphics}

\begin{document}

\preprint{APS/123-QED}

\title{Strongly interacting spin-orbit coupled Bose-Einstein condensates in one dimension}

\author{Siddhartha Saha}
\affiliation{%
Department of Physics and Astronomy, Center for Materials Theory, Rutgers University, Piscataway, NJ 08854 USA
}%
\author{E. J. K\"onig}%
\affiliation{%
Department of Physics and Astronomy, Center for Materials Theory, Rutgers University, Piscataway, NJ 08854 USA
}%
\author{Junhyun Lee}%
\email{junhlee@umd.edu}
\affiliation{%
Department of Physics, Condensed Matter theory center and the Joint Quantum Institute,
University of Maryland, College Park, MD 20742, USA
}%
\author{J. H. Pixley}%
\email{jed.pixley@physics.rutgers.edu}
\affiliation{%
Department of Physics and Astronomy, Center for Materials Theory, Rutgers University, Piscataway, NJ 08854 USA
}%

\date{\today}

\begin{abstract}
We theoretically study  dilute superfluidity of spin-1 bosons with antiferromagnetic interactions and synthetic spin-orbit coupling (SOC) in a one-dimensional lattice. Employing a combination of density matrix renormalization group and quantum field theoretical techniques we demonstrate the appearance of a robust superfluid spin-liquid phase in which the spin-sector of this spinor Bose-Einstein condensate remains quantum disordered even after introducing quadratic Zeeman and helical magnetic fields. Despite remaining disordered, the presence of these symmetry breaking fields lifts the perfect spin-charge separation and thus the nematic correlators obey power-law behavior. 
We demonstrate that, at strong coupling, the SOC induces a charge density wave state that is not accessible in the presence of linear and quadratic Zeeman fields alone. In addition, the SOC induces oscillations in the spin and nematic expectation values as well as the bosonic Green's function. These non-trivial effects of a SOC are suppressed under the application of a large quadratic Zeeman field. We discuss how our results could be observed in experiments on ultracold gases of $^{23}$Na in an optical lattice. 
\end{abstract}

\maketitle

\section{Introduction}

Ultracold gases of spin-1 bosons offer an exciting platform to understand the interplay of superfluidity and magnetism~\cite{Kawaguchi-2012,StamperKurnUedaRMP2013}.  Depending on the species of atom, the spin dependent interactions can either be ferromagnetic (as in the case of $^{87}$Rb) or antiferromagnetic (as in the cases of $^{23}$Na), which induces ferromagnetic and polar superfluidity, respectively~\cite{Ho-1998,Ohmi-1998}. As a result, in addition to the condensate breaking the U$(1)$ charge symmetry of the system, the SU$(2)$ spin symmetry can also be broken due to the spinful hyperfine interactions. With the development of artificial gauge fields in ultracold atoms, it is now possible to couple the internal hyperfine spin states to their momentum through an engineered spin-orbit coupling (SOC)~\cite{GalitskiSpielman2015,Cooper-2019}. This has now been realized in gases of fermions~\cite{Wang-2012} or bosons~\cite{Lin-2009,Lin-2011,stuhl2015,Campbell-2016,Valdes-2017} with a SOC in one and two dimensions~\cite{Huang-2016,Wu-2016,Sun-2017,Song-2017}. In bosonic gases this induces superfluid order at a non-zero momentum that is dictated by the SOC~\cite{Ueda-12,Li2012,mott_2d_cole,Li2013,Hickey2014,Martone-2014, Lan2014,Pixley-2016,Hurst-2016,YanScarola2017}, while symmetry protected topological phases are possible~\cite{Nonne2013symmetry,Hou2018} and have been observed in Fermi gases~\cite{Song-2017}. This opens an interesting avenue to explore intertwined order between superfluidity, magnetism, and topology in ultracold gases. 

By introducing an optical lattice, both dimensionality and the strength of correlations can be controlled with great accuracy.
This allows experimentally realizing phenomena in one-dimension (1D) where strong correlations are significantly enhanced. For example, Luttinger liquid physics has been observed in strongly interacting fermionic gases \cite{PaganoFallaini2014,hulet2018}. In addition, 1D SOC is much easier to realize experimentally as compared to its 2D analog. This experimental prospect therefore requires a detailed theoretical understanding of the strongly correlated problem. Fortunately in 1D, the existence of powerful analytical and numerical techniques make this understanding possible. 

Recently, significant theoretical progress has been made in understanding ultracold gases with a  SOC using mean field theories~\cite{Li2012,mott_2d_cole,Li2013,Hickey2014,Martone-2014,Lan2014,Hurst-2016} and variational wavefunctions~\cite{Ueda-12,NatuCole2015,Pixley-2016}. 
One can also view the SOC induced ``hopping'' between different hyperfine states as a ``synthetic dimension'' that carry topological edge currents in the superfluid regime~\cite{Po-2014,Celi-2014,mancini2015,stuhl2015,Zeng-2015,Barbarino-2015,Hurst-2016}, which has also been explored in 1D ladder models~\cite{Orignac-2001,Dhar-2012,Dhar-2013,Petrescu-2013,Piraud-2015}. In the strongly correlated regime~\cite{Imambekov2003,RizziFazio2005}, the SOC spin-1 Bose-Hubbard model at the odd integer filled Mott lobes can be mapped to an insulating quantum spin-1 magnet in a helical magnetic field which tunes a quantum phase transition~\cite{PixleyDasSarma2017,ZhouZhang2019}. However, on the other hand, the strongly correlated superfluid regime of \emph{dilute} bosons in 1D in the presence of a SOC has not yet received much attention, despite the potentially rich magnetic phenomena due to the spin-1 nature of the problem, which extends beyond the spin-1/2 case~\cite{Po-2014,ColeSau2019}.

Here, we present a comprehensive study of strongly interacting SOC-ed polar superfluidity  in 1D. Using a field theoretic framework, we develop  the theory for strongly interacting superfluidity in the  spin-1 Bose Hubbard model as the complexity of the problem is increased  to include a quadratic Zeeman field, a transverse magnetic field, and then finally a SOC. In each case, we verify our theoretical predictions using precise density matrix renormalization group (DMRG) calculations. As a result, we are able to isolate and determine the effect of each of these perturbations on the polar superfluid behavior of spin-1 bosons in a 1D  optical lattice. As we show below, in the absence of any perturbing fields the spin-1 Bose-Hubbard model at a fixed dilute filling displays a transition in the excitation spectrum; at sufficiently large interactions the system remains gapless but forms a molecular superfluid phase as the single particle excitations gap out and the two-particle excitations become gapless. We choose to avoid this extreme interaction limit and focus on the experimentally relevant regime with gapless single-particle excitations. First, introducing a quadratic Zeeman field and a transverse magnetic field (which can be considered as a SOC with zero wave vector) in this regime we interestingly find that the spin sector remains quantum disordered in a remarkably robust spin-liquid phase. We determine the Luttinger parameter in the charge sector as well as nematic correlations, which inherit the density-density response due to the gapped spin-liquid sector. In the presence of a full SOC, the ground state displays  superfluidity at zero and non-zero momenta concomitant with  the existence of a strong coupling charge density wave oscillating at the bosonic particle density. This imprints strong density oscillations in the nematic correlation function and the von Neumann entanglement entropy. Lastly, we discuss how these phases can be observed in experiments on ultracold gases of $^{23}$Na in an optical lattice.

The remainder of the paper is organized as follows: In Sec.~\ref{sec:model} we discuss the model and the DMRG approach we have used. In Sec.~\ref{sec:Analytics} we present the results of our field theoretical analysis and in Sec.~\ref{sec:MainResults} we verify the physical predictions of the field theory using DMRG. We discuss the implications of our results and their experimental realization in Sec.~\ref{sec:discussion}. The detailed derivation of the field theory is exposed in Appendix~\ref{app:fieldtheory} (effective field theory Hamiltonian), Appendix~\ref{app:Vortex} (contribution of phase slips), Appendix~\ref{app:DeltaKc} (reduction of Luttinger parameter).

\section{Model and Methods}
\label{sec:model}

We focus on the spin-1 Bose Hubbard model in the presence of a quadratic Zeeman field and a SOC. This is given by the following lattice Hamiltonian in 1D:
\begin{subequations}\label{eq:H0}
\begin{equation}
    H = H_{\rm kin} + H_{\rm loc} + \delta H ,
\end{equation}
where
\begin{align}
H_{\rm kin} &= -  t \sum_{j}\left [ b^\dagger_{j} b_{j+1} +  b^\dagger_{j+1} b_{j}\right], \label{eq:Hkin} \\
H_{\rm loc} &=\sum_{j} \left [\frac{g_0}{2} :\hat n^2_j: + \frac{g_2}{2} :\hat {\mathbf S}^2_j: - \mu \hat n_j \right], \label{eq:Hloc} \\
\delta H &=\sum_{j}\left [ \mathbf h_j \cdot \hat{\mathbf S}_j + q b_j^{\dag}(S_z)^2b_j\right]. \label{eq:HPert}
\end{align}
\end{subequations}
Here we introduced three component bosonic onsite creation and annihilation operators $b^\dagger_j = (b^\dagger_{x,j},b^\dagger_{y,j},b^\dagger_{z,j}),  b_j = (b_{x,j},b_{y,j},b_{z,j})^T$, the number operator $\hat n_j = b^\dagger_j b_j$, and the spin operator $\hat{\mathbf S}_j =  {b_j}^\dagger \mathbf S b_j$, where $\mathbf S = (S_x,S_y,S_z)$ and $(S_a)_{bc} = - i \epsilon_{abc}$ are the spin-1 matrices. This basis is related to the hyperfine eigenbasis
by $b_{x,j} = (i b_{-1,j} - i b_{1,j}) /\sqrt{2}, b_{y,j} = (b_{1,j} + b_{-1,j}) /\sqrt{2}, b_{z,j} =   b_{0,j}$ ($b_{m_z,j}$ annihilates a boson with magnetic quantum number $m_z \in \{-1,0,1\}$).

\begin{figure}[t]
\includegraphics[width=\columnwidth]{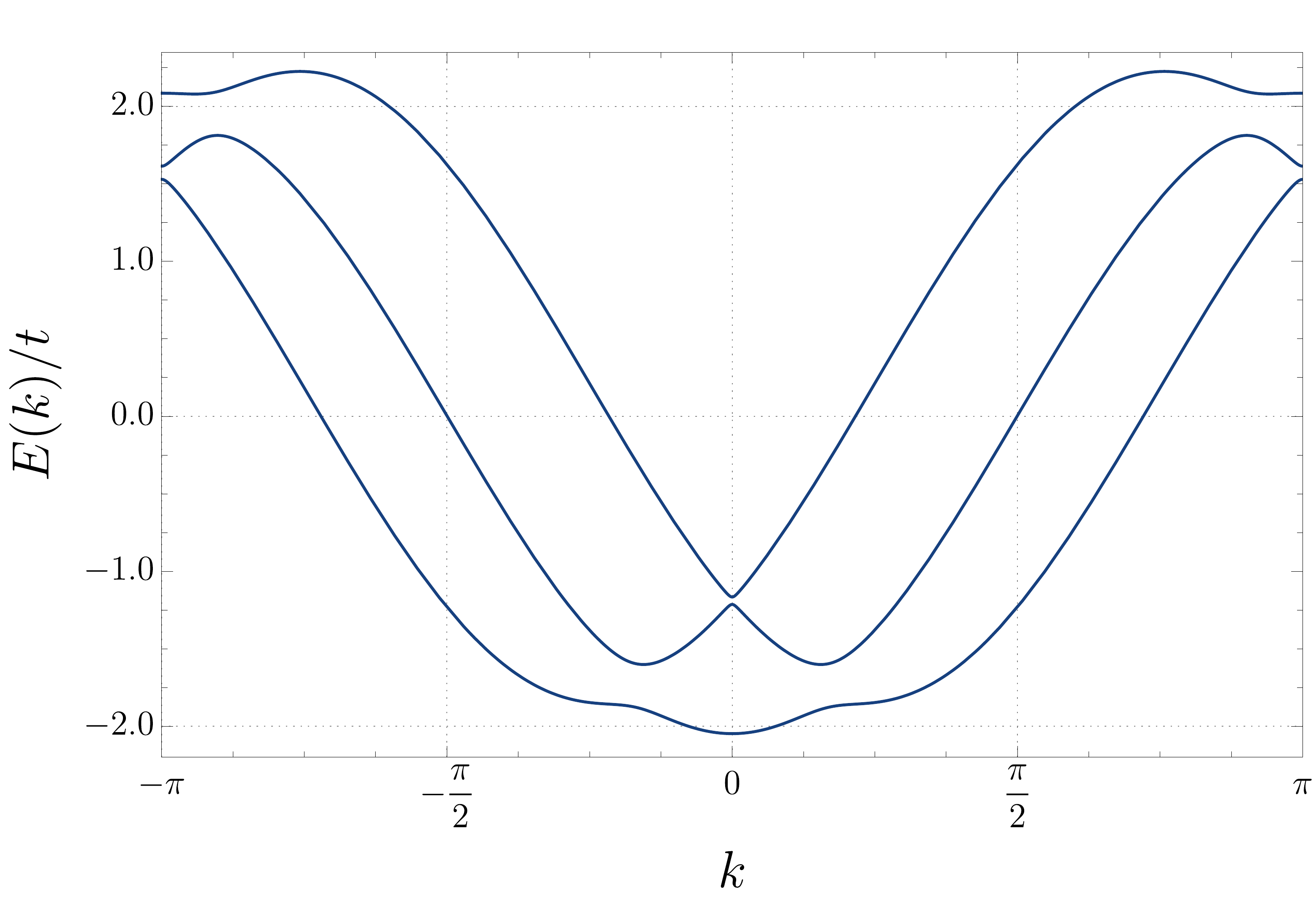}
\caption{Single particle dispersion of spin-1 bosons in the co-rotating frame for $\Theta = \pi/4$ and $q = 0.2 t$. A Zeeman field of $h = 0.2 t$ opens gaps at the level crossings between different $m_z$ states. }
\label{fig:DispersionSpin1}
\end{figure}

The parameters of the Hamiltonian contain the hopping strength $t$ 
that we take as the unit of energy,
on site repulsion $g_0>0$, antiferromagnetic spin exchange interaction $g_2>0$, chemical potential $\mu$, quadratic Zeeman field $q$, and a helical Zeeman field that gives rise to a SOC:
\begin{equation}
\mathbf h_j = h(\cos(\Theta j), \sin(\Theta j),0).\label{eq:h}
\end{equation} 
Here $h$ is the strength of the helical field and $\Theta$ is the pitch of the spiral, i.e., the SOC wave vector. Unless it is explicitly restored, we set the lattice constant $a$  to one. 
We consider both $\Theta \neq 0$ where $\mathbf h_j$ induces SOC, and $\Theta = 0$ where $\mathbf h_j$ is merely a transverse field.
Upon transforming into a co-rotating frame 
\begin{equation}
b_j \rightarrow e^{i \Theta j S_z} b_j,
\label{eqn:rotateL}
\end{equation}
all terms are invariant except for the helical Zeeman field ${\bf h}_j \rightarrow (h,0,0)$ and the kinetic term 
\begin{equation}
H_{\rm kin} \rightarrow - t \sum_{j} \left [b_j^\dagger e^{i \Theta S_z} b_{j + 1} + b_{j+1}^\dagger e^{-i \Theta S_z} b_{j} \right ].
\end{equation}
The spin-orbit coupling and $\Theta$ being its wave vector is manifest in this co-rotating frame. The corresponding single particle dispersion is shifted depending on the hyperfine eigenstate, see Fig.~\ref{fig:DispersionSpin1}.   
We also define the nematic operator $\hat N_{ab,j} = {b_j}^\dagger N_{ab} b_j$ with $N_{ab} = \delta_{ab} \mathbf 1 - \lbrace S_a, S_b \rbrace/2$ to probe nematic order, which shows non-trivial behavior in the polar superfluid phase~\cite{CarusottoMueller2004,Mueller2004}.

We solve the Hamiltonian in Eq.~\eqref{eq:H0} with two complementary methods, namely analytical field theoretical calculations and DMRG simulations. The results of the field theoretic analysis are presented below and the detailed derivations are provided in the Appendices. The DMRG directly simulated Eq.~\eqref{eq:H0} on a 1D lattice with open boundary conditions. We focus on a dilute filling of $\rho_0 = 1/5$ and work in the strong interaction regime so that we can truncate the local bosonic Hilbert space. 
For all of the results presented here we consider a truncated bosonic Hilbert space to at most two bosons per site. We have checked that in this strongly correlated dilute regime the particle number fluctuations are always small making this approximation very accurate. Note that we also verify that the truncation of the local Hilbert space to two bosons per site in the regime of strong coupling is valid analytically  in the derivation of the low energy field theory given below.
In the numerical calculations, we used 
a field strength of $h = 0.1 t$ and a SOC wave vector $\Theta = \pi/10$ on a system of size $L=200$ unless stated otherwise. We monitor the convergence of the DMRG by specifying a truncation error of $10^{-10}$, and a maximum number of $1000$ states were kept to obtain the ground state within the truncation error. Lastly, the DMRG calculations are performed using the ITensor library~\cite{itensor}. 

\section{Effective field theory}
\label{sec:Analytics}

In this paper we focus on the strong coupling regime $g_{0}, g_2 \gg t$, where the interaction energies parametrically exceed the bandwidth. For clarity and completeness, we also discuss analytical results in the opposite limit~\cite{KoenigPixley2018} to provide a complete understanding of the problem.
The analytical strong coupling calculations~\cite{PowellSachdev2007} are derived in the  dilute limit of small superfluid density corresponding to $0<\mu + 2t \ll t$ and perturbatively in $\delta H$ [Eq.~\eqref{eq:HPert}]. 

Both in the strong and weak coupling limits, Eq.~\eqref{eq:H0} in the lab frame maps to the following Hamiltonian density for the bosonic three spinor fields $\psi(x)$ in the continuum
\begin{eqnarray}
\mathcal H &=& \frac{\nabla \bar \psi \nabla \psi}{2 m} - \tilde \mu \bar \psi \psi + \frac{\tilde g_0}{2} (\bar \psi \psi)^2 + \frac{\tilde g_2}{2} (\bar \psi \mathbf S \psi)^2 \notag \\
&+& \bar \psi [\mathbf h(x) \cdot \mathbf S + q S_z^2] \psi. \label{eq:Fieldtheory}
\end{eqnarray}
The parameters of this theory depend non-trivially on the microscopic parameters of Eq.~\eqref{eq:H0}. The weak and strong coupling asymptotes of this functional dependence are compared in Table~\ref{tab:Parameters} and a derivation of Eq.~\eqref{eq:Fieldtheory} for the strong coupling limit is given in Appendix ~\ref{app:fieldtheory}. We highlight that onsite eigenstates of Eq.~\eqref{eq:Hloc} with up to only two bosons are involved in the derivation; three or more boson eigenstates enter the derivation only for higher order terms. This analytically demonstrates that in the dilute limit the constraint we have imposed on the local Hilbert space in our DMRG studies is a very accurate approximation for Eq.~\eqref{eq:H0}.

\begin{table}[t]
\begin{tabular}{|c||c|c|}
\hline
quantity & weak coupling & strong coupling\\
\hline \hline
$m$ & ${1}/({2t a^2})$ & ${1}/({2t a^2})$ \\ \hline
$\tilde \mu$ & $\mu$ & $2t + \mu$ \\ \hline
$\tilde g_0$ & $g_0 a$ & ${4t}{a} \left (1 - \frac{2t}{g_0-2g_2 + 4t} - \frac{8}{3} \frac{t}{g_0+g_2 + 4t} \right)$\\ \hline 
$\tilde g_2$ & $g_2 a$ &$ {4t}{a} \left ( \frac{2t}{g_0-2g_2 + 4t} - \frac{4}{3} \frac{t}{g_0+g_2 + 4t} \right)$ \\ \hline
\end{tabular}
\caption{Parameters entering the field theory, Eq.~\eqref{eq:Fieldtheory}, as determined from the microscopic Hamiltonian, Eq.~\eqref{eq:H0}, in the weak coupling limit $g_{0,2} \ll t$ and strong coupling limit $0<\mu + 2t \ll t \ll g_{0,2}$. Within our perturbative calculations $\mathbf h(x)$ and $q$ are unchanged. Here, we have restored the lattice constant (denoted as $a$) in order to make both units of energy and length manifest.}
\label{tab:Parameters}
\end{table}

\subsection{Low-energy field theory}

Without symmetry breaking terms (i.e. $h=q=0$) the theory displays a $[\mathbf U(1)_{\rm charge} \times \mathbf O(3)_{\rm spin}]/\mathbb Z_2$ symmetry under $\psi \rightarrow e^{i \vartheta} O \psi$ (here, $O^T O = \mathbf 1$). This symmetry is spontaneously broken to $\mathbf O(2)$ on the mean field level, where the field takes the form $\psi_{\rm MF} = \sqrt{\rho_0} e^{i \vartheta} \hat n$ ($\hat n \in \mathbb S_2$ and $\rho_0 = \tilde \mu/\tilde g_0$).

The low energy field theory of Goldstone modes is obtained from Eq.~\eqref{eq:Fieldtheory} in the rotating frame 
\begin{equation}
\psi \rightarrow e^{i \Theta x S_z} \psi
\label{eqn:rotateC}
\end{equation}
 by Gaussian integration of longitudinal, massive fluctuations about the mean field solution. There are two kinds of longitudinal fluctuations: the total density $\delta \rho(x,\tau) = \rho(x,\tau) - \rho_0$ with gap $ \Lambda_c = \rho_0 \tilde g_0$ 
and massive spin fluctuations with gap $ \Lambda_s = \rho_0 \tilde g_2$ 
(see Ref.~\cite{KoenigPixley2018} for details). 
Integrating out the massive spin excitations leads to an effective Lagrangian density in the lab frame (LF)
\begin{subequations}\label{eq:LF}
  \begin{align}
  \mathcal L_{\textrm {LF}} &= i \rho  \dot \vartheta + \frac{\rho}{2m} {\vartheta'}^2 + \frac{\tilde g_0 \delta \rho^2}{2}+ \frac{\vert \dot{\hat n} \vert^2}{2 \tilde  g_2} + \frac{\rho}{2 m} \vert \hat n ' \vert^2  \label{eq:Lunpert} \\
& +\rho q \hat n S_z^2 \hat n - \frac{1}{2\tilde g_2} \hat n [\mathbf h(x) \cdot \mathbf S]^2\hat n.
  \label{eq:Lpert}
  \end{align}
\end{subequations}

It is customary~\cite{GiamarchiBook} to relabel field integration variables $\delta \rho \rightarrow - \phi'/\pi$ to make the Luttinger liquid nature of the first three terms in Eq.~\eqref{eq:Lunpert} apparent. The last two terms in Eq.~\eqref{eq:Lunpert} correspond to a non-linear sigma model (NL$\sigma$M) in the spin sector. We define the dimensionless stiffness and velocity in charge and spin sector as $K_{c,s} = \pi\sqrt{\rho_0/[m \tilde g_{0,2}]}$ and $v_{c,s} = \sqrt{ \rho_0 \tilde g_{0,2}/m}$, respectively. In the weak coupling regime, the stiffnesses $K_{c,s}$ are both large, while in the strong coupling regime they can be small (see also Fig.~\ref{fig:Kc} below). For example, as $g_0\rightarrow \infty$ the Luttinger parameter  $K_{c} \rightarrow \pi \sqrt{(\rho_0a)/2 + (\rho_0a)^2 - 4 (\rho_0a)^3}$.  Finally, the second line, Eq.~\eqref{eq:Lpert} contains the leading perturbative corrections due to Eq.~\eqref{eq:HPert}. 
The result in the rotating frame (RF), is obtained via Eq.~\eqref{eqn:rotateC} that amounts to the replacements 
\begin{eqnarray}
\mathbf{ h}(x) &\rightarrow& h (1,0,0), 
\\
q &\rightarrow& \epsilon \equiv q + \frac{\Theta^2}{2m},
\\
\mathcal{L}_{\textrm{RF}} &\rightarrow& \mathcal{L}_{\textrm{LF}}- i \frac{\rho}{m}\Theta \hat n' S_z \hat n.
\end{eqnarray}
 
The integral over the superfluid phase $\vartheta = \vartheta_{\rm smooth} + \vartheta_{\rm vortex}$ incorporates both smooth fluctuations and phase slips (i.e. space-time vortices). While the smooth part enters in the form of Eq.~\eqref{eq:LF}, the summation over vortex configuration leads to an additional term of the form (see Appendix~\ref{app:Vortex}) 
\begin{equation}
\mathcal L_{\rm vortex} = -y \cos[2(\pi \rho_0 x - \phi)]. 
\label{eq:Lvortex}
\end{equation}
Here, $y$ is the fugacity (Boltzmann weight) of the vortex which is typically $\ln(y) \sim -K_c$. This term is highly oscillatory and produces a contribution to the action that averages to zero unless $\rho_0$ is an integer. For non-integer $\rho_0$,  
while this term is not relevant in the renormalization group sense, it is responsible for imprinting density oscillations in various observables that are significantly enhanced by a SOC as we demonstrate below.
 
\subsection{Unperturbed theory: $h,q=0$}

Before analyzing the implications of the symmetry breaking terms due to the quadratic Zeeman field and the SOC, we briefly discuss the ``unperturbed theory,'' i.e., the spin-1 Bose-Hubbard model defined in Eqs.~\eqref{eq:Hkin},\eqref{eq:Hloc} which lead to the Lagrangian Eq.~\eqref{eq:Lunpert}, \eqref{eq:Lvortex}~\cite{PowellSachdev2007,EsslerTsvelik2009}.
In the dilute limit $0<\rho_0 \ll 1$ of major interest in this paper, the charge sector is a 1D superfluid, i.e., $K_c >1$. In this case the cosine in Eq.~\eqref{eq:Lvortex} wildly oscillates in real space and is ineffective. Contrary, at integer filling, e.g., $\rho_0  \in \mathbb{Z}$, no such oscillations occur and the system undergoes a superfluid to Mott insulating transition as $K_c$ drops below $2$. 
 
Unlike the various scenarios in the charge sector, the spin sector is always quantum disordered in the absence of symmetry breaking terms (Mermin-Wagner theorem)~\cite{EsslerTsvelik2009}. Since the primary order parameter $\psi$ does not display off-diagonal long range order but $\psi^T\psi$ does, this is an example of quantum vestigial order \cite{FernandesSchmalian2019}. The spin-liquid gap $\Delta_{\rm SL}$ in the sigma model part of Eq.~\eqref{eq:Lunpert} is of order $\Lambda_s e^{- 2 K_s}$ in the weak coupling regime and of the order $\Lambda_s / K_s$ at strong coupling. 

Employing the strong coupling parameters of Tab.~\ref{tab:Parameters}, and fixed superfluid density $\rho_0 = \tilde \mu /\tilde g_0$, the chemical potential $\mu (\rho_0)$ drops below the lower band edge $-2t$ when 
\begin{equation}
g_2 = \frac{1}{12} \left(\sqrt{81 {g_0}^2+252{g_0}t-188t^2}-3 {g_0}-2t\right). \label{eq:MoleculeCond}
\end{equation} 
Note that this condition is $\rho_0$ independent. When $g_2$ exceeds this line, on-site ``molecules'' of two bosons with $S=0$ form. This can be viewed as the strong coupling limit of the aforementioned vestigial order. At large $g_0/t$, Eq.~\eqref{eq:MoleculeCond} reproduces the simple condition $2g_2 = g_0 + 2t$ at which the local 2-boson $S = 0$ configuration becomes energetically advantageous to the single boson state, cf. Eq.~\eqref{eq:Hloc} with $\mu = -2 t$ (for a table of the eigenstates and energies, see Tab.~\ref{tab:States} in App.~\ref{app:fieldtheory}.). 

\begin{figure}
\ig[width=0.47\linewidth]{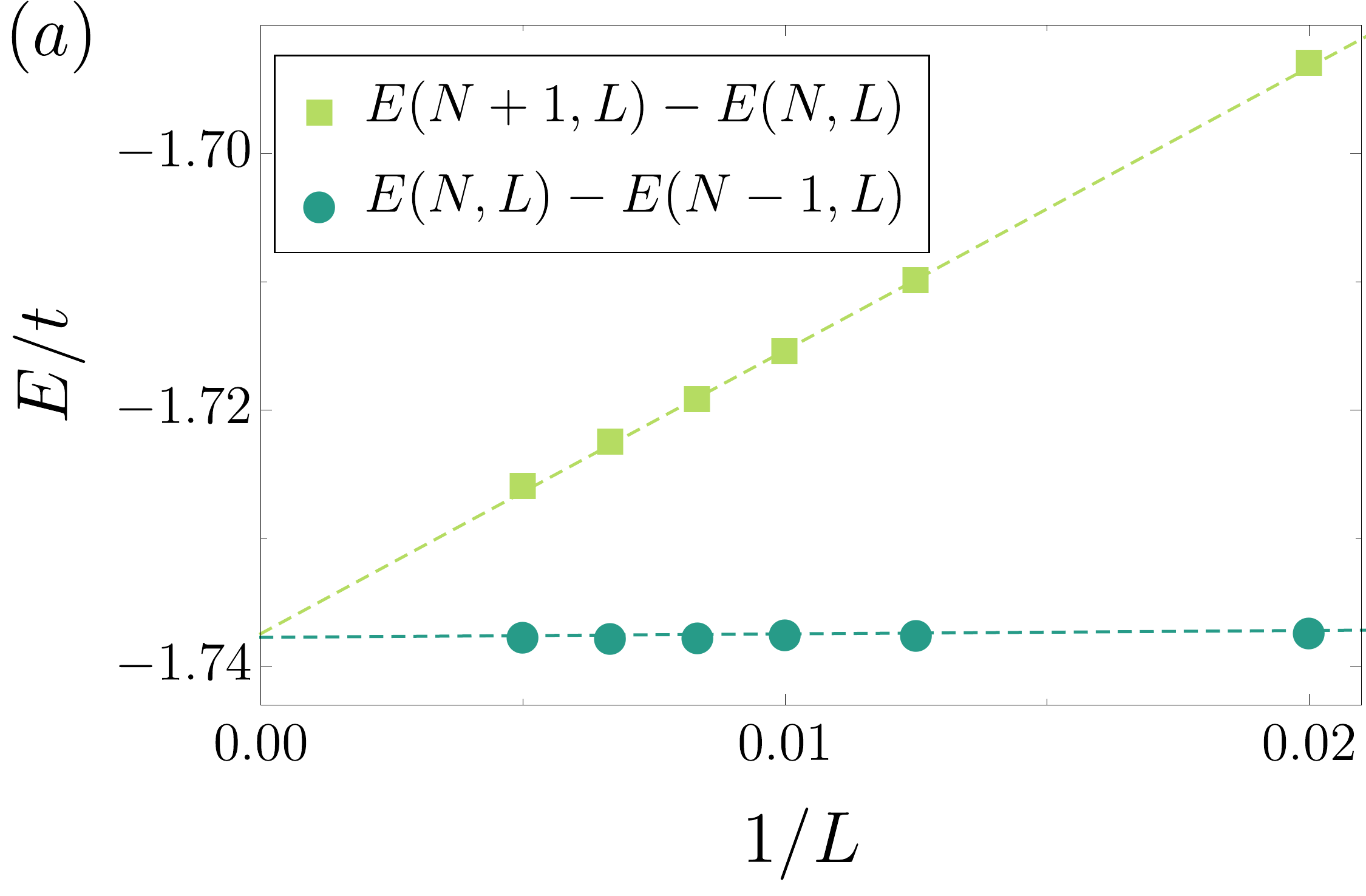} 
\ig[width=0.47\linewidth]{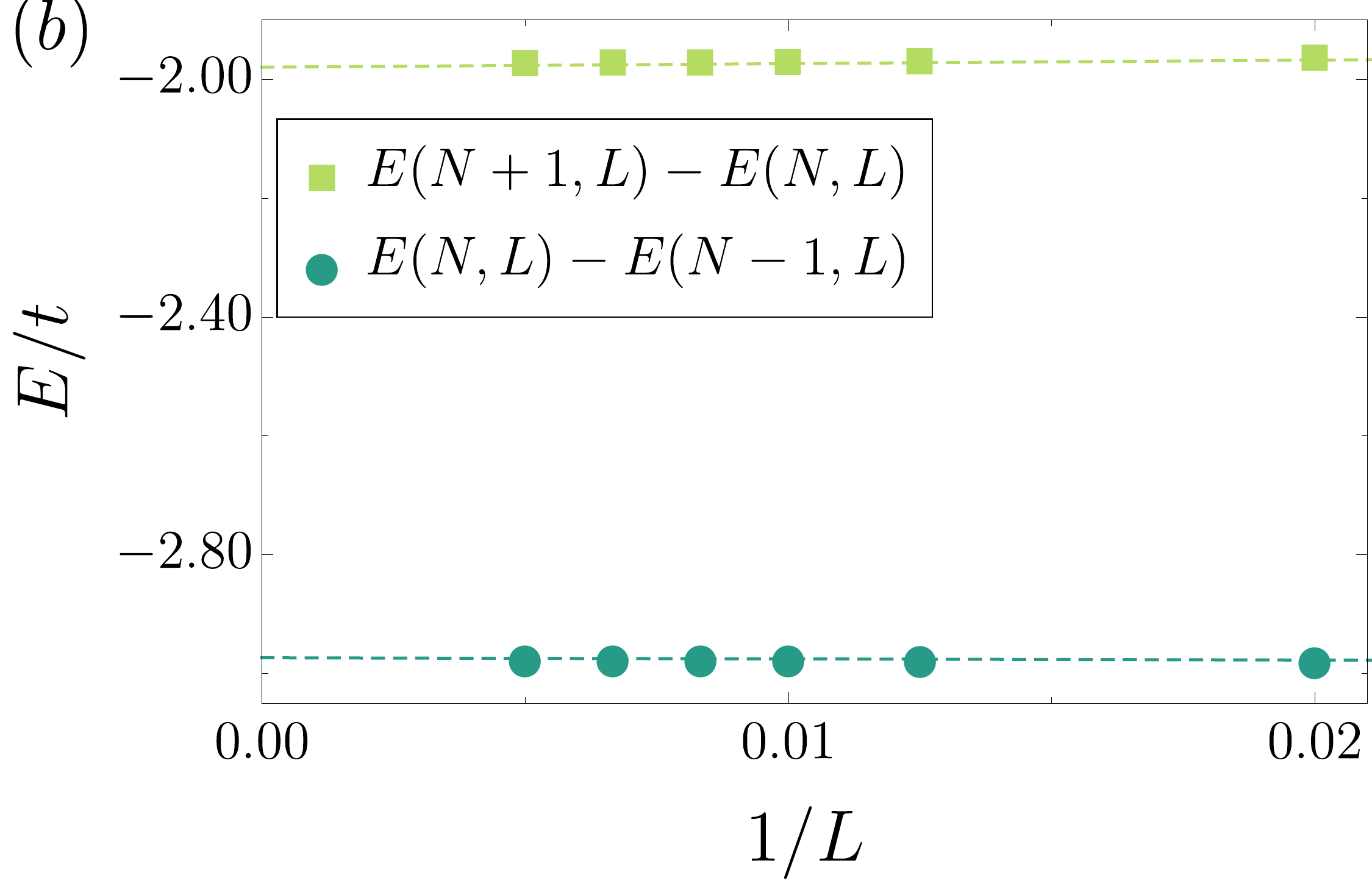} 
\\
\ig[width=0.47\linewidth]{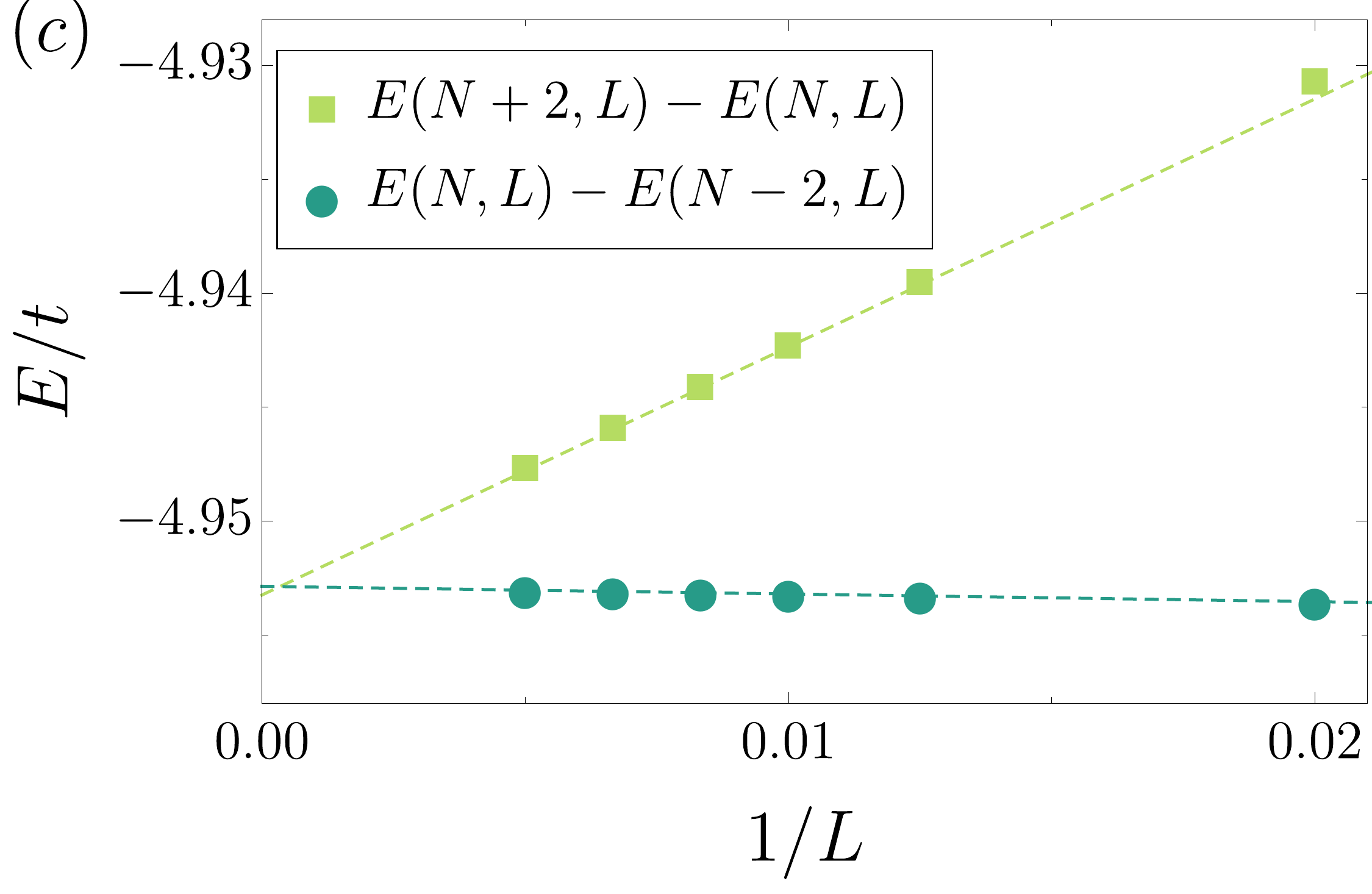}
\hspace{10pt}
\ig[width=0.42\linewidth]{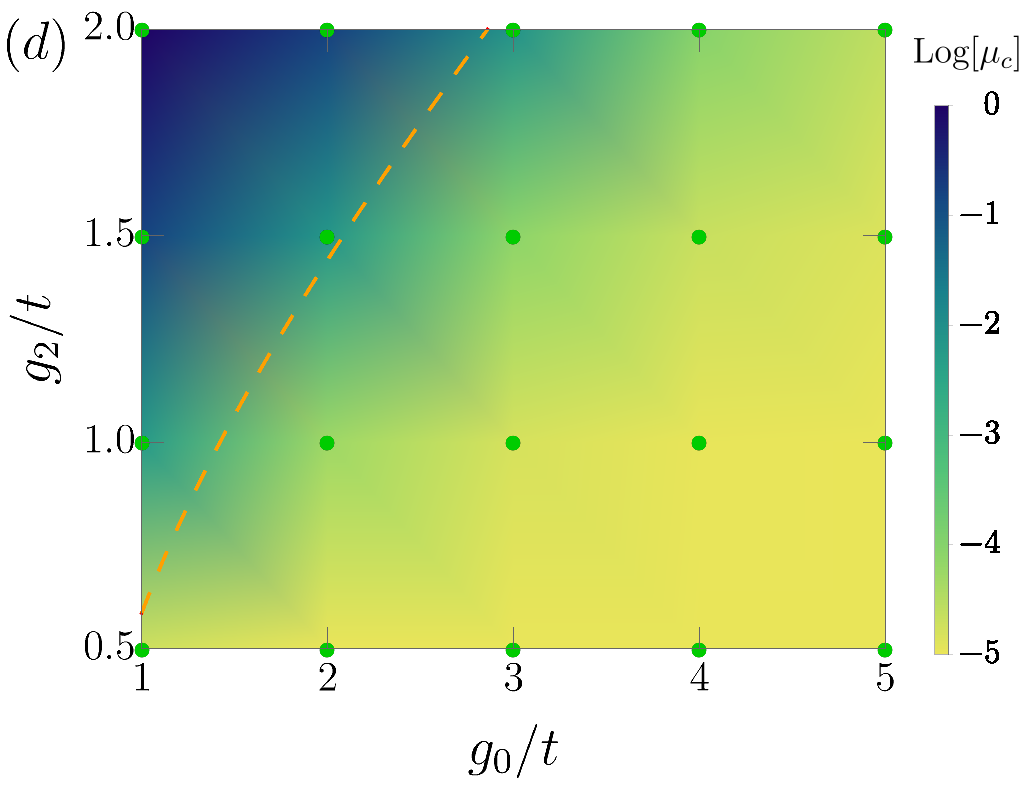}
\caption{(a-c) One- and two-particle gaps extracted from the ground state energy calculations of the finite size DMRG. The energy gap for the thermodynamic limit is obtained by the extrapolation to $1/L \rightarrow 0$. (a) One-particle gap versus $1/L$ for $g_{2}=0.5t$, $g_{0}=5.0t$, (b) One-particle gap versus $1/L$ for $g_{2}=2.0t$, $g_{0}=1.0$t, and (c) Two-particle gap for $g_2 = 2.0t$, $g_0 = 1.0t$.
(d) Phase diagram of the unperturbed model in the plane of density-density interaction constant $g_0/t$ and spin-spin interaction parameter $g_2/t$. The color code represents the numerically obtained single particle gap $\mu_c$. Below the orange line defined by Eq.~\eqref{eq:MoleculeCond} $\mu_c$ vanishes: provided $\hat n$ establishes long range-order the system is a nematic superfluid. Above the line, a molecular phase occurs, here the spin-1 bosons form bound states on each site. The green dots represent the position of actual data points and the plot is constructed by interpolation of the data. The naturally occuring ratio of coupling constants for $^{23}$Na ($g_2/g_0 \approx 1/32$) is well in the phase of vanishing single particle gap.} 
\label{fig:bosonic2} 
\end{figure}

We now verify these expectations using DMRG. In our  DMRG calculations we always have a finite-size gap that we can use to determine the nature of one- and two-particle excitations. To compute the charge gap in the thermodynamic limit and observe the ``molecular phase'' transition, we calculate the length dependence of the finite-size charge excitation gap in the $n$-particle sector following Ref.~\cite{Arcila2016}, focusing on $n=1$ and 2. The finite-size chemical potential to add or remove $n$ particles via the difference in ground state energy is:
\begin{eqnarray}
\mu_{n+}(N,L) &=& E(N+n,L) - E(N,L),
\nonumber\\
\mu_{n-}(N,L) &=& E(N,L) - E(N-n,L),
\end{eqnarray}
where $E(N,L)$ is the energy of the system of size $L$ with $N$ particles. The nature of the gaps in the single ($n=1$) and double ($n=2$) particle sectors follow from the dependence of $\mu_{n\pm}(N,L)$ as a function of $L$ while fixing the density $\rho \equiv N/L$, as shown in Fig.~\ref{fig:bosonic2} (a), (b), and (c). For gapless excitations we fit the chemical potential~\cite{Arcila2016} to $\mu_{n\pm} \sim a_{n\pm} + b_{n\pm}/L$ with $a_{n+} = a_{n-}$ and have  $\mu_{n+} - \mu_{n-} \rightarrow 0$ as $L \rightarrow \infty$. Whereas in the presence of a finite charge gap, we fit the chemical potential~\cite{Arcila2016} to $\mu_{n\pm} \sim a_{n\pm} + b_{n\pm}/L^2$ with $a_{n+} > a_{n-}$ and find that  $ \mu_{n+} - \mu_{n-} >0$ in the thermodynamic limit. We find that for $g_2 \ll g_0$ [Fig.~\ref{fig:bosonic2} (a)] the system is in a robust superfluid phase with a gapless single particle sector. The single particle excitations become gapped in the opposite limit of $g_2 \gg g_0$ [Fig.~\ref{fig:bosonic2} (b), (c)], however the two-particle (i.e. molecular) excitations remain gapless which is in excellent agreement with the field theoretic analysis. The numerically calculated single particle gap together with the analytical phase boundary are summarized in the $g_2-g_0$ phase diagram in Fig.~\ref{fig:bosonic2} (d). 
 
\begin{figure}
\includegraphics[scale=.6]{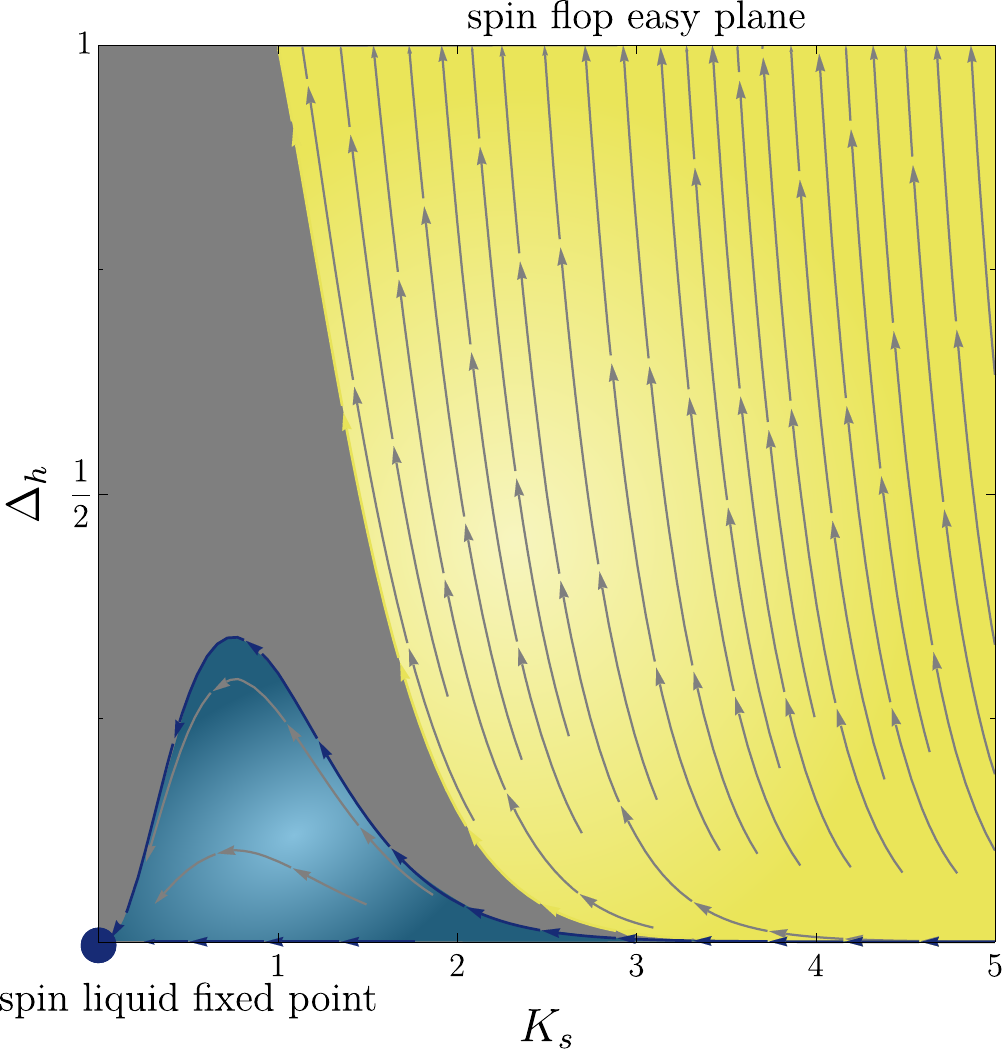}
\caption{Weak coupling RG flow, Eq.~\eqref{eq:RG_NLSM}, which illustrates the separation of a spin flop easy plane phase from a spin liquid. Strictly, the RG equations are inapplicable at $K_s \sim 1$, leaving details in the gray transition region unknown.} 
\label{fig:RG_NLSM}
\end{figure}

\subsection{Review: Weak coupling theory }

As we explained in the previous section, the weak coupling regime implies an exponentially small gap ($\Delta_{\rm SL} \sim \Lambda_s e^{- 2 K_s}$) in the quantum disordered spin sector. Therefore, the system is very susceptible to symmetry breaking perturbations and a moderate spiral Zeeman field $\Delta_h \equiv v_s h^2/[\Lambda_s \tilde g_2] \gtrsim \Delta_{\rm SL}$ is sufficient to drive the spin sector into an easy plane (spin flop) phase \cite{KoenigPixley2018}. This is nicely illustrated within weak coupling renormalization group (perturbative in $1/K_s$ and $\Delta_h$) with flow equations:
\begin{eqnarray}
\frac{d K_s}{d l} &=& - \frac{1}{2},\nonumber \\
\frac{d \Delta_h}{d l} &=& \left(2 - \frac{3}{2 K_s}\right) \Delta_h.\label{eq:RG_NLSM}
\end{eqnarray}
Here, $l$ is the running logarithmic scale. 
Depending on the relative magnitude of $\Delta_h$ and $\Lambda_s e^{-2K_s}$ the system either flows into a spin liquid phase (for small $\Delta_h$) or into an easy plane where $\hat n \perp (1,0,0)$ [see Fig.~\ref{fig:RG_NLSM}].

The easy plane model at sufficiently large $h$ contains two different nematic phases and a spin disordered phase in the parameter space spanned by $K_s$ and $\epsilon$. When the renormalized $K_s \vert_{l = \ln(\Lambda_s/\Delta_h)}$ is larger than 2, there is a direct transition between the two nematic states at $\epsilon = 0$. The critical theory is characterized by a spin-charge separated line of pairs of Luttinger liquid fixed points. On the contrary, the transition is indirect as a function of $\epsilon$ at smaller $K_s$ with an intermediate spin liquid state. In the regime where the transition is split, the critical state between nematically ordered and disordered states is a rather exotic $c = 3/2$ conformal field theory \cite{SitteGarst2009,AlbertonAltman2017,han2019}. It consists of a spin-charge locked pair of a Luttinger liquids in charge space and a Majorana (Ising) critical state in spin space where $v_c = v_s$ at the critical fixed point. Such a field theory attracted substantial attention recently since it represents a rather simple example of supersymmetric field theories~\cite{HuijseBerg2015} and is related to topological superconductivity~\cite{RuhmanAltman2015,KaneStern2017}.

\subsection{Strong coupling limit}

In contrast to the weak coupling case, $K_s \sim 1$ at strong coupling. According to the RG estimate from Eq.~\eqref{eq:RG_NLSM} $\Delta_h \gtrsim \Lambda_s/K_s$ would be needed to drive the system into the easy plane. However, in this regime the low-energy many body theory [Eq.~\eqref{eq:LF}, \eqref{eq:Lvortex}] is not applicable. 
Physically, when $h$ is that large, the single particle spin polarizing term in Eq.~\eqref{eq:HPert} is larger than the many-body interaction terms, Eq.~\eqref{eq:Hloc}. The system is then close to the conventional BEC ground state of fully polarized (i.e., essentially spinless) bosons, instead of being in the vicinity of the spin-nematic BEC. 

Since we are interested in the non-trivial regime when many-body effects dominate over $h$, we assume $\Delta_h < \Lambda_s/K_s$ and the spin sector is always spin disordered in the remainder of the paper. At time scales beyond $1/\Delta_{\rm SL}$, it is justified to integrate out $\hat n$ from the low-energy many body theory to obtain an effective Luttinger liquid action of the charge excitations, which is valid at largest length/time scales:
\begin{eqnarray}
\mathcal L_{\rm SC} &=& - i\frac{\phi' \dot \vartheta}{\pi} + \frac{1}{2\pi} \left [ v_cK_c (\vartheta ')^2 + \frac{v_c}{K_c} (\phi')^2 \right] \notag \\
&&- y \cos [2 (\pi \rho_0 x- \phi)]. \label{eq:LLuttinger}
\end{eqnarray}
Somewhat counterintuitively, spin-orbit coupling (the quadratic Zeeman splitting) has an indirect impact on the charge sector, as it reduces (enhances) the Luttinger parameter $K_c$. In App.~\ref{app:DeltaKc} we derive the correction to $v_c/K_c$ due to the fifth (sixth) term in~Eq.~\eqref{eq:LF}, proportional to $\rho \vert \hat n'\vert^2$ ($q \rho \hat n S_z^2 \hat n$). Using a discretization of the field theory on the scale of the coherence length $\xi_s \sim v_s/\Delta_{SL}$ we integrate gapped fluctuations in the spin sector and obtain
\begin{align}
K_c^{\rm eff}(q, \Theta, h) &= \frac{K_{c}}{\sqrt{1 + f(q, \Theta, h){K_c}/{v_c}}},\label{eq:Kofq}\\
v_c^{\rm eff}(q, \Theta, h) &= v_c {\sqrt{1 + f(q, \Theta, h){K_c}/{v_c}}}.\label{eq:vofq}
\end{align}
where
\begin{align}
f(q, \Theta, h) &= \Big \lbrace \sin^2(\Theta \xi_s) \left [\frac{K_s h}{\Delta_{\rm SL}}\right]^4 \frac{\alpha v_s/K_s}{\left (\delta +q \frac{K_c^2 K_s  \tilde g_0}{\Delta_{\rm SL} \tilde g_2}\right )^2} \notag\\
&- \beta \frac{v_s}{K_s} - \gamma \frac{K_s v_s q^2}{\Delta_{\rm SL}^2} \Big \rbrace \Big / \left (\delta + q\frac{K_c^2 K_s  \tilde g_0}{\Delta_{\rm SL} \tilde g_2}\right ),
\end{align}
and $\alpha, \beta, \gamma, \delta$ are non-universal numerical coefficients. Note that the $\Theta$ induced suppression of $K_c$ can be suppressed when $q/\Delta_{\rm SL} \gtrsim \tilde g_2/(\tilde g_0 K_c^2 K_s)$ (which is still much smaller than unity).
We test this prediction in Sec.~\ref{sec:so_LL} numerically and find that the SOC wave vector drives a charge density wave by making $K_c (\Theta) <1$. 

\section{Observables}
\label{sec:MainResults}

In this section we determine the consequences of our field theoretic results on physical observables such as the nematicity, entanglement, and correlation functions. 
We verify this by the finite size DMRG calculations on the SOC $S=1$ Bose-Hubbard model in the lab frame [Eq.~\eqref{eq:H0}], which shows excellent agreement with the field theory results.
We reiterate the parameter regime of the numerical calculation which is in the strong coupling limit [$t \ll g_{0,2}$], dilute filling [$\rho = 1/5$], and we use $h = 0.1t$, $\Theta = \pi/10$ on a $L=200$ lattice.

\subsection{Effects of homogeneous fields $q,h \neq 0$ and $\Theta =0$}
\label{sec:homogfields}
To understand the effect of the symmetry breaking field and the SOC separately, we begin by analyzing the situation without the SOC, i.e. $\Theta = 0$, but with nonzero fields [$q, h \neq 0$]. Note that especially $h \neq 0$ but $\Theta = 0 $ leads to a homogeneous transverse magnetic field [Eq.~\eqref{eq:h}], and this allows us to build up our intuition for this  case before moving onto the effect of a full SOC. The main result for this is that the  model remains ``stuck'' in the spin liquid phase despite tuning the degeneracy lifting quadratic Zeeman and transverse fields, if we stay in the non-trivial regime at which many-body effects are dominant. To demonstrate this we first analyze the nematic order parameter $\langle N_{zz} - N_{yy} \rangle $, which should vanish linearly as $q\rightarrow 0$ in the spin liquid phase~\cite{KoenigPixley2018}. In addition, we use the entanglement entropy to determine the number of gapless modes and show that it is independent of the fields. This also substantiates the evidence for the gapped spin liquid phase since the only gapless excitations result from the charge sector of the theory. 
We furthermore study various correlators and the Luttinger parameter (of the charge sector) which provides a comprehensive understanding of the model. 

\subsubsection{Nematicity tensor}

\begin{figure}
\centering
\ig[width=.95\linewidth]{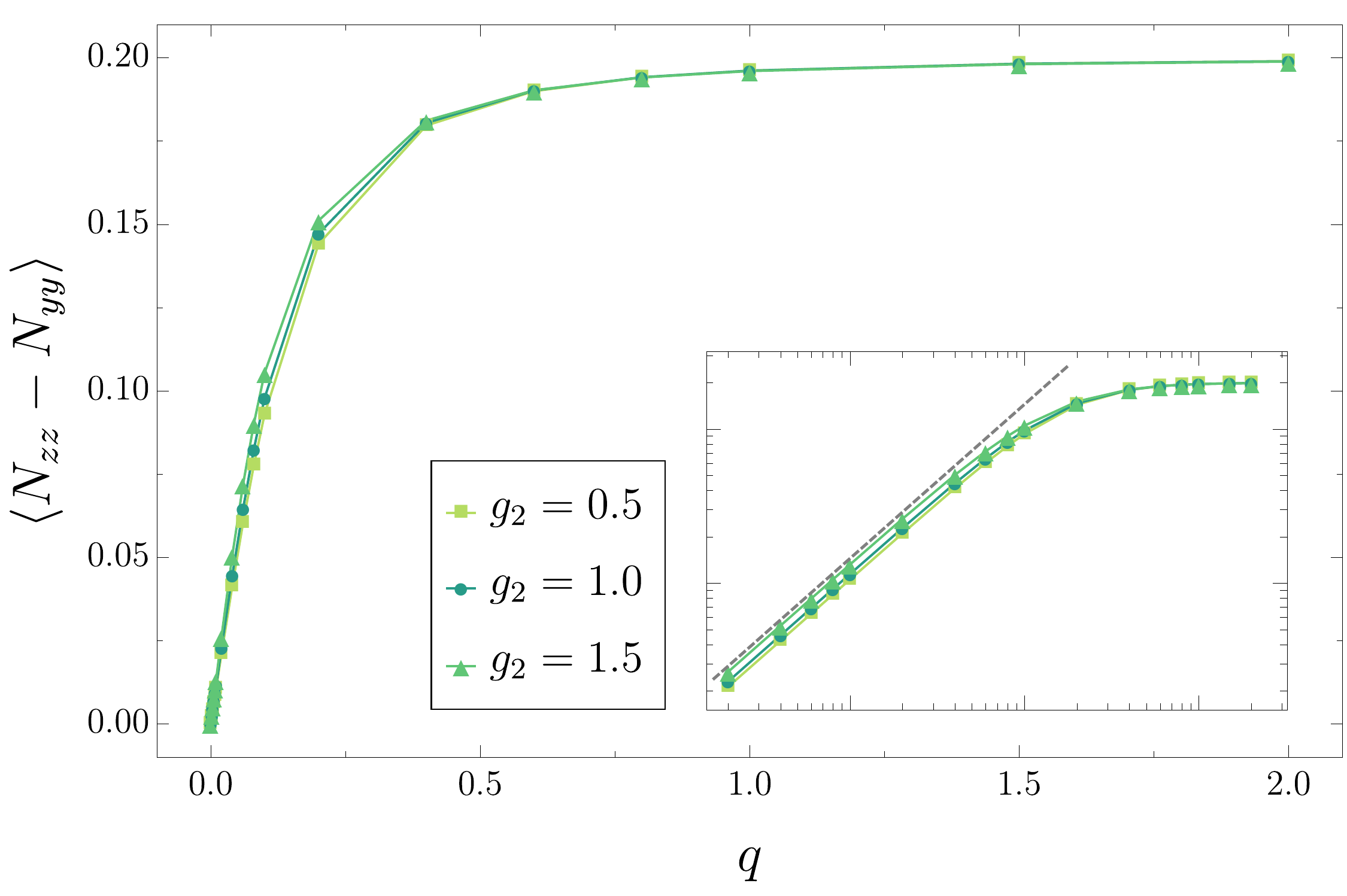} 
\caption{The expectation value of the nematicity tensor $\langle N_{zz} - N_{yy}\rangle$ as function of $q$ at $g_0 = 5.0t$ and $\Theta = 0$. The inset is the same plot in a log-log scale to show the power law behavior, which holds for a wide range of $g_2$. The gray dashed line is a guide to the eye which has a slope of 1, indicating $\langle N_{zz} - N_{yy}\rangle\sim q$ as $q\rightarrow 0$. }
\label{fig:DeltaN1}
\end{figure}

While all spin and nematic correlators are short ranged, the presence of a (quadratic) Zeeman field induces a finite expectation value of the nematicity tensor. Even in the spin disordered phase, the linear field $h$ implies $\hat n \perp (1,0,0)$ locally. Therefore, the only non-trivial expectation value of the nematicity tensor is $\langle N_{zz} - N_{yy} \rangle $, with $\langle N_{zz} + N_{yy} \rangle =\rho_0$ being fixed by our normalization convention. 

In the quantum disordered spin liquid phase, the field theory expectation~\cite{KoenigPixley2018} is that $\langle N_{zz} - N_{yy} \rangle \sim q $, since in any (quantum or thermally) disordered phase the expectation value of the order parameter vanishes linearly as a function of its conjugate variable. In Fig.~\ref{fig:DeltaN1} we numerically demonstrate this behavior for a number of parameters quite clearly. This serves as a strong numerical evidence for the system robustly remaining a spin-liquid in the presence of the fields. 

\subsubsection{Entanglement entropy}
\label{sec:ee0}

\begin{figure}
\centering
\ig[width=.95\linewidth]{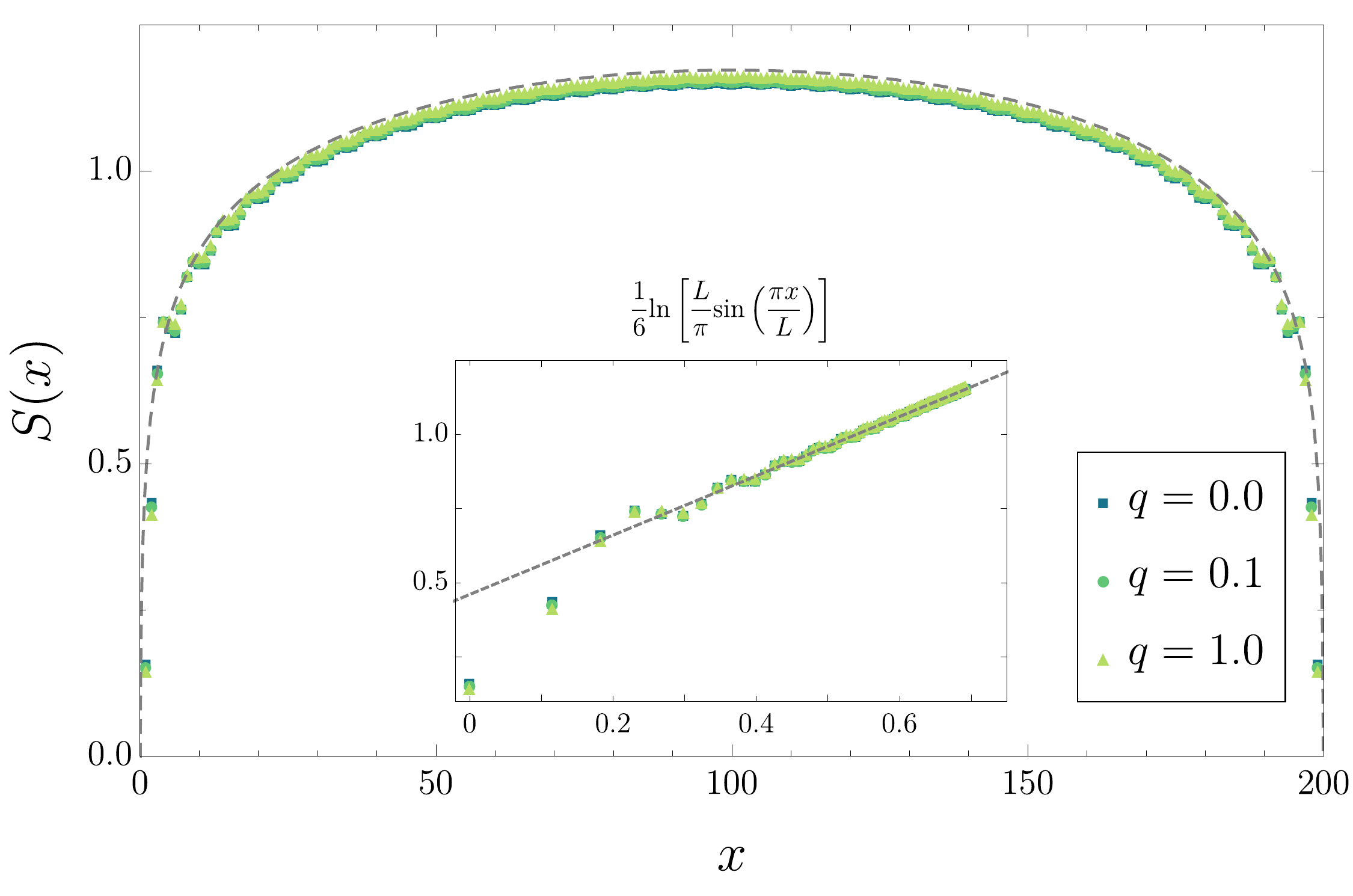} 
\caption[]{The entanglement entropy as a function of the bipartite position $x$, for $g_0 = 5.0t $, $g_2 = 1.0t$, and $\Theta = 0$. The inset is the same data with the horizontal axis as $\frac{1}{6}\ln(\frac{L}{\pi}\sin(\frac{\pi x}{L}))$, the slope determines the central charge per Eq.~\eqref{eq:EE}. The gray dashed lines are guide to the eye which corresponds to $c=1$ result while explicit linear fits gave $c = 1.002$, 1.003, and 1.004 for $q=0.0t$, $0.1t$, and $1.0t$ respectively. }\label{fig:CentralCharge} 
\end{figure}

Another evidence for the gapped spin-liquid would be added if we can observe the nonzero spin gap. 
An indirect method to detect the gap is by counting the gapless modes, or calculating the central charge, of the system.
For the current case of $\Theta = 0$ we have an algebraically ordered superfluid in the charge sector, i.e., a Luttinger liquid with $K_c >1$, which is known to contribute central charge $c=1$ to the system.
Considering the spin sector, the system is clearly in a gapped spin-liquid phase in the unperturbed regime ($h, q = 0$) with the gap $\Delta_{\rm SL} \sim \Lambda_s/K_s$. 
Due to this spin-liquid gap, the spins do not contribute to the central charge and thus the total central charge will be $c=1$. 
And if the system remains in this gapped spin-liquid after turning on the fields ($h, q \neq 0$), the central charge will as well remain $c=1$. 

To extract this numerically, we analyze the dependence of von Neumann entanglement entropy $\mathcal S(x)$ as a function  of the position $x$ of the bipartition. 
We fit $\mathcal{S}(x)$ to the well known result from conformal field theory with open boundary conditions~\cite{HolzheyWilzcek1994,Korepin2004,CalabreseCardy2004}  
\beq \mathcal{S}(x) = \frac{c}{6}\ln(\frac{L}{\pi}\sin(\frac{\pi x}{L})) + d, \label{eq:EE}\eeq 
where $c$ is the central charge and $d$ is a nonuniversal constant.
We calculate this for a number of parameters in Fig.~\ref{fig:CentralCharge} together with the fit to the form of Eq.~\eqref{eq:EE}. 
We consistently obtain $c \approx 1$ up to $q$ in the order of $t$ which adds strong evidence that the ground state of the model across this parameter regime has a spin-sector that remains in a gapped spin-liquid phase.

\subsubsection{Bosonic correlators and Luttinger parameter}

\begin{figure*}
\centering
\ig[width=0.32\linewidth]{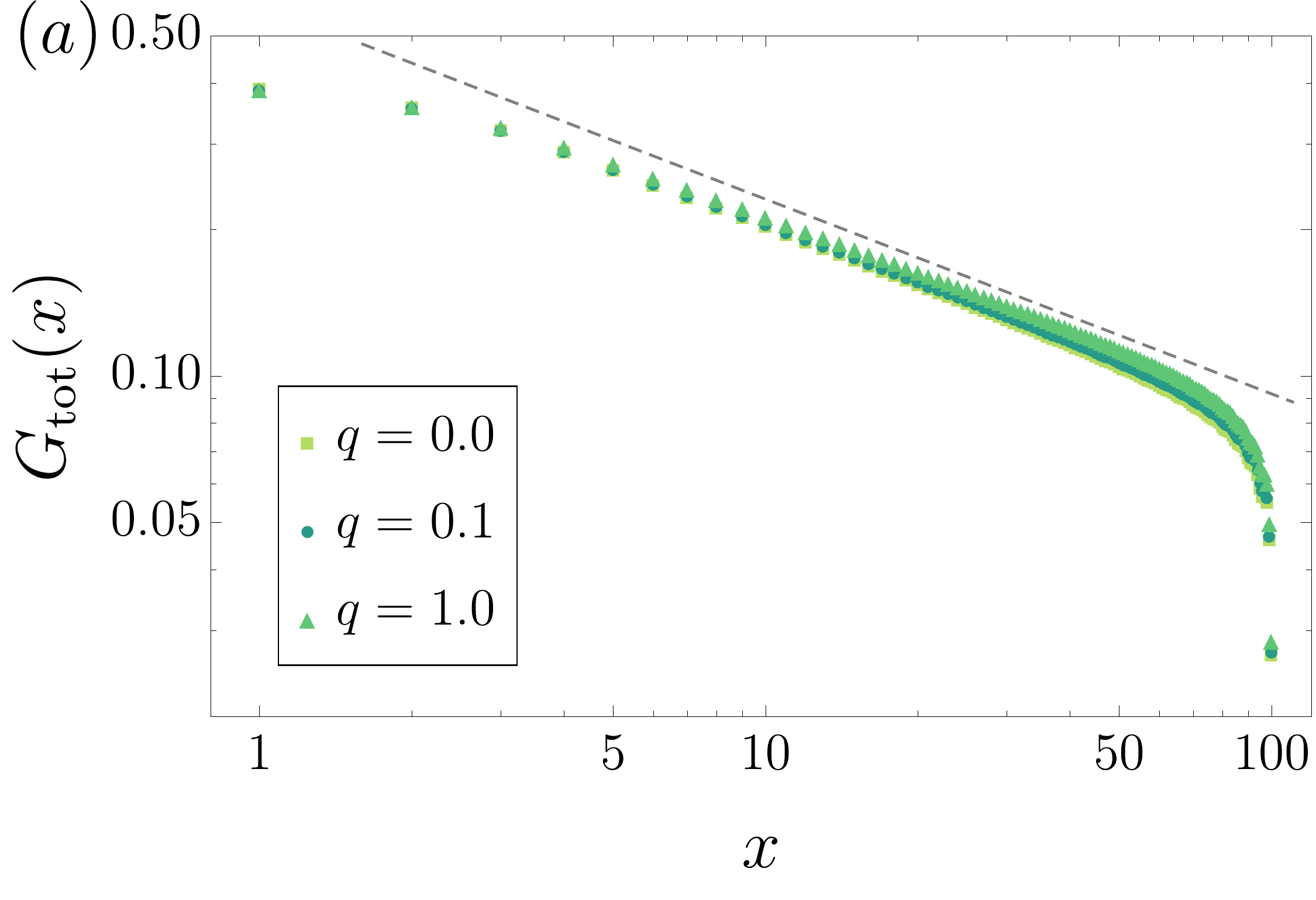} 
\ig[width=0.32\linewidth]{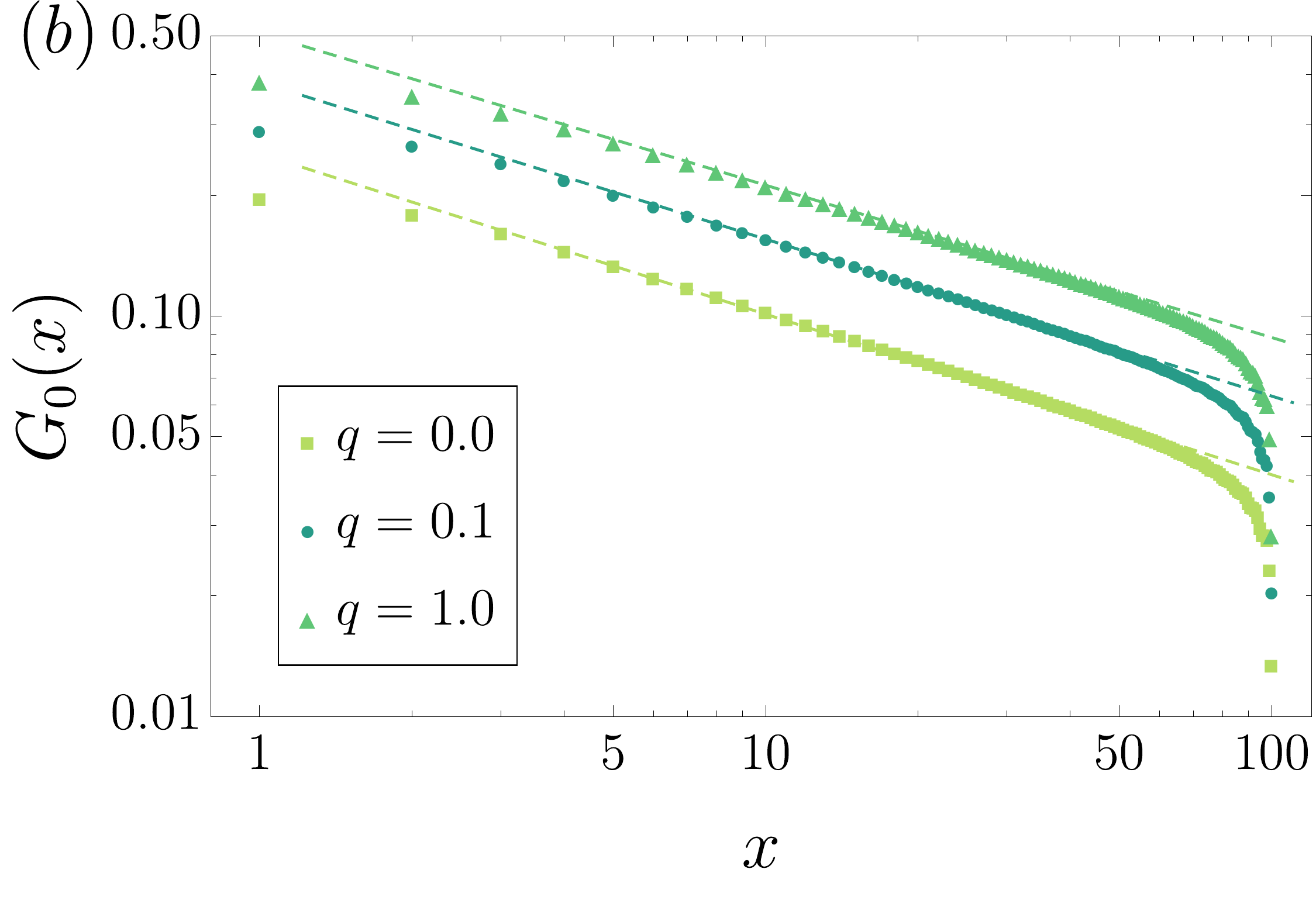}
\ig[width=0.32\linewidth]{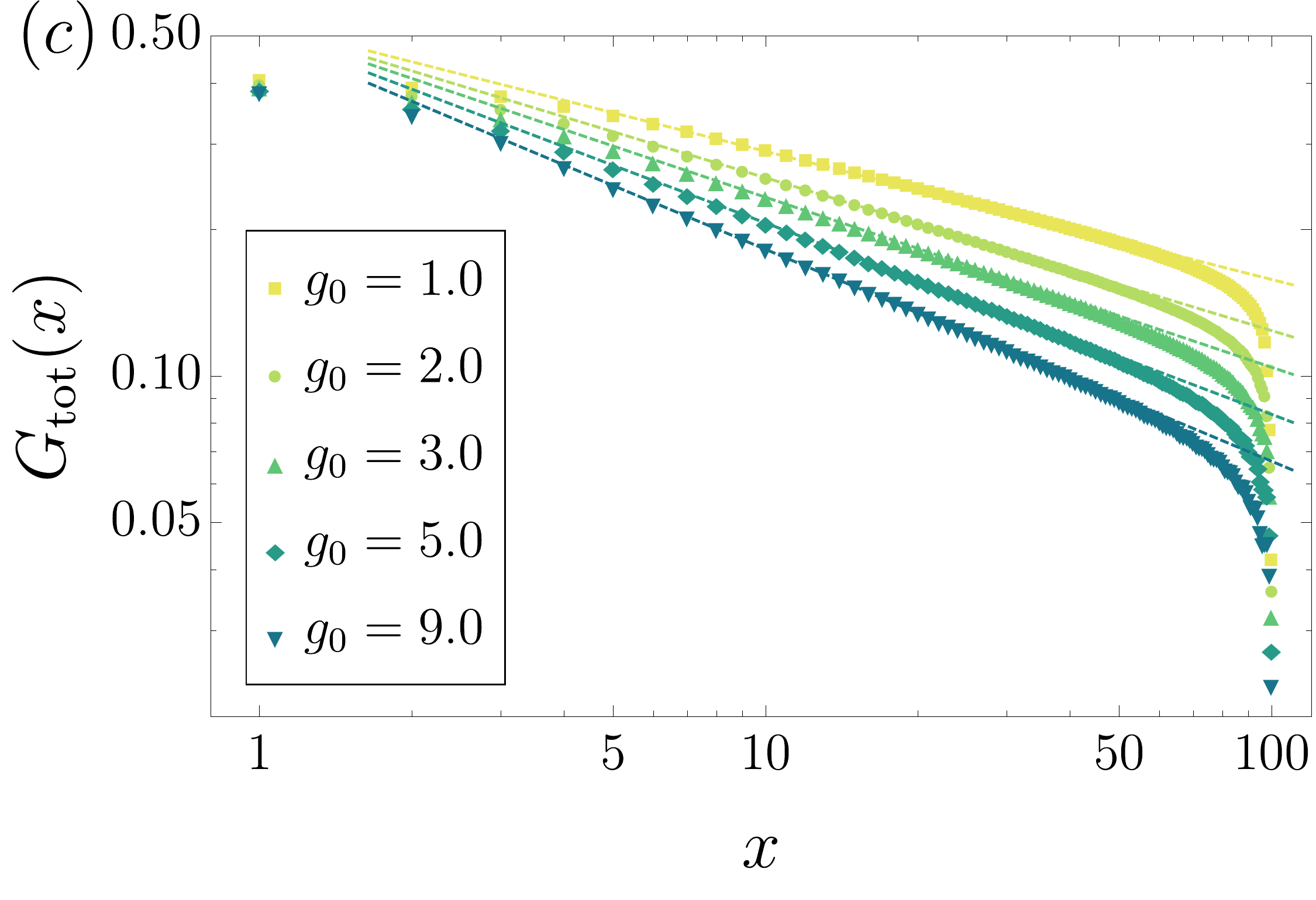}
\caption[]{Log-log plot of (a) $G_{\textrm{tot}}(x)$ versus $x$ and (b) $G_{0}(x)$ versus $x$ for $g_0 = 5.0t$, $g_2 = 1.0t$, and $\Theta = 0$. (c) Log-log plot of $G_{\textrm{tot}}(x)$ versus $x$ for $g_2 = 0.5t$, $q = 0.0t$, and $\Theta = 0$.The gray dashed line in (a) is a guide to the eye with a slope of $-0.4$ while the dashed lines in (b) and (c) are linear fits of each $q$ and $g_0$ for intermediate values of $x$. These linear fits in log-log scale show the power-law behavior of the correlators from which we extract the Luttinger parameter. } \label{fig:G0} 
\end{figure*}

We now investigate the correlation functions at $\Theta = 0$ and study the behavior of the Luttinger parameter.
We begin by studying bosonic correlators $G_0(x) = \langle b_{z,j+x} b^\dagger_{z,j} \rangle$ of the spin-0 projection of the bosonic field (note that $S_z (0,0,b_{z,j})^T=0$) as well as the total Green's function 
$G_{\rm tot}(x) = \text{tr}[\langle b_{j+x} b^\dagger_{j}\rangle ]$.
Since our numerics have open boundary conditions, we calculate the correlation functions at the center of the chain and set $j = L/2$ to avoid boundary effects as much as possible. 
Since the spin sector is gapped, the correlators are dictated by the Luttinger liquid charge sector and behave as: 
\begin{equation}
G_{\alpha}(x) \sim x^{-1/(2K_c)}, 
\label{eq:G}
\end{equation}
for both $\alpha = 0, {\rm tot}$.
This power-law behavior is demonstrated in Fig.~\ref{fig:G0} for various parameters. 

\begin{figure}[t]
\centering
\ig[width=0.95\linewidth]{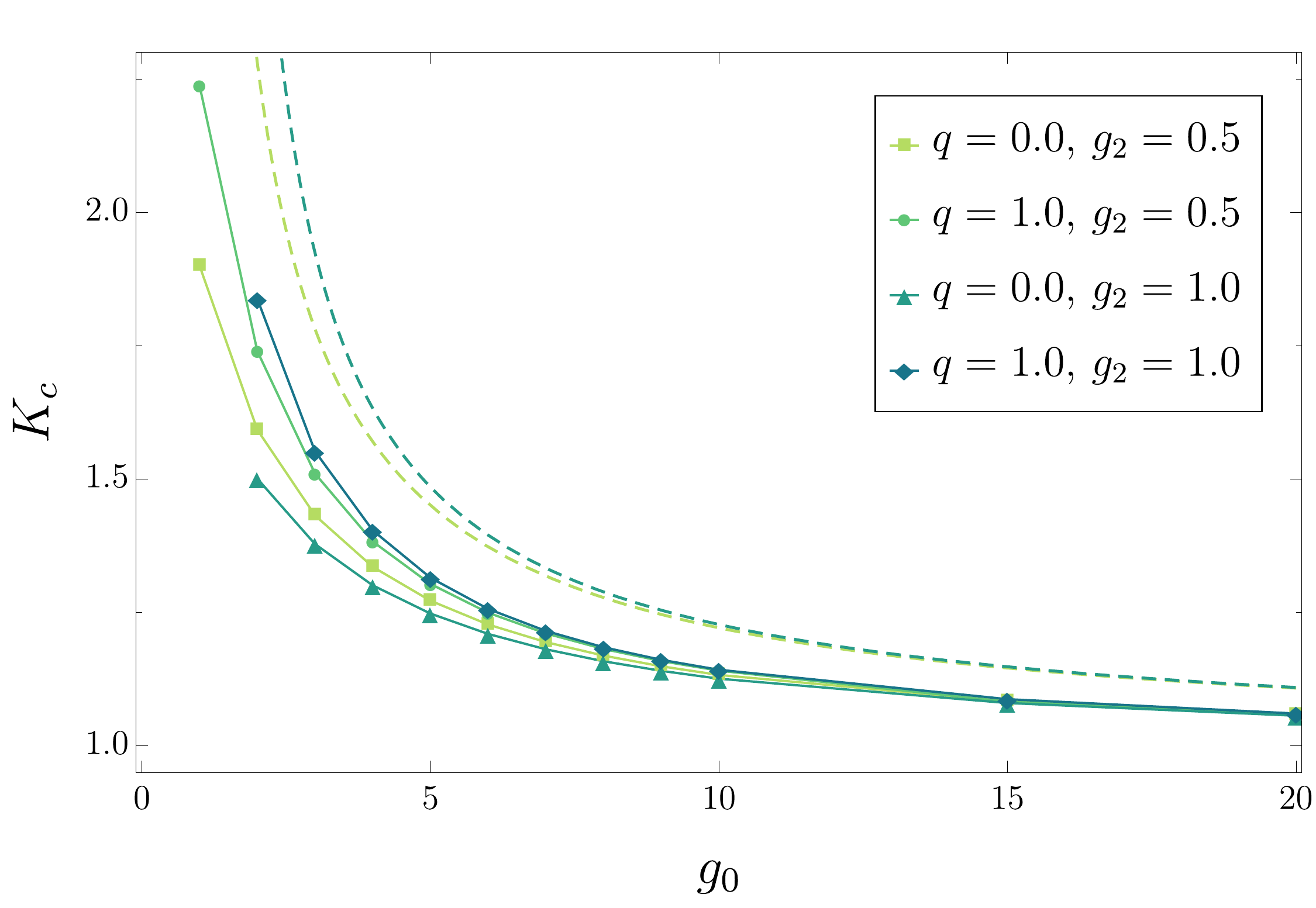} 
\caption{The Luttinger parameter $K_c$ in the charge sector for $g_2 = 0.5t$ and $g_2 = 1.0t$ obtained from numerical calculations, while $g_0 = 5.0t$ is fixed. The dashed curves denote the plot for $q = 0$ and $K_c$ in the strong coupling limit from analytical calculations. Note that $K_c >1$ for a wide range of parameters in this limit.}
\label{fig:Kc}
\end{figure}

From the power-law fit of the correlations we extract the Luttinger parameter $K_c$ for various values of $g_{0}, g_2,$ and $q$.
We find $K_c$ has a very weak dependence on $q$, whereas the $g_0$ and $g_2$ dependence is prominent. This is in agreement with the analytical results, according to which the $q$ dependence enters only via weak fluctuation corrections, Eq.~\eqref{eq:Kofq}.
To understand this we make a comparison between the numerics and the analytical expectation $K_c = \pi\sqrt{\rho_0/[m \tilde g_0]}$ for $q=0$, based on Tab.~\ref{tab:Parameters}.
The result is presented in Fig.~\ref{fig:Kc}, which demonstrates good qualitative agreement between the two.
Moreover, this shows $K_c > 1$ in the wide parameter regime of strong coupling. 

\subsubsection{Nematic correlators}

\begin{figure}[t]
\centering
\vspace{3pt}
\includegraphics[width=.98\linewidth]{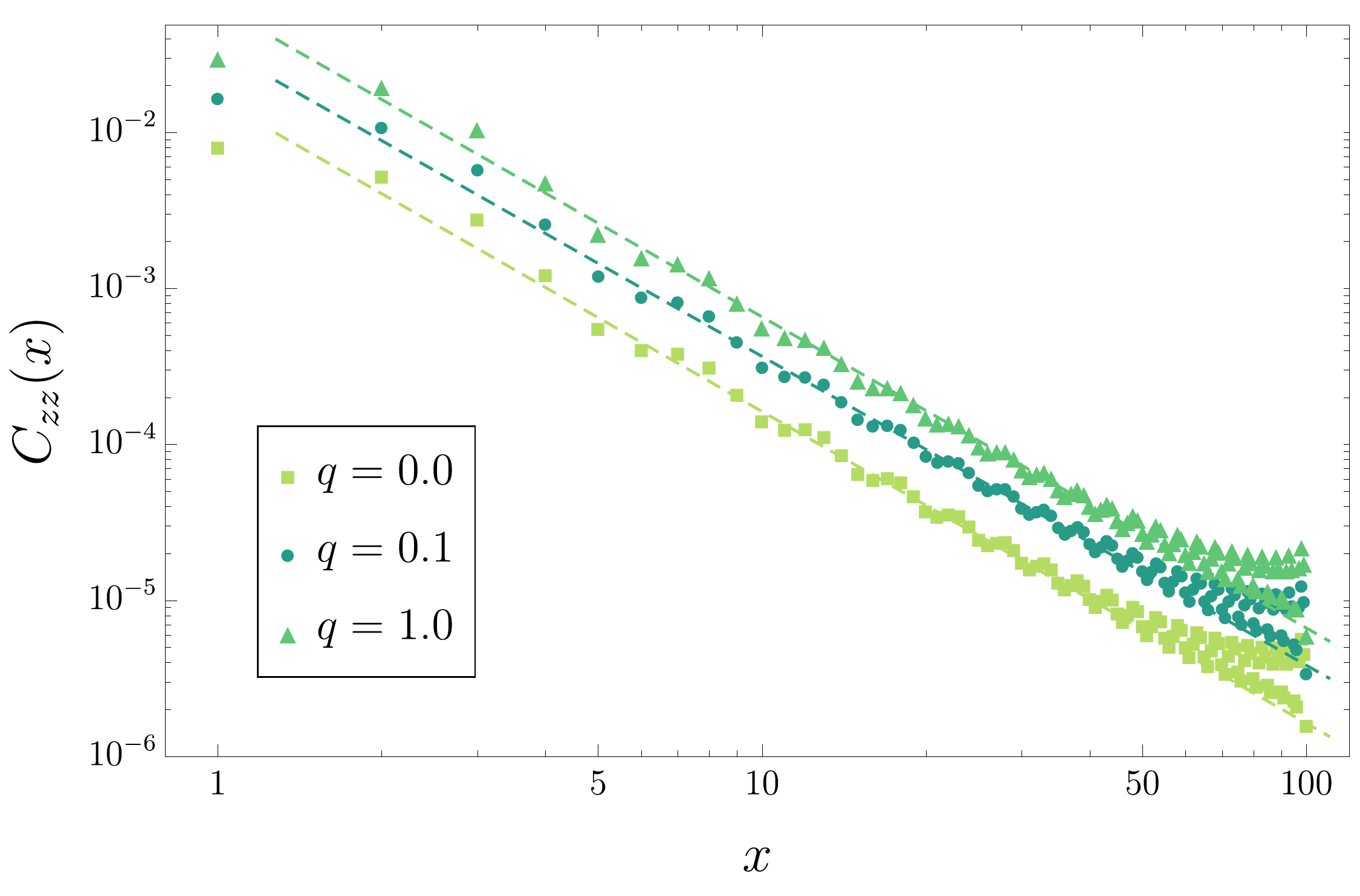}
\caption{Nematic correlator $C_{zz}(x)$ in log scale for $g_0 = 5.0t$, $g_2 = 0.5t$, $h = 0.1t$, and $\Theta = 0$. The dashed lines indicate linear fits in the log-log plot, of which the slope is close to $-2$ (precise values are $-1.998$, $-1.980$, and $-1.994$ respectively for $q=0.0t$, $0.1t$, and $1.0t$). In the strong coupling regime under investigation, the behavior is entirely due to the power-law decay of density-density correlations, see Eq.~\eqref{eq:NematicCorrelator}. In the present case of $\rho_0 = 1/5$, $K_c$ is larger than 1 and the long range asymptotics is given by $C_{zz} \simeq x^{-2}$, which is in good agreement with the numerical calculation. }
\label{fig:NematicCorrel}
\end{figure}

We now turn to the behavior of the nematic correlation function, and compute the connected correlation function that is defined as
$C_{zz}(x) = \langle :N_{zz}(x):: N_{zz}(0) : \rangle$, where $:N_{zz}(x):\, = N_{zz}(x) - \langle N_{zz}(x) \rangle$ denotes normal ordering. Within the field theory description, we can understand the individual charge and spin contributions to $C_{zz}(x)$ by introducing source fields $q \rightarrow q + \delta q (x)$ in Eq.~\eqref{eq:LF}, \eqref{eq:Lvortex} and appropriately differentiating with respect to $\delta q(x)$, before taking the limit $\delta q(x) \rightarrow 0$ at the end. This generates a vertex $\rho(x) \hat n(x) S_z^2 \hat n(x)$ whose relation to $C_{zz}(x)$ is given by:
\begin{equation}
C_{zz}(x) = \langle :\rho(x) \hat n(x) S_z^2 \hat n(x): : \rho(0) \hat n(0) S_z^2 \hat n(0): \rangle.
\label{eqn:czz}
\end{equation}
Thus, the nematic correlation function receives a contribution from the spin $\hat n(x)$ and the charge $\rho(x)$ sectors of the field theory.

Since the spin sector is gapped, integrating out $\hat n$ we obtain:
\begin{equation}
C_{zz} (x) \propto \langle \delta \rho(x) \delta \rho(0) \rangle \propto \frac{K_c}{2\pi^2 x^2} + \mathcal C \frac{\cos(2\pi \rho_0  x)}{ x^{2K_c}},  \label{eq:NematicCorrelator}
\end{equation}
where $\delta \rho(x)  = \rho(x) - \rho_0$ and $\mathcal C$ is a non-universal constant  $\mathcal C \propto y^2$ [see Eq.~\eqref{eq:Lvortex} and below for definition of $y$]. This result has important implications from the value of the Luttinger parameter.
When the Luttinger parameter obeys $K_c >1$ as in $\Theta = 0$ [Fig.~\ref{fig:Kc}], the asymptotic power law regime for $x \gg 1$ is dominated by the $1/x^2$ contribution, while the second term stemming from Eq.~\eqref{eq:Lvortex} is subdominant and only generates weak oscillations in the amplitude. On the other hand, $K_c <1$ implies the oscillatory second term dominates $C_{zz}$ and thus the ground state will be in a charge density wave state with a wave vector $Q_{\rm CDW} = 2\pi \rho_0$. 
For $\Theta = 0$, we calculate $C_{zz}$ in Fig.~\ref{fig:NematicCorrel} which shows a $1/x^2$ power-law decay, consistent with $K_c >1$ from the previous section. 
A $q$-independent weak oscillation with a wave vector $2\pi \rho_0$ is also apparent from the data.

\subsection{Effect of the spin-orbit coupling: $h, q, \Theta \neq 0$}

We move on from the spatial uniform transverse magnetic field and now consider the effect of a SOC on the strong coupling superfluidity of polar spin-1 bosons, by considering $\Theta \neq 0$. 
As we consider a regime with a robust gapped spin-liquid phase, the physics with the SOC is very rich and a correlated charge density wave state also appears.

\subsubsection{Spin and nematic texture}

\begin{figure}
\centering
\ig[width=0.48\linewidth]{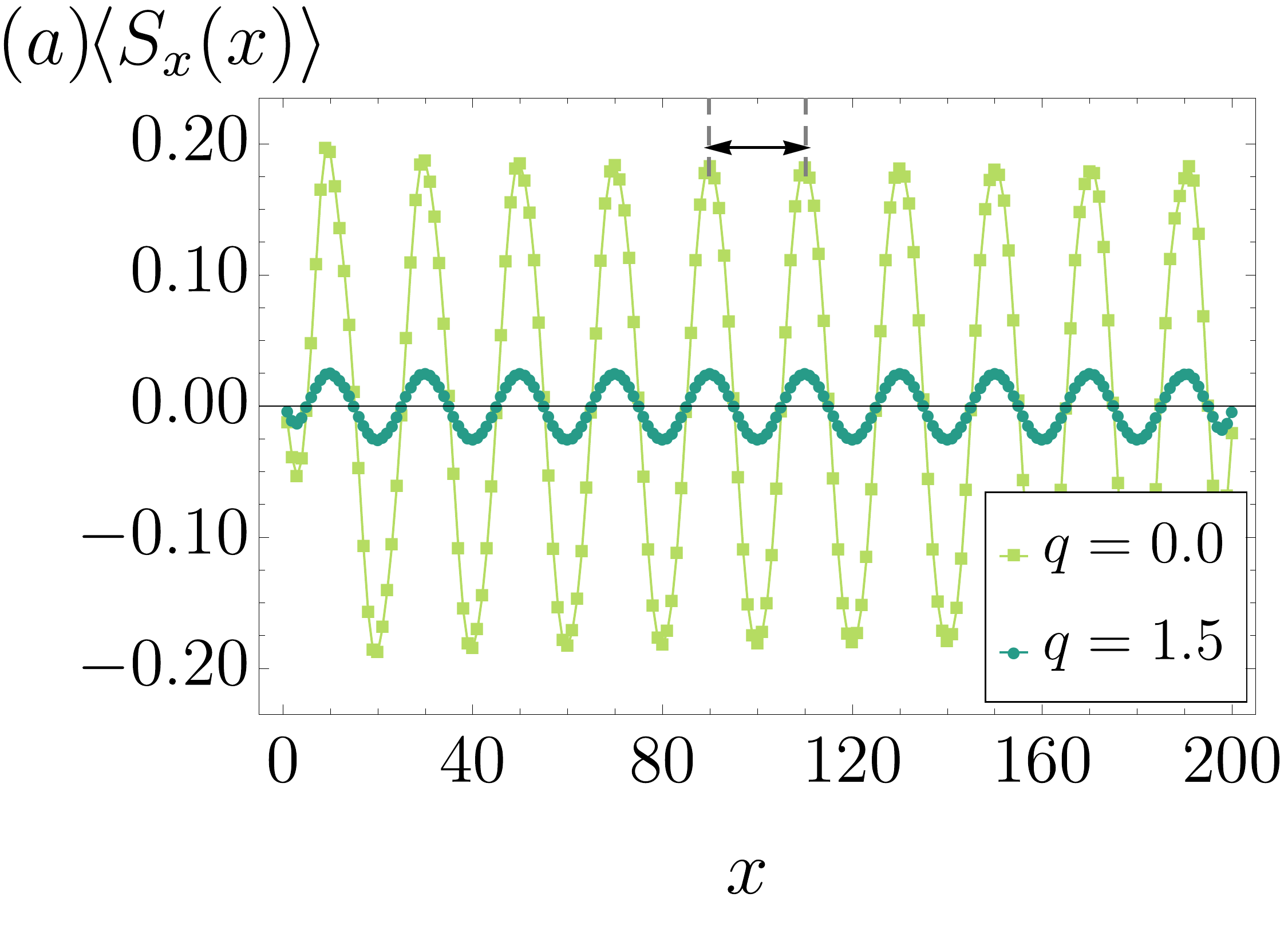} 
\ig[width=0.48\linewidth]{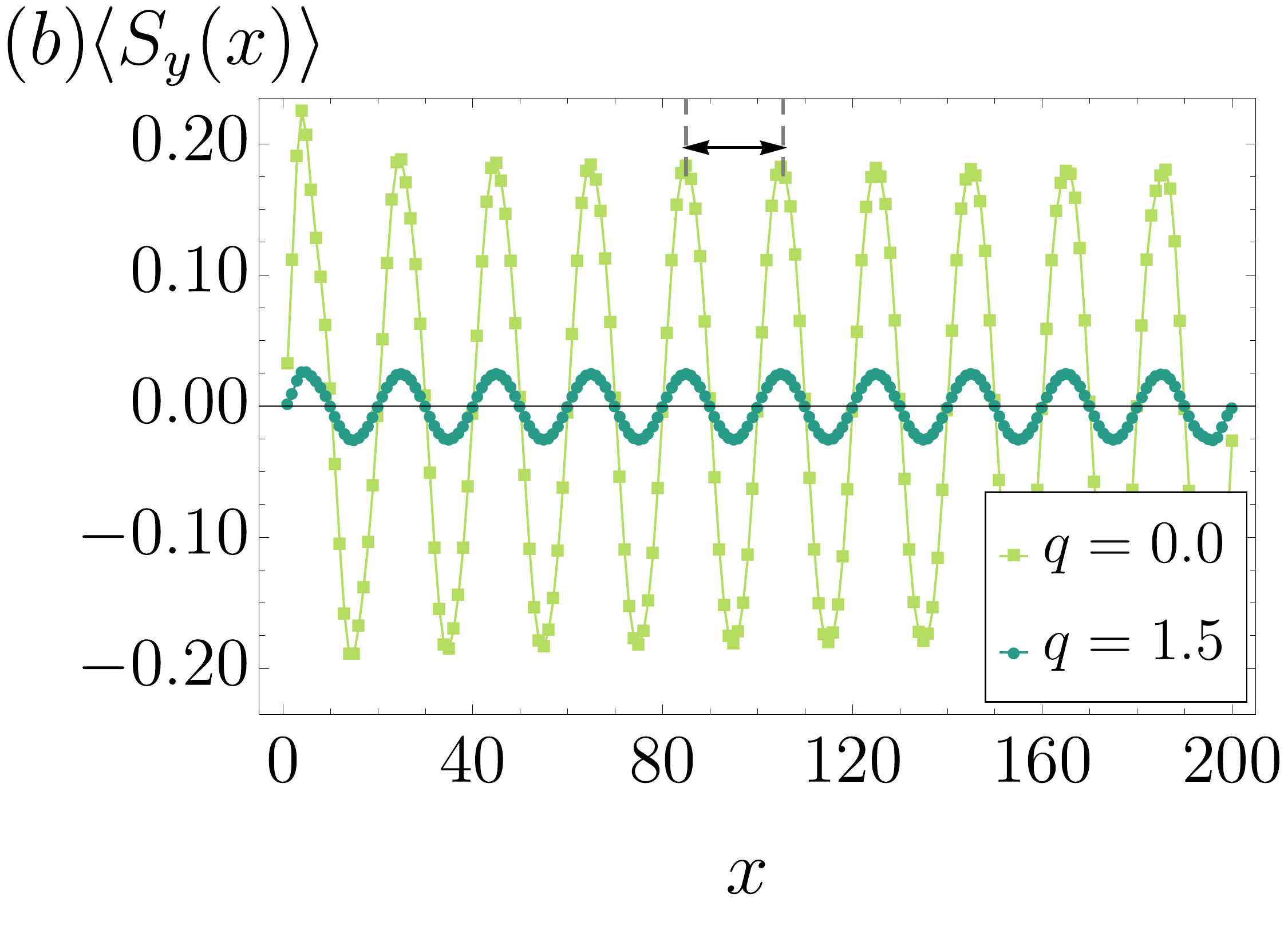}
\caption[]{Spin expectation values (a) $\langle S_x(x)\rangle$ and (b) $\langle S_y(x)\rangle$ for $g_0 = 5.0t$, $g_2 = 0.5t$, $\Theta = \pi /10$, and two values of $q=0.0t$, $1.5t$. Note the oscillation wavelength $2\pi/ \Theta$ is indicated as an arrow between two maxima, and the nonzero $q$ suppresses the oscillation. } \label{fig:Graph35} 
\end{figure}

\begin{figure}[t]
\centering
\ig[width=0.48\linewidth]{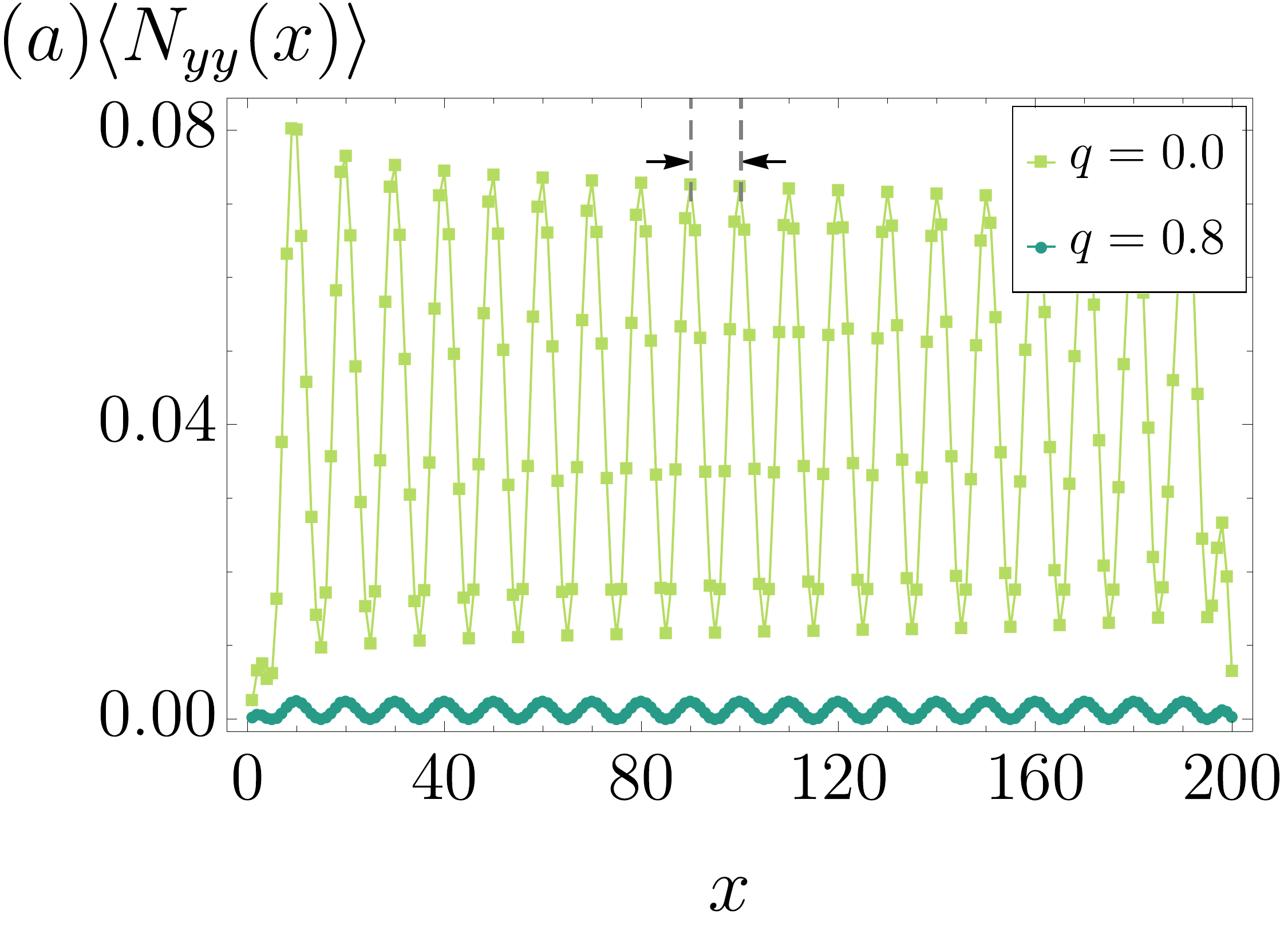}
\ig[width=0.48\linewidth]{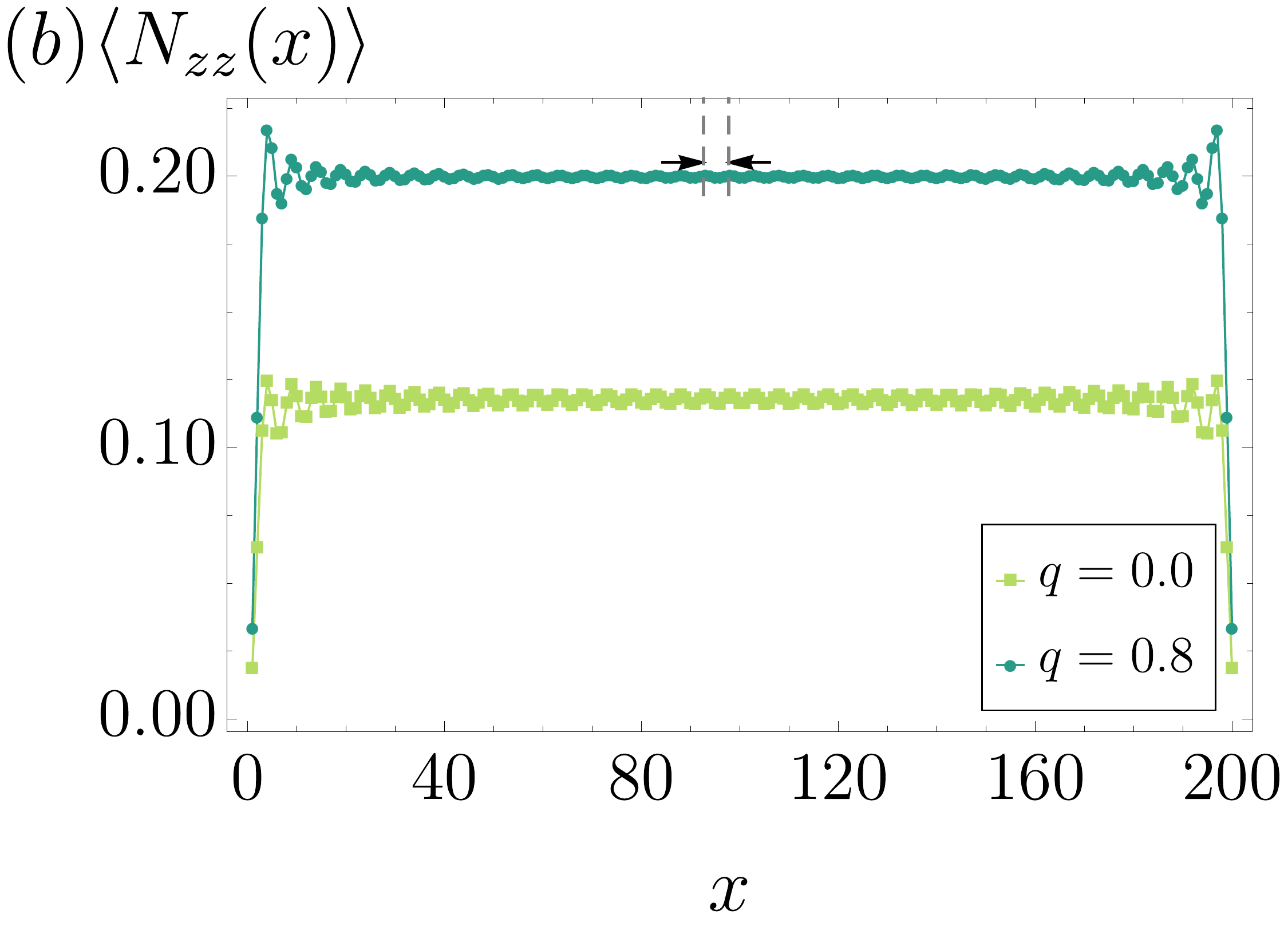}
\caption[]{Elements of the nematicity tensor (a) $\langle N_{yy}(x) \rangle$ and (b) $\langle N_{zz}(x) \rangle$ for $g_0 = 5.0t$, $g_2 = 0.5t$, $\Theta = \pi /10$ and two values of $q=0.0t$, $0.8t$.  Note the oscillation wavelengths (a) $2\pi/2\Theta = 10$ and (b) $1/\rho_0 = 5$ indicated in the figures which are both different from the wavelength of Fig.~\ref{fig:Graph35}.  } \label{fig:Graph37} 
\end{figure}

As in the transverse field case, we start our analysis with the spin and nematic expectation values. 
In the lab frame, the average spin component will try to locally anti-align with the magnetic field along the chain.
As a result of the finite SOC (with wave vector $\Theta = \pi/10$ in the DMRG), the spin expectation values $\langle S_{x}(x) \rangle$ and $\langle S_{y}(x) \rangle$ follow the pattern of the helical magnetic field. 
Explicit forms are given by:
\begin{align}
	\langle S_{x}(x) \rangle &= -A_x\cos(\Theta x), \nonumber\\
	\langle S_{y}(x) \rangle &= A_y\sin(\Theta x ). 
\end{align}
$A_x$ and $A_y$ are the amplitudes for each spin expectations. 
This functional form can be understood analytically by considering the transformation of the $\hat S_{x}$ and $\hat S_{y}$ operators from the lab frame to the rotating frame using Eq.~\eqref{eqn:rotateC}.
On the other hand, the spin component perpendicular to the field is suppressed due to $g_2>0$ and we find $\langle S_{z}(x) \rangle \approx 0$.
In Fig. \ref{fig:Graph35} we show plots of $\langle S_{x}(x) \rangle$ and $\langle S_{y}(x) \rangle$ for two different values of $q$, showing oscillations at the wavelength of the SOC. 
The oscillation is suppressed by the quadratic Zeeman field as expected. 

Upon the unitary transformation from the lab to rotating frame of the bosonic operators, the nematicity tensors of components $\langle N_{xx}(x) \rangle$ and $\langle N_{yy}(x) \rangle$ pick up a contribution from the spatially dependent phase factor 
that is not present for $\Theta = 0$. 
The functional forms are obtained as 
\begin{subequations}
\begin{align}
	\langle N_{xx}(x) \rangle &= A_{xx} - B_{xx}\cos(2\Theta x+\phi_{xx}),  \\
	\langle N_{yy}(x) \rangle &= A_{yy} + B_{yy}\cos(2\Theta x+ \phi_{yy}),
\end{align}
\label{eq:NOscillations}
\end{subequations}
with amplitudes $A$, $B$, and phase $\phi$.
On the other hand, $\langle N_{zz} (x)\rangle$ remains invariant under the unitary transformation to the rotating frame and does not acquire any oscillatory behavior due to the SOC. 
Rather, the oscillations occur from the charge density modulation with a wave vector of $2\pi \rho_0$:
\beq
\langle N_{zz}(x) \rangle = A_{zz} + B_{zz} \cos(2\pi \rho_0 x + \phi_{zz}).
\eeq
This can be understood by considering the $\langle N_{zz} (x)\rangle$ being generated through the vertex $\rho(x) \hat n S_z^2 \hat n$ [see Eq.~\eqref{eqn:czz}]. 
In Fig. \ref{fig:Graph37} we show the nematic expectation values of $\langle N_{yy}(x) \rangle$ and $\langle N_{zz}(x) \rangle$.
We observe oscillations, which are suppressed with the quadratic Zeeman field, with different wavelengths originating from the SOC and charge density, respectively. 

\begin{figure}[t]
\centering
\ig[width=.95\linewidth]{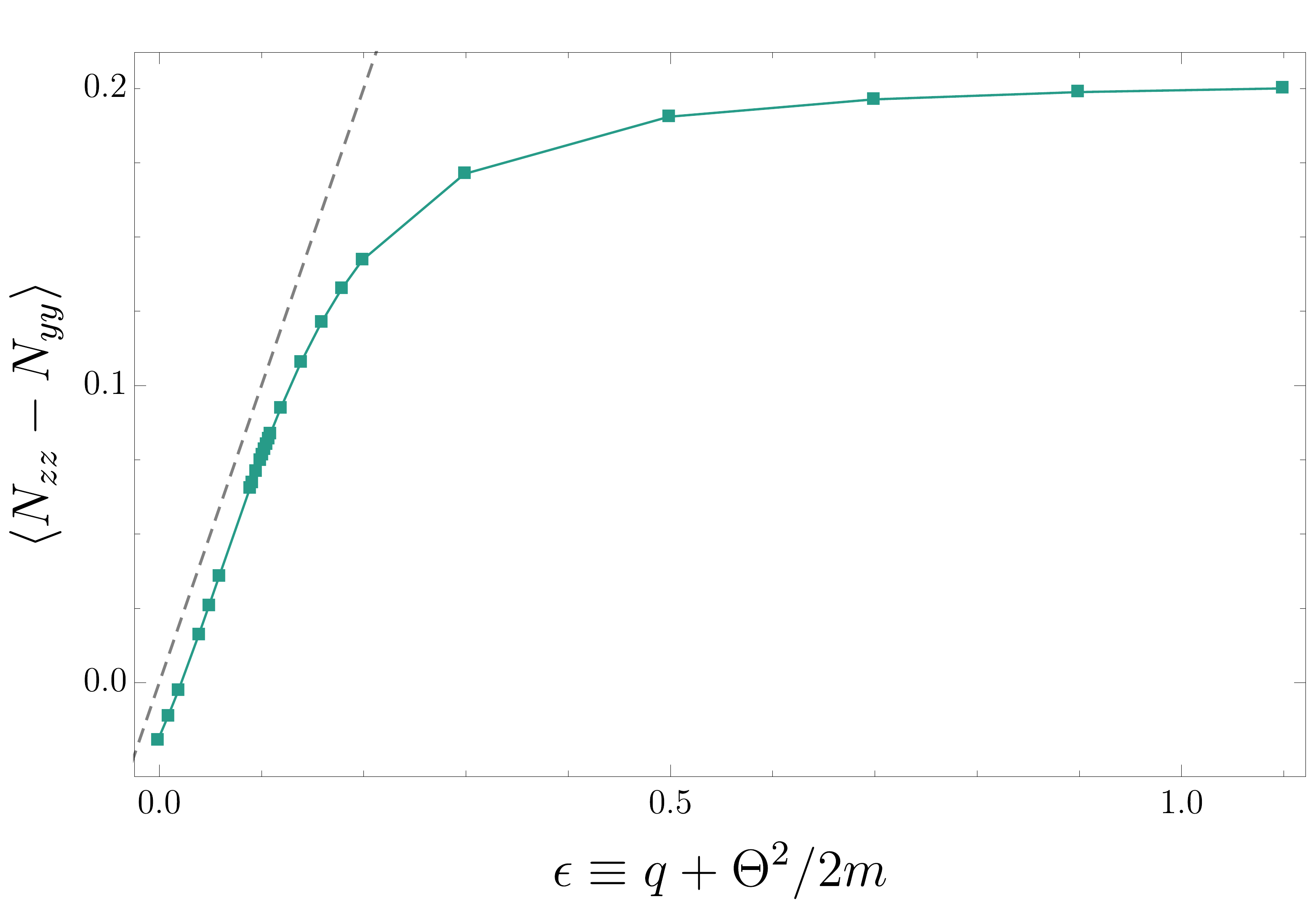}
\caption[]{Relative nematicity $\langle N_{zz} -  N_{yy}\rangle$ as a function of $\epsilon \equiv q + \Theta^2/2m$ for $g_0 = 5.0t$, $g_2 = 0.5t$, and $\Theta$. The gray dashed linear line is a guide to the eye with a slope of 1, showing $\langle N_{zz} -  N_{yy}\rangle\sim \epsilon$. } \label{fig:NzzNyyTheta} 
\end{figure}

To determine whether the model remains in the spin-liquid phase we use these functional forms to extract an estimate of the difference of the nematic expectation values $\langle N_{zz} -  N_{yy}\rangle$. However, since they both oscillate at different periods we first fit the data to the functional forms given in Eq.~\eqref{eq:NOscillations}, and then determine $\langle N_{zz} -  N_{yy}\rangle$ via the following procedure: we evaluate $\langle N_{zz}\rangle$ by averaging $\langle N_{zz}(x)\rangle$ over the lattice (we exclude some sites at the boundary during averaging to avoid boundary effects), we extract $A_{yy}$ from the fit of $\langle N_{yy}(x) \rangle$ to the functional form above and use $\langle N_{zz}\rangle - A_{yy}$ as a proxy for $\langle N_{zz} -  N_{yy}\rangle$. 
 We expect that $\langle N_{zz} -  N_{yy}\rangle$ vanish linearly in the spin liquid regime like $\langle N_{zz} -  N_{yy}\rangle\sim \epsilon$~\cite{KoenigPixley2018}, where $\epsilon = q + \Theta^2/2m$. As shown in Fig.~\ref{fig:NzzNyyTheta} we find good agreement with this vanishing linearly with $\epsilon$, however due to the oscillation periods being distinct this leads to a non-perfect estimate of $\langle N_{zz} -  N_{yy}\rangle$ and shifts the zero away from $\epsilon=0$. 

\subsubsection{Entanglement entropy}

\begin{figure}[t]
\centering
\ig[width=.95\linewidth]{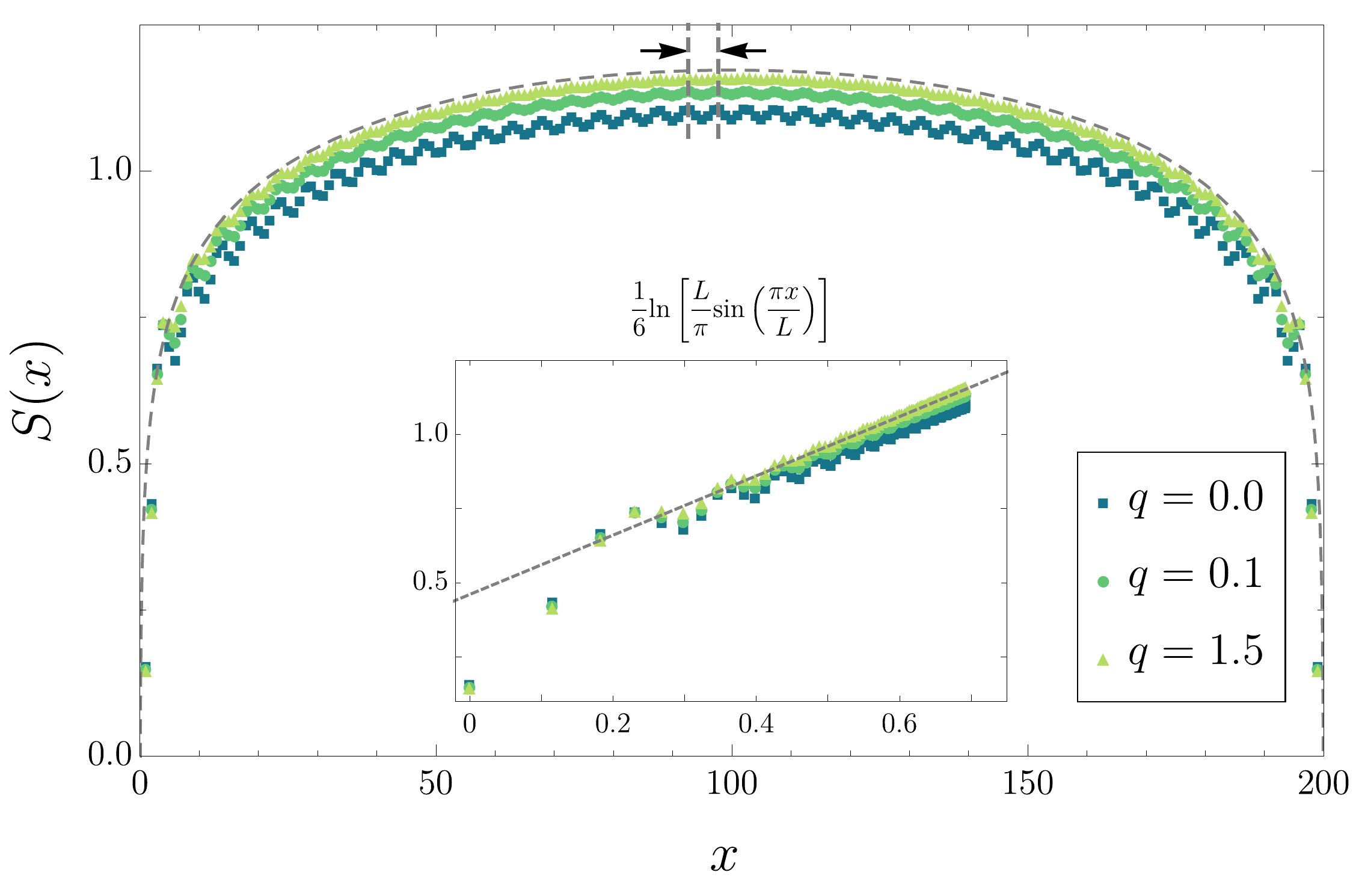} 
\caption[]{The entanglement entropy as a function of the bipartite position $x$ for $g_0 = 5.0t$, $g_2 = 0.5t$, and $\Theta = \pi/10$. The inset is the same data with the scaled horizontal axis to obtain the central charge from the slope (Eq.~\eqref{eq:EE}). The gray dashed lines are guide to the eye which correspond to $c=1$, while explicit linear fits gave $c = 0.995$, $1.002$, and $1.004$ for $q = 0.0t$, $0.1t$, and $1.5t$ respectively.} 
\label{fig:SETheta} 
\end{figure}
We again look at the entanglement entropy and calculate the central charge for additional evidence of the spin gap. 
As shown in Fig.~\ref{fig:SETheta}, we find that entanglement entropy is very weakly affected by a quadratic Zeeman field and obtain a central charge $c\approx 1$ from the linear fit of $S(x)$ versus $\log (\frac{L}{\pi}\sin(\frac{\pi x}{L}))$ [see Eq.~\ref{eq:EE}], 
which is in excellent agreement with the expectation that the spin sector remains gapped and the only gapless modes are due to the superfluidity in the charge sector. This also is in agreement with our results for $\Theta=0$ [Sec.~\ref{sec:ee0}], thus we conclude the model remains in the spin-liquid phase even in the presence of a full SOC. 

However, comparing with the case of $\Theta=0$, we find that the oscillations in the entanglement entropy are much larger in the case of nonzero SOC. These oscillations occur with a period given by $1/\rho_0$ and are thus due to the oscillation in the charge density. As we demonstrate below, the SOC induces a charge density wave of period $1/\rho_0$ due to the Luttinger liquid in the charge sector having $K_c <1$. [See also Eq.~\eqref{eq:NematicCorrelator} and the discussion below]

\subsubsection{Bosonic correlators and Luttinger parameter} 
\label{sec:so_LL}

\begin{figure}[t]
\centering
\ig[width=0.95\linewidth]{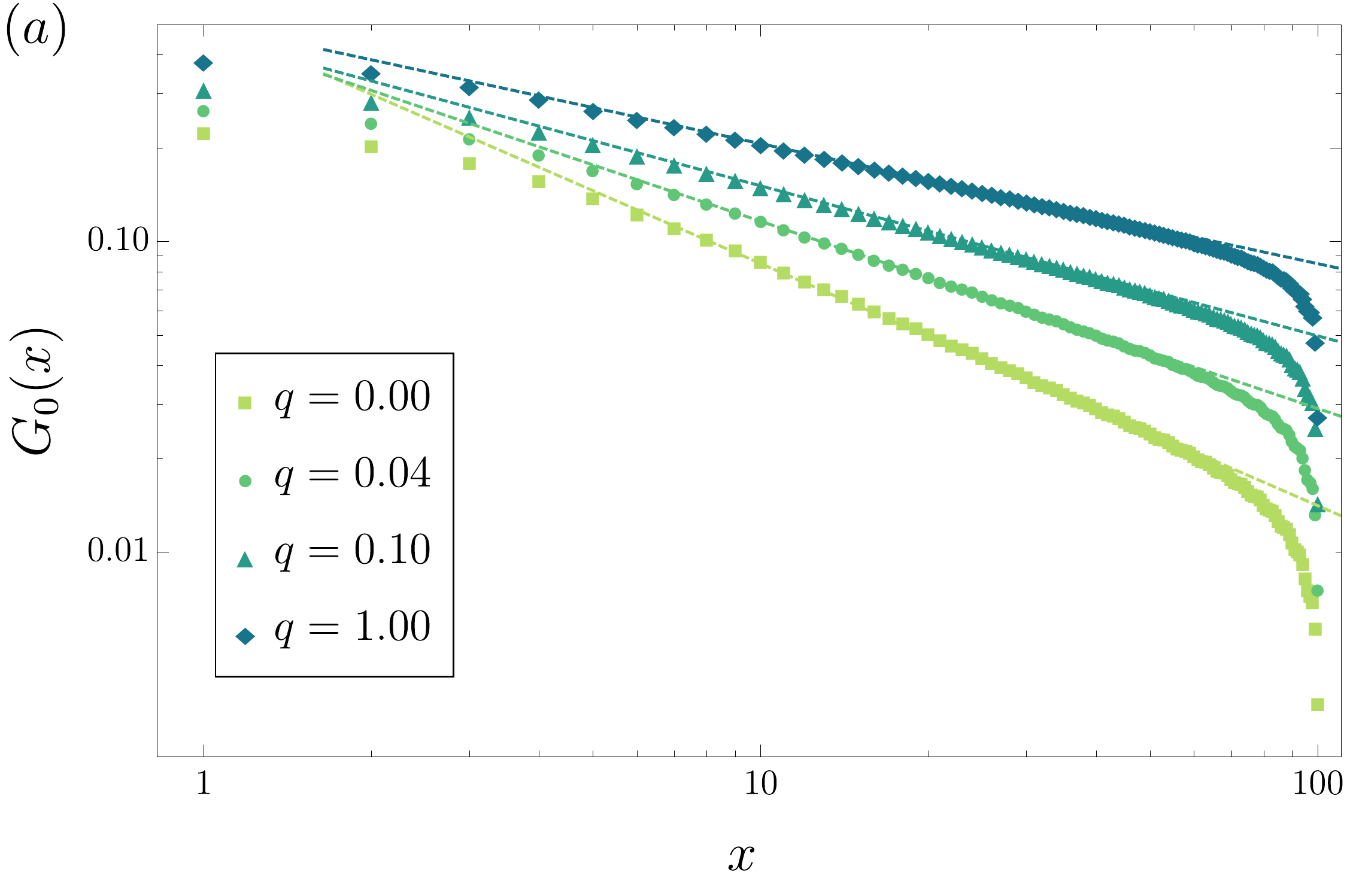}\\
\ig[width=0.95\linewidth]{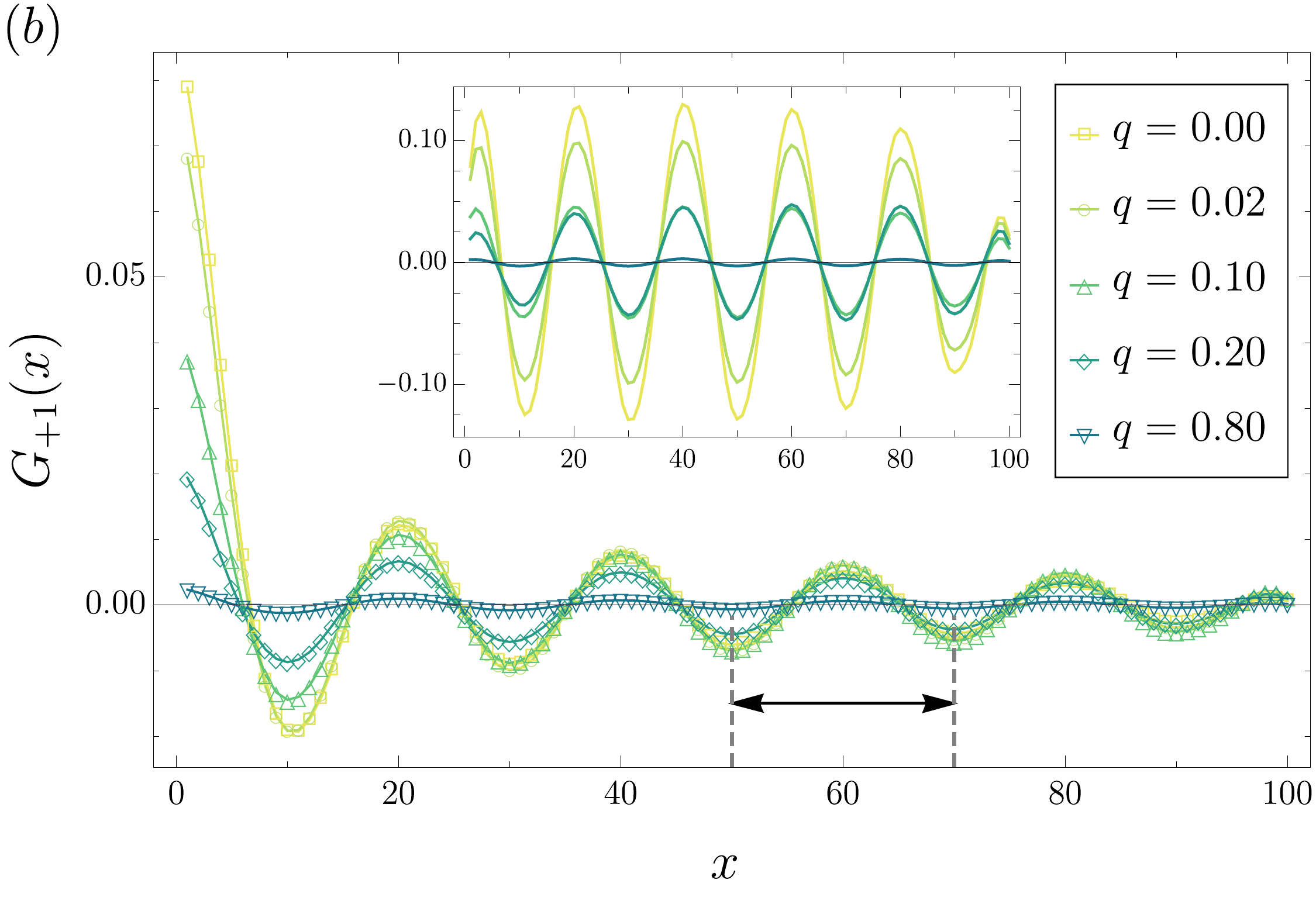} 
\caption[]{(a) Log-log plot of spin-0 bosonic correlator $G_0 (x)$ for $g_0 = 5.0t$, $g_2 = 0.5t$, and $\Theta = \pi/10$. The dashed lines are linear fits in the log-log plot indicating power law behavior ($G_0 (x) \sim x^{-0.78}$, $x^{-0.60}$, $x^{-0.48}$, and $x^{-0.39}$ for $q=0.00t$, $0.04t$, $0.10t$, and $1.00t$). (b) Plot of spin-(+1) bosonic correlator $G_{+1} (x)$ for $g_0 = 5.0t$, $g_2 = 0.5t$, and $\Theta = \pi/10$. The inset is the modified data $x^{1/(2K_c)}G_{+ 1}(x)$ showing the oscillatory part in Eq.~\eqref{eqn:G1}, where the data points are omitted for clarity. 
} 
\label{fig:GTheta} 
\end{figure}

\begin{figure}
\centering
\ig[width=0.95\linewidth]{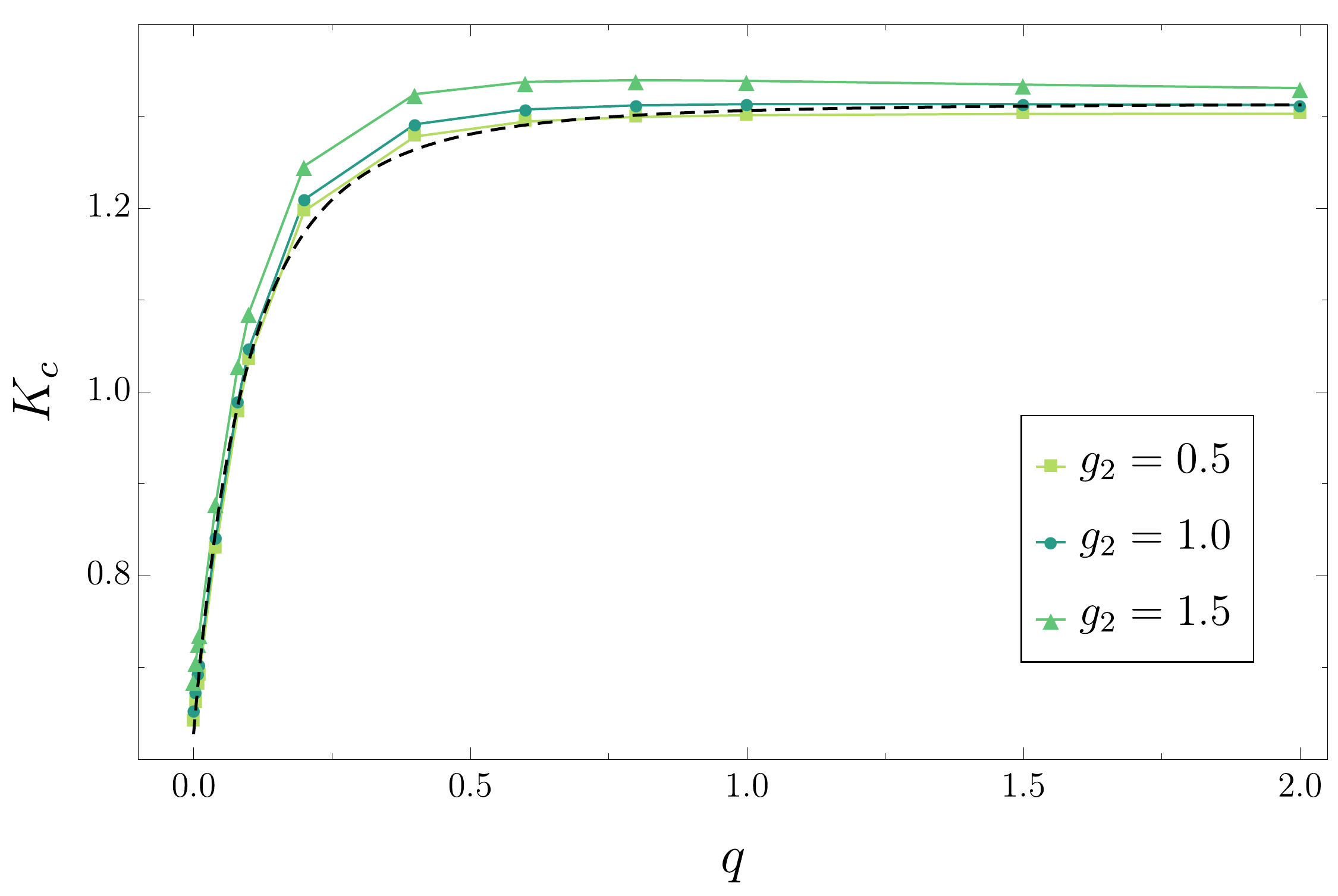}
\caption{$K_c$ extracted from power law fits plotted as a function of $q$, with fixed $g_0=5.0t$. The black line indicates a fit for $g_2 = 0.5t$ data against Eq.~\eqref{eq:deltaKcsimple}. The fitted parameters are $A \sin^2(\Theta\xi_s) = 0.645$ and $B = 3.48$.}
 \label{fig:KcTheta} 
\end{figure}

We now turn to the bosonic Green function of each spin state, see Fig.~\ref{fig:GTheta}. For the spin-0 component these are given by 
$G_{0}(x) = \langle b_{0}^{\dg}(x)b_{0}(0) \rangle$ and the spin-($\pm 1$) component of the bosonic correlator is 
$G_{\pm 1}(x) = \langle b_{\pm1}^{\dg}(x)b_{\pm1}(0) \rangle$. 
Applying the transformation from the rotating frame to the lab frame allows us to deduce the functional form of $G_{\alpha}(x)$. Since the spin-0 component is unaffected by this transformation, the form remains as in Eq.~\eqref{eq:G}:
\begin{equation}
G_{0}(x) \sim x^{-1/(2K_c)}. 
\label{eqn:G0}
\end{equation}
The DMRG results for $G_0 (x)$ are presented in Fig.~\ref{fig:GTheta}(a) and we extract the Luttinger parameter $K_c$ from a fit to the power-law form. 
Interestingly, distinct from the case with $\Theta=0$, we now find that $K_c$ strongly depends on the quadratic Zeeman field. As shown in Fig.~\ref{fig:KcTheta}, our data fits remarkably well to a simplified variant of the field theoretical result Eq.~\eqref{eq:Kofq}
\begin{equation}
K_c^{\rm eff} \sim K_c- \frac{A \sin^2(\Theta\xi_s)}{(1 + B q)^3}, \label{eq:deltaKcsimple}
\end{equation}
with two fitting parameters $A \sin^2(\Theta \xi_s)$ and $B$. 

In contrast to the spin-0 Green function, the spin-$(\pm 1)$ components do alter as we transform to the lab frame
\beq 
G_{\pm 1}(x) \sim \frac{\cos(\Theta x + \alpha)}{x^{1/(2K_c)}}.
\label{eqn:G1}
 \eeq 
This suggests that the power-law form is identical to the spin-0 case but it acquires an oscillatory component due to the SOC, consistent with the data shown in Fig.~\ref{fig:GTheta} (b).  To demonstrate this, we first extract $K_c$ from $G_0(x)$ using Eq.~\eqref{eqn:G0} and then plot $x^{1/(2K_c)}G_{+ 1}(x)$ in the inset, which does not decay and oscillates with a period $2 \pi /\Theta$ thus confirming the functional form in Eq~\eqref{eqn:G1}. Lastly, the positive quadratic Zeeman field strongly suppresses $G_{\pm 1}(x)$ as expected.

The extracted Luttinger parameter in the charge sector $K_c$ for a finite $\Theta$ as a function of the quadratic Zeeman field is given in Fig.~\ref{fig:KcTheta} for various values of $g_2$. This demonstrates that the finite SOC leads to $K_c <1$ in small $q$, which  induces a charge density wave state due to the functional form of the charge correlation function [see Eq.~\eqref{eq:NematicCorrelator} and the discussion below]. By applying a large quadratic Zeeman field, the effect of SOC and thus the charge density wave is suppressed inducing a crossover from $K_c <1$ to $K_c >1$. The proximate charge density wave regime is the reason that the Luttinger parameter is so sensitive to tuning $q$ in contrast to the limit of $\Theta = 0$. The charge density wave can be clearly seen in the nematic correlation function, which we now turn to.

\begin{figure}
\centering
\includegraphics[width=0.95\linewidth]{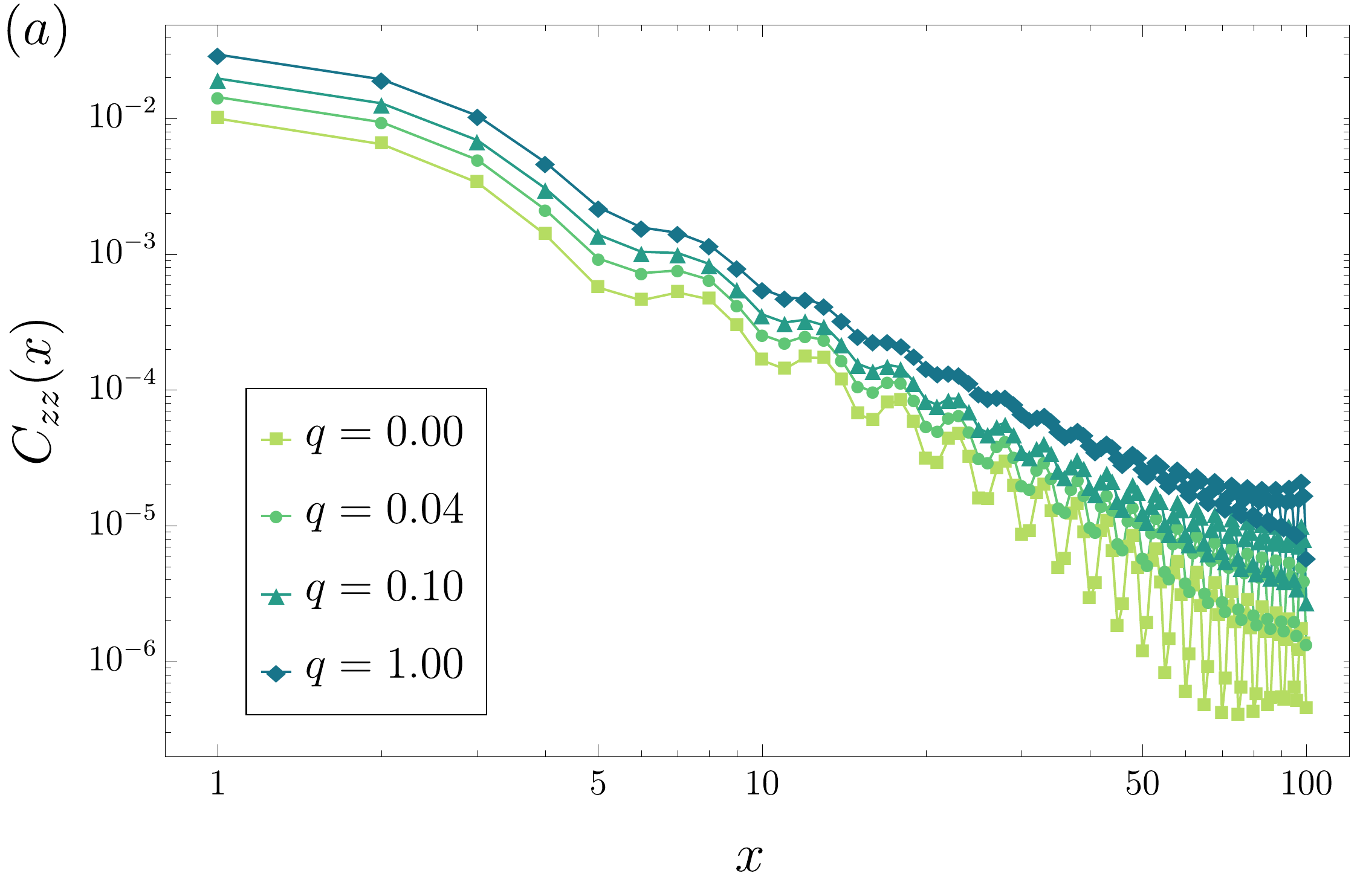} \\
\includegraphics[width=0.95\linewidth]{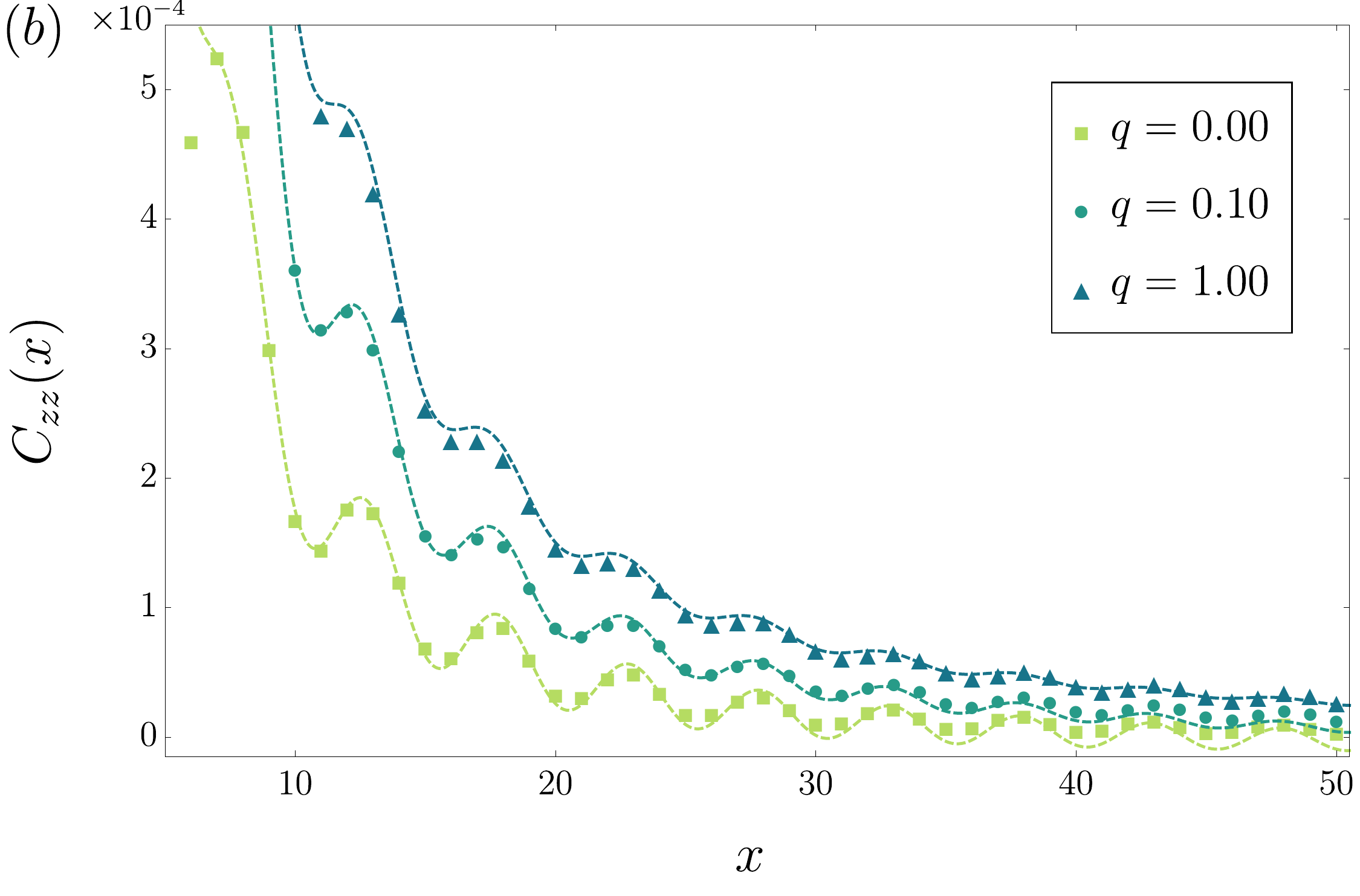}
\caption{(a) Log-log plot and (b) linear plot of the nematic correlator $C_{zz}(x)$ for $g_0 = 5t$, $g_2 = 0.5 t$, $h = 0.1 t$, and $\Theta = \pi/10$. We can observe the power-law decay from density-density correlations in (a) as in Fig.~\ref{fig:NematicCorrel} (Eq.~\eqref{eq:NematicCorrelator}). However, with nonzero SOC $K_c$ is reduced below $1$ and the long range asymptotics is given by $C_{zz} \simeq \cos(2\pi \rho_0 x)x^{-2K}$ which is seen as enhanced oscillations for large $x$. The dashed lines in (b) are fitted results to Eq.~\eqref{eq:NematicCorrelator}.}

\label{fig:NematicCorrelSOC}
\end{figure}


\subsubsection{Nematic correlators}
As a result of the SOC driving $K_c < 1$, we expect that the nematic correlation function in Eq.~\eqref{eq:NematicCorrelator} is dominated by the oscillating term with a power law given by $2 K_c$. 
We demonstrate this by plotting the nematic correlator $C_{zz}(x)$ for a number of different values of $q$ in the presence of SOC in Fig.~\ref{fig:NematicCorrelSOC}(a). 
For $q=1.0$, we can check from Fig.~\ref{fig:KcTheta} that $K_c > 1$ and $C_{zz}(x)$ show similar behavior as in $\Theta = 0$ case. 
However, as we decrease $q$ to the regime where $K_c <1$ in Fig.~\ref{fig:KcTheta}, we find that oscillations enhance as well as the power of the decay changes. 
If we use the $K_c$ value extracted from $G_0 (x)$ [Fig.~\ref{fig:KcTheta}] to Eq.~\eqref{eq:NematicCorrelator}, we find excellent agreement between the numerics and the functional form, which is shown in Fig.~\ref{fig:NematicCorrelSOC}(b). 
This also confirms the emergence of a charge density wave from SOC with the wave vector $Q_{\rm CDW}=2\pi \rho_0$.
Thus, we reach one of our main conclusions: In the presence of large interactions a SOC induces a strong coupling charge density wave phase in dilute polar superfluids.

\section{Discussion}
\label{sec:discussion}

In summary, we have presented a combined numerical and analytical study of polar spin-1 lattice bosons at non-integer filling in one dimension under the influence of spin-orbit coupling and quadratic Zeeman field. Complementary to the previous study at weak coupling \cite{KoenigPixley2018}, we here concentrated on the limit when interaction effects are stronger than the kinetic energy. Our main finding, which is supported by the excellent agreement between analytics and  numerics, is that in this regime the spin-liquid gap is substantial and therefore the perturbative inclusion of symmetry breaking terms is insufficient to restore the algebraic nematic order. At the same time, the robustness of the spin-liquid phase does not render the spin sector entirely innocuous: we have demonstrated that spin-orbit coupling is capable of tuning the charge sector into a charge density wave by reducing the Luttinger parameter $K_c$ below unity. 

A qualitative explanation of this reduction of $K_c$ may be understood in the limit of large helical background magnetization $h$ and negligible quadratic Zeeman field $q$. We first discuss this limit in the case of vanishing spin-orbit wave vector $\Theta = 0$. Then, only the $b_{x}$ boson is of importance and our model displays conventional BEC of spinless bosons. We repeat that the superfluid stiffness is $K_c \sim \sqrt{t}  \rho_0 a/\sqrt{\tilde \mu}$. The first factor accounts for the intuitive increase in stiffness with increasing hopping strength while the remaining factors stem from the on-site mean field solution and are independent of the kinetics. Now we return to $\Theta \neq 0$, in the presence of the such a SOC the  BEC has a rotating on-site polarization. Therefore, the overlap of 
neighboring single-particle 
wavefunctions of adjacent sites is substantially weakened due to the spin dependent hopping and the numerator in $K_c$ is reduced.

Lastly, we conclude with the experimental realization of our strong coupling theory using ultracold gases of the polar spin-1 boson $^{23}$Na. A natural generalization of the experimental setup in Ref.~\cite{Jacob-2012} by including a one-dimensional optical lattice, should be able to straightforwardly realize the spin liquid phase we have uncovered in the limit of no spin orbit coupling in Sec.~\ref{sec:homogfields}. The ability to tune the quadratic Zeeman field across the nematic transition in the weak coupling limit implies such a transition can also be studied here. A clear cut signature of the spin liquid regime would be given by the difference in nematic expectation values~\cite{Zibold-2016} vanishing linearly with decreasing quadratic Zeeman field (as in Fig.~\ref{fig:DeltaN1}). The realization of
our newly discovered strong coupling charge density wave
phase that is induced by spin orbit coupling is in principle also possible within existing experimental setups. However, it requires long coherence times for $^{23}$Na atoms in miscible $F = 1$ hyperfine states -- a requirement which so far has been challenging due to strong magnetic noise. We are hopeful that the most recent experimental breakthrough in  shielding techniques has overcome this bottleneck~\cite{FarolfiFerrari2019}. Thus, we expect that a spin-orbit coupling can be induced in polar spin-1 bosons in the near future and the non-trivial  predictions of our theory can be tested. In particular, the strong coupling charge density wave can be observed either directly, through measuring the charge response via single-site imaging techniques~\cite{BakrGreiner2009,ShersonBloch2010} and Bragg scattering~\cite{MiyakeWeld2011,hartHulet2015}, or indirectly, using nematic tensor components~\cite{Zibold-2016}. 

\section*{Acknowledgments}
We thank R. Fernandes and J. Schmalian for pointing out the relationship to vestigial order. EJK is supported by DOE Basic Energy Sciences grant DE-FG02-99ER45790. JL is supported by NSF-PFC at the JQI. 
\appendix

\section{Derivation of effective field theory}
\label{app:fieldtheory}

\begin{table}[h]
\begin{tabular}{|c||l|l|l|}	
\hline
& n = 0 & n =1 & n = 2  \\
\hline \hline
S = 0 & $E_{0} = 0,$ &  &  $E_{2, 0} = -2\mu + g_0 - 2g_2,$ \\
 & $ \ket{\underline{0}}$ &  &  $ \ket{2,0}= \frac{1}{2} b_a^\dagger \lambda^{(0)}_{ab} b_b^\dagger \ket{\underline{0}}$ \\
\hline
S = 1 &  & $E_{1} = -\mu,$ & \\
 &  & $ \ket{1,a} = b_a^\dagger \ket{\underline{0}}$ & \\
\hline
S = 2 & & & $E_{2, 0} = -2\mu +  g_0+ g_2, $  \\
 & & & $ \ket{2,2,i}=\frac{1}{2} b_a^\dagger \lambda^{(i)}_{ab} b_b^\dagger \ket{\underline{0}}$  \\
\hline
\end{tabular}
\caption{Table of lowest eigenvalues and corresponding eigenstates of Eq.~\eqref{eq:Hloc}. $\lambda^{(i)}$ are the symmetric Gell-Mann matrices (i.e. $i = 0,1,3,4,6,8$), and we introduced $\lambda^{(0)} = \sqrt{\frac{2}{3}} \mathbf 1$}
\label{tab:States}
\end{table}

In this appendix we derive the Hamiltonian density of the effective field theory, Eq.~\eqref{eq:Fieldtheory}.

\subsubsection{Solution of local problem and molecular phase}

The weak coupling limit of Eq.~\eqref{eq:Fieldtheory} follows trivially from the continuum limit of Eq.~\eqref{eq:H0}. Therefore, this section focuses on the strong coupling limit, where we perturb about local eigenstates, Tab.~\ref{tab:States}. To determine the latter, note that $:\hat {\mathbf S}^2: = \hat {\mathbf S}^2 - 2 \hat n$ and $ : \hat n^2: = \hat n ( \hat n-1)$. Eigenvalues follow from  $\hat n \rightarrow n$ and $\hat {\mathbf S}^2 \rightarrow S (S+1)$ for conserved quantum numbers $n$ and $S$. The structure of eigenstates follows from
\begin{eqnarray}
:\hat {\mathbf S}^2: b^\dagger_c b^\dagger_d \ket{\underline{0}} &=& - \epsilon_{a b c'} \epsilon_{a' b ' c'} b^\dagger_a b_b b^\dagger_{a'} b_{b '} b^\dagger_c b^\dagger_d \ket{\underline{0}} \notag \\
&=&2(b^\dagger_c b^\dagger_d -  b^\dagger_a b^\dagger_a \delta_{c d}) \ket{\underline{0}}.
\end{eqnarray}

\subsubsection{Derivation of effective continuum field theory in the strong coupling limit}

As a first step, we decouple the hopping term
\begin{eqnarray}
H_{\rm kin} &=& - t(b^\dagger_i b_{i+1} + h.c.) 
\equiv - \vec b^\dagger { \underline t} \vec b\notag \\
&= & \bar \Psi_i (\underline t)^{-1}_{ij} \Psi_j  + [ b^\dagger_i \Psi_i + \bar \Psi_i  b_i] .
\end{eqnarray}
Note that $\bar \Psi_i (\underline t)^{-1}_{ij} \Psi_j \equiv \sum_k \bar \Psi(k) [2t \cos(k)]^{-1} \Psi(k) \simeq \bar \Psi_i (\underline t)^{-1}_{ij} \Psi_j \simeq \frac{1}{2t} \sum_k \bar \Psi(k)  \left[ 1 +\frac{k^2}{2}\right ] \Psi(k) $. This also demonstrates that the matrix $\underline t$ is positive definite in the infrared limit of interest~\cite{SachdevBook}.

The overall strategy is to derive an effective action for $\psi = \Psi/(E_1 \sqrt{a})$. To this end, we express the non-local term $\delta H_\Psi = [ b^\dagger_i \Psi_i + \bar \Psi_i  b_i]$ in the basis of $\lbrace \ket {\underline 0 }, \ket{a}, \ket{i} \rbrace$, where $\ket{i} = \frac{1}{2} b^\dagger_a \lambda^{(i)}_{ab} b_b^\dagger \ket{\underline 0}$

\begin{equation}
\delta H_\Psi = \left (\begin{array}{ccc}
0 & \bar \Psi_a & 0 \\ 
\Psi_a & 0 & (\bar \Psi \lambda^{(i)})_a \\ 
0&( \lambda^{(i)}  \Psi)_a  & 0
\end{array}  \right) .
\end{equation}
The matrix elements of $\delta H = b^\dagger \delta hb$, of Eq.~\eqref{eq:Hloc} in the single particle sector are obviously given by the matrix form of $\delta h$.

\subsubsection{Effective Action}

We begin with the derivation of the effective action $S[\psi]$ by focusing only on quadratic terms of the kind
\begin{eqnarray}
S^{(2)} &=& -  \Big \langle\frac{1}{2} \left [ \int d\tau \;\left ( b^\dagger_a \Psi_a + \bar \Psi_a  b_a\right) \right]^2 \Big \rangle_{S_{\rm loc}[b]} \\
&=&- \int d\tau d \tau' \bar \Psi_a (\tau) G_{ab}(\tau-\tau') \Psi_b(\tau').
\end{eqnarray}
The Green's function is $G_{ab} (\tau) = \langle T[b_a (\tau) b_b^\dagger (0)] \rangle]$, so that for $\tau >0$ 
\begin{eqnarray}
G_{ab} (\tau)
&=&\delta_{ab}\frac{ e^{-E_1 \vert \tau\vert}[\theta (\tau) + e^{- \beta E_1} \theta(- \tau)]}{1 + 3 e^{-\beta E_1}} \theta (\tau).
\end{eqnarray}
In the limit $E_1/T \rightarrow +\infty$, this leads to
\begin{eqnarray}
S^{(2)} &=& - \int d\tau \int_0^\infty d \Delta\tau \bar \Psi_a (\tau) e^{- E_1 \Delta \tau} \Psi_a(\tau - \Delta \tau) \notag \\
&=& -\int d \tau \frac{1}{E_1}\Bar \Psi_a [\Psi_a -\frac{1}{E_1} \dot \Psi_a + \frac{1}{E_1^2} \ddot{\Psi}_a] \\
&\simeq & \int d \tau d x \bar \psi \dot \psi - \frac{\dot{\bar \psi} \dot \psi}{\mu} - \tilde \mu \bar \psi \psi.
\end{eqnarray}
In addition to the conventional time derivative term $\bar \psi \dot \psi$ there is a term with two derivatives. However, in the interesting regime of time scales $\tau \Lambda_c \gg 1$ it is suppressed and henceforth omitted.

We now determine all other static terms in Eq.~\eqref{eq:Fieldtheory}. To this end, it is sufficient to consider time independent field configurations.
The bare partition function is
\begin{equation}
Z = 1 + 3 e^{-\beta E_1} + e^{-\beta E_{2,0}} + 5 e^{-\beta E_{2,2}}.
\end{equation}
We will consider sufficiently large $E_1/T,E_{2,0}/T,E_{2,2}/T \rightarrow \infty$, and will only keep the contribution of occupied states if the contribution of empty states vanishes.
We will further use the following identities:
\begin{align}
(\bar \Psi \bar\Psi^T)( \Psi^T \Psi) &= (\bar \Psi \Psi)^2 - (\bar \Psi \mathbf S \Psi)^2,\\
\sum_{i = 1,3,4,6,8} (\bar \Psi \lambda^{(i)} \bar\Psi^T)( \Psi^T \lambda^{(i)} \Psi) & = 
 \frac{4}{3} (\bar \Psi \Psi)^2 + \frac{2}{3} (\bar \Psi \mathbf S \Psi)^2.
\end{align}
We obtain the following perturbative correction to the ground state energy
\begin{align}
E &= - \frac{(\bar \Psi \Psi)}{E_1}  + \frac{(\bar \Psi \delta h \Psi)}{E_1^2}  - \frac{(\bar \Psi \Psi)^2- (\bar \Psi \mathbf S \Psi)^2}{E_{(2,0)}(E_1)^2} \notag \\
&- \frac{2}{3}\frac{2(\bar \Psi \Psi)^2 + (\bar \Psi \mathbf S \Psi)^2}{E_{(2,2)}(E_1)^2} + \frac{(\bar \Psi \Psi)^2}{E_1^3}.
\end{align}
Restoring slow time dependence of fields and $\int d \tau E = \delta S$ leads to the remaining terms in Eq.~\eqref{eq:Fieldtheory}.
Collecting all terms and rescaling $\Psi \rightarrow \psi$ leads to the identification of parameters of the field theory
\begin{eqnarray}
\tilde m &=& 2t/E_1^2, \\
\tilde \mu &=& E_1 \left [1-\frac{E_1}{2t}\right ],\\
\tilde g_0 &=& 2a^2 E_1 \left [1 - \frac{E_1}{E_{2,0}} - \frac{4 E_1}{3 E_{2,2}} \right], \\
\tilde g_2 &=& 2a^2 E_1 \left [\frac{E_1}{E_{2,0}} - \frac{2 E_1}{3 E_{2,2}} \right],
\end{eqnarray}
which leads to the expressions in Tab.~\ref{tab:Parameters} of the main text (the leading order in $\mu/t + 2\ll  1$ is presented there).\\

\section{Phase slips}
\label{app:Vortex}

In this appendix we derive the effective action of phase phase slips, Eq.~\eqref{eq:Lvortex}.
It is sufficient to consider the first three terms of Eq.~\eqref{eq:Lunpert} for the sake of this derivation. 
As mentioned in the main text, we introduce the field $\phi$ by means of $\delta \rho = - \phi'/\pi$. 
Amongst all boundary conditions of the fields, the important one is $\vartheta(x,\beta) = \vartheta(x,0) + 2\pi f(x)$ where $f(x) \in \mathbb Z \forall x $ is a piecewise constant function.
In order to introduce vortices in $\vartheta$ we split $\partial_\mu \vartheta = \partial_\mu \vartheta_{\rm reg} + A_\mu$, where the gauge potential accounts for vortices $\epsilon_{\mu \nu} \partial_\mu A_\nu = 2\pi \sum_{i} n_i \delta(\vec x - \vec x_i)$. It is convenient to choose a ``Landau'' gauge in which $A_\mu = (0, 2\pi \sum_i n_i \delta(\tau - \tau_i) \theta(x - x_i))$. Note that, contrary to usual Berezinskii-Kosterlitz-Thouless physics, also non-neutral configurations $\sum_i n_i \neq 0$ are consistent with the periodic boundary conditions and kept. To avoid double counting, we keep only vortices of $n_i = \pm 1$ but allow them to sit on top of each other (i.e. effectively creating double vortices). Furthermore, avoiding double counting also implies that we do not count permutations of equivalent sets of vortex positions $\lbrace \vec x_i \rbrace$ twice. 

We then obtain the Lagrangian as $\mathcal L_{\rm reg} + \delta \mathcal L$ where
\begin{align}
\mathcal L_{\rm reg} &=  - \frac{i}{\pi} \phi' \dot \vartheta + \frac{1}{2\pi} [{u}{K} (\vartheta')^2+\frac{u}{K} (\phi')^2], \\
\delta \mathcal L &= {i}\rho A_\tau \Rightarrow \delta S = -2 i \sum_{n_i} (\pi \rho_0 x_i - \phi(x_i, \tau_i)) .
\end{align}
We have dropped the $reg$ subscript in $\vartheta_{\rm reg}$. The total amplitude in a sector of a total of $N = n + \bar n$ vortices, where $n$ ($\bar n$) is the number of vortices with positive (negative) winding is
\begin{widetext}
\begin{eqnarray}
A_N &=& e^{-S_{\rm reg}}  \sum_{n = 0}^\infty \sum_{\bar n = 0}^\infty y^{n + \bar n} \underbrace{\delta_{n - \bar n, N}}_{= \int d\theta e^{i \theta (n - \bar n - N)}} {\frac{1}{n! \bar n!}} \int \prod_{i = 1}^n d^2x_i \prod_{\bar i = 1}^{\bar n} d^2x_{\bar i} e^{ 2i [\pi \rho_0 x_{i} - \phi(\vec x_{ i})]} e^{- 2i \int d \tau [\pi \rho_0 x_{\bar i}  - \phi(\vec x_{\bar i})]} \\
&=& e^{-S_{\rm reg}} \int d\theta  e^{i N \theta} e^{ \int d^2x \; y e^{i \theta} \cos( 2 [\pi \rho_0 x - \phi(\vec x)])}.
\end{eqnarray}
\end{widetext}
The combinatorial factor is the number of possibilities ${\frac{(n + \bar n)!}{n! \bar n!}} $ to arrange $n$ vortices with positive winding if there are $n+\bar n$ vortices in total divided by the number of configurations with equivalent spatial ordering  ${{(n + \bar n)!} }$. The Boltzmann weight of a vortex is denoted $y$. Summation over $N$ leads to a Dirac function $\delta (\theta)$, so that the overall theory is given by the effective Lagrangian, Eq.~\eqref{eq:Lvortex}.

For the calculation of density correlators perturbatively in $y$ we may integrate $\vartheta$ and obtain
\begin{equation}
\mathcal L_{\rm eff} = \frac{1}{2\pi} [\frac{1}{vK} (\dot \phi)^2+\frac{v}{K} (\phi')^2] - y \cos (2 [\pi \rho_0 x - \phi]).
\end{equation}
In this part of the appendix, the index $c$ in $K_c,v_c$ is suppressed.

For the derivation of the density correlator, we will use the following intermediate results ($v =1$)
\begin{widetext}
\begin{eqnarray}
\langle \phi(\vec x) \phi(\vec x') \rangle &=& G(\vec x - \vec x') = -\frac{ K}{2} \ln(\vert \vec x - \vec x'\vert),\\
\langle \partial_x \phi(\vec x) \cos(2\pi \rho_0 x' - 2\phi(\vec x')) \rangle &=& \partial_{x'} G(\vec x - \vec x') \sin(2 \phi(\vec x') - 2\pi \rho_0 x' ),\\
\langle \sin(2 \phi(\vec x') - 2\pi \rho_0 x' ) \sin(2 \phi(\vec y') - 2\pi \rho_0 y' ) \rangle &\propto & e^{4 G(\vec x' - \vec y')} \cos (2\pi \rho_0 (x' - y')).
\end{eqnarray}
The density density correlator thus contains the following correction to leading order in $y$
\begin{eqnarray}
\langle \rho(\vec x) \rho(\vec y) \rangle & \ni &y^2 \int d^2x' d^2y' \partial_{x'} G(\vec x - \vec x') \partial_{y'} G(\vec y - \vec y') e^{4G(\vec x' - \vec y')} \cos (2\pi \rho_0 (x' - y')), \\
&=& y^2 K^2 \int d^2x' d^2 y' \frac{x'}{[x']^2 + [\tau_x']^2}\frac{y'}{[y']^2 + [\tau_y']^2} \frac{1}{\vert \Delta \vec x' + \Delta \vec x \vert^{2 K}} \cos (2\pi \rho_0 \Delta x [ 1 + \Delta x'/\Delta x]) .
\end{eqnarray}
\end{widetext}
Here, $\Delta \vec x' = (\Delta x', \Delta \tau') = (x'- y',\tau_x' - \tau_y')$. We rescale all integration variables, e.g. $\vec x' \rightarrow \vec x'/\Delta x$ and
\begin{eqnarray}
\int_{-\infty}^\infty d x \cos [2\pi \alpha (1 +  x)] f(x) 
&\stackrel{\alpha \rightarrow \infty}{\propto} & \frac{\cos(2\pi \alpha)}{\alpha^2}.
\end{eqnarray}
This implies for the density correlator\cite{GiamarchiBook}
\begin{equation}
{\langle \rho(x) \rho(0) \rangle = \rho_0^2 + \frac{K}{2\pi^2 x^2} + \text{const.} \times \frac{\cos(2\pi \rho_0  x)}{ x^{2K}}.} 
\end{equation}
This concludes the derivation of Eq.~\eqref{eq:NematicCorrelator}

\section{SOC induced suppression of $K_c$}
\label{app:DeltaKc}

In this appendix we determine the SOC induced  corrections to $K_c/v_c$ in the strong coupling limit, {Eq.~\eqref{eq:Kofq} and \eqref{eq:vofq}}. 
Since the NL$\sigma$M sector is gapped one may integrate out the spin sector, and the term $\rho \vert \hat n'\vert^2/2m$ leads to additional terms $(\phi')^2$ i.e. to a renormalization of $K_c$. Here we estimate these terms by evaluation of $\langle \vert \hat n' (x, \tau)\vert^2 \vert \hat n' (x', \tau')\vert^2 \rangle$. 

In view of the short range correlations in spin space, $\hat n$ decays on the scale $\xi_s$ and we discretize the field theory in segments of length $\xi_s$. The spin sector of the Goldstone theory, Eqs.~\eqref{eq:Lunpert}, \eqref{eq:Lpert}, is then
\begin{align}
S_2^{\rm spin} &= \sum_i \int d\tau \xi_s\Big  [ \frac{1}{\tilde g_2} \vert \dot{\hat n}_i \vert^2 - \frac{\rho_i(\tau)}{m \xi_s^2} {\hat n_i \hat n_{i+1}} \notag \\
&+ \rho_i(\tau) q \hat n_i S_z^2 \hat n_i - \frac{1}{2 \tilde g_2} \hat n_i (\mathbf h (x_i) \cdot \mathbf S)^2 \hat n_i \Big ]. \label{eq:SspinApp}
\end{align}
In Hamiltonian formulation~\cite{SachdevBook}, the time derivative term becomes $H_{\rm top} = \frac{1}{2 I} \sum_i \mathbf L_i^2$, i.e.~it is a sum over tops with moment of inertia $I =\xi_s/\tilde g_2$. The energy levels $l(l+1)/2I$ have eigenstates given by spherical harmonics $\braket{\Omega \vert l,m} = Y_{l}^m(\Omega)$, where $\Omega$ is the solid angle parametrizing the target manifold of the sigma model. 

We need to calculate 
\begin{equation}
G_{i}(\tau, \tau') = \langle \hat n_{i,\alpha}(\tau) \hat n_{i, \alpha'}(\tau') \rangle \langle \hat n_{i+1,\alpha}(\tau) \hat n_{i+1, \alpha'}(\tau') \rangle
\end{equation}
to evaluate the dominant correction to $K_c$ given by
\begin{equation}
\delta S = - \sum_i \int d \tau d \tau' \frac{\rho_i(\tau) \rho_i(\tau')}{2m^2 \xi_s^2} G_{i}(\tau, \tau').
\end{equation}
For $\tau' > \tau$ we obtain (we momentarily suppress the index $_i$)
\begin{equation}
\langle \hat n_{\alpha}(\tau) \hat n_{\alpha'}(\tau') \rangle = \sum_{m, m'} M_{\alpha m} [e^{ - H_1 (\tau' - \tau)}]_{mm'} M_{m' \alpha}.
\end{equation}
Here we introduced
\begin{eqnarray}
M_{\alpha m} &:=& \braket{0\vert \hat n_\alpha \vert 1,m} \notag \\
&=&\frac{1}{\sqrt{3}} \left ( \begin{array}{ccc}
i/\sqrt{2} & 0 & i/\sqrt{2} \\ 
1/\sqrt{2} & 0 & -1/\sqrt{2} \\ 
0 & 1 & 0
\end{array} \right)_{\alpha m}.  \label{eq:Overlap}
\end{eqnarray}
Moreover, $H_1$ is the projection of the Hamiltonian to the space of $l = 1$ states (i.e. of the states with dominant contribution)
\begin{eqnarray}
H_1 &\stackrel{p = 0}{=}& \frac{\tilde g_2}{\xi_s} + \frac{\xi_s\rho_0 q}{5} \left (\begin{array}{ccc}
4 & 0 & 0 \\ 
0 & 2 & 0 \\ 
0 & 0 & 4
\end{array} \right ) \notag \\
&& - \frac{\xi_s h^2}{10\tilde g_2} \left (\begin{array}{ccc}
3 & 0 & - e^{2i \Theta x} \\ 
0 & 4 & 0 \\ 
- e^{-2i \Theta x} & 0 & 3
\end{array} \right ).
\end{eqnarray}
We further use 
\begin{equation}
U(x) = \frac{1}{\sqrt{2}}\left(
\begin{array}{ccc}
 e^{i \Theta x} & 0 & -e^{i\Theta  x} \\
 0 & \sqrt{2} & 0 \\
 e^{-i\Theta  x} & 0 & e^{-i \Theta  x} \\
\end{array}
\right)
\end{equation}
to diagonalize $H \rightarrow \tilde g_2/\xi_s  + [\xi_s \rho_0 q/5 ]\text{diag}(4,2,4) - [\xi_s h^2/(5 \tilde g_2)] \text{diag}(1,2,2) = \text{diag}(E_{k =1},E_{k =3},E_{k =2})$. Then
\begin{eqnarray}
G_i(0, \tau) \vert_{\tau > 0} &=& e^{- (E_k + E_l) \tau} (U^\dagger (x_i) M^T M U^*(x_{i + 1}))_{kl} \notag \\
&&(U^T (x_i) M^T M U(x_{i + 1}))_{kl} \notag \\
&=& \frac{\delta(\tau-\eta)}{9} \Big [\cos^2 (\Theta \xi_s) \left (\frac{1}{2 E_1}+ \frac{1}{2 E_2} \right)\notag \\
&& + 2 \frac{\sin^2 (\Theta \xi_s)}{E_1+E_2}+  \frac{1}{2 E_3} \Big],
\end{eqnarray}
and we used that at long time scales $ e^{- (E_k + E_l) \tau} \simeq \delta(\tau-\eta)/[E_k +E_l]$ (the limit $\eta \rightarrow 0$ is to be understood). Note that, by symmetry, an analogous result holds for $G_i(\tau, 0)\vert_{\tau > 0}$.

We now restore the continuum limit for smoothly varying $\rho = \rho_0 - \phi'/\pi$ and, by identification, we recognize
\begin{eqnarray}
\delta \left (\frac{v_c}{K_c}\right ) &=& - \frac{2}{9 m^2 \pi \xi_s^3} \Big [\cos^2 (\Theta \xi_s) \left (\frac{1}{2 E_1}+ \frac{1}{2 E_2} \right)\notag \\
&& + 2 \frac{\sin^2 (\Theta \xi_s)}{E_1+E_2}+  \frac{1}{2 E_3} \Big].
\end{eqnarray}
We can use that ($\xi_s = v_s/\Delta_{\rm SL}$ and $v_s = \sqrt{\rho_0 \tilde g_2/m}$)
\begin{equation}
\frac{1}{m^2 \xi_s^2 \tilde g_2} \sim \frac{\Delta_{\rm SL}^2}{m \rho_0 \tilde g_2^2} \sim \frac{\Delta_{\rm SL}^2}{\Lambda_s^2} \frac{\rho_0}{m} \sim \frac{v_s}{K_s}.
\end{equation}
Here we also used $K_s v_s \sim \rho_0/m$ and $\Delta_{\rm SL} = \Lambda_s/K_s$ at strong coupling.

By the same token, one may estimate the correction to the $K_c/v_c$ by the $q$ term in Eq.~\eqref{eq:SspinApp}. It yields a correction
\begin{equation}
\delta S \sim - \sum_i\int d \tau \frac{\xi_s^2 q^2}{\tilde g_2/\xi_s} (\phi')^2,
\end{equation}
and thus ($K_s = \pi v_s/\tilde g_2$)
\begin{eqnarray}
\delta \left (\frac{v_c}{K_c} \right) = -\frac{\xi_s^2 q^2}{\tilde g_2} \sim - \frac{v_s^2}{\Delta_{\rm SL}^2}\frac{q^2}{\tilde g_{2}} \sim - K_s v_s \frac{q^2}{\Delta_{\rm SL}^2}.
\end{eqnarray}
In total we obtain a correction
\begin{eqnarray}
\delta \left (\frac{v_c}{K_c} \right) &=&\Big \lbrace \mathcal C_1 \left [-{3} + \frac{\sin^2(\Theta \xi_s) /{25}}{\left (1 + \frac{K_c^2 K_s q \tilde g_0}{\Delta_{\rm SL} \tilde g_2}\right )^2} \left ( \frac{K_s h	}{\pi \Delta_{\rm SL}}\right)^4 \right] \frac{v_s}{K_s} \notag \\
&&- \mathcal C_2 \frac{K_s v_s q^2}{\Delta_{\rm SL}^2} \Big \rbrace  \Big / {\left (1 + \frac{K_c^2 K_s q \tilde g_0}{\Delta_{\rm SL} \tilde g_2}\right )}.
\end{eqnarray}
Here we used ($\tilde \mu/\Lambda_s = \tilde g_0/\tilde g_2$)
\begin{equation}
\frac{\xi_s^2 \rho_0 q}{\tilde g_2} \sim \frac{v_s^2}{\Delta_{\rm SL}^2} \frac{\rho_0 q}{\tilde g_2} \sim \frac{q \tilde \mu}{\Delta_{\rm SL}^2} \frac{\rho_0}{m \tilde g_0} \sim \frac{q}{\Delta_{\rm SL}} \frac{\tilde g_0}{\tilde g_2} K_c^2 K_s
\end{equation}
such that (perturbatively in small $h \ll \Delta_{\rm SL}$)
\begin{equation}
1 + \frac{E_1}{E_2} - \frac{4}{1 + \frac{E_2}{E_1}} \simeq \left ( \frac{K_s h}{	\pi \Delta_{\rm SL}}\right)^4 \frac{1/50}{\left (1 + \frac{K_c^2 K_s q \tilde g_0}{\Delta_{\rm SL} \tilde g_2}\right )^2}.
\end{equation}
This is the origin of Eq.~\eqref{eq:Kofq} in the main text.

\bibliography{Spin1DMRGv1}

\begin{thebibliography}{72}%
\makeatletter
\providecommand \@ifxundefined [1]{%
 \@ifx{#1\undefined}
}%
\providecommand \@ifnum [1]{%
 \ifnum #1\expandafter \@firstoftwo
 \else \expandafter \@secondoftwo
 \fi
}%
\providecommand \@ifx [1]{%
 \ifx #1\expandafter \@firstoftwo
 \else \expandafter \@secondoftwo
 \fi
}%
\providecommand \natexlab [1]{#1}%
\providecommand \enquote  [1]{``#1''}%
\providecommand \bibnamefont  [1]{#1}%
\providecommand \bibfnamefont [1]{#1}%
\providecommand \citenamefont [1]{#1}%
\providecommand \href@noop [0]{\@secondoftwo}%
\providecommand \href [0]{\begingroup \@sanitize@url \@href}%
\providecommand \@href[1]{\@@startlink{#1}\@@href}%
\providecommand \@@href[1]{\endgroup#1\@@endlink}%
\providecommand \@sanitize@url [0]{\catcode `\\12\catcode `\$12\catcode
  `\&12\catcode `\#12\catcode `\^12\catcode `\_12\catcode `\%12\relax}%
\providecommand \@@startlink[1]{}%
\providecommand \@@endlink[0]{}%
\providecommand \url  [0]{\begingroup\@sanitize@url \@url }%
\providecommand \@url [1]{\endgroup\@href {#1}{\urlprefix }}%
\providecommand \urlprefix  [0]{URL }%
\providecommand \Eprint [0]{\href }%
\providecommand \doibase [0]{http://dx.doi.org/}%
\providecommand \selectlanguage [0]{\@gobble}%
\providecommand \bibinfo  [0]{\@secondoftwo}%
\providecommand \bibfield  [0]{\@secondoftwo}%
\providecommand \translation [1]{[#1]}%
\providecommand \BibitemOpen [0]{}%
\providecommand \bibitemStop [0]{}%
\providecommand \bibitemNoStop [0]{.\EOS\space}%
\providecommand \EOS [0]{\spacefactor3000\relax}%
\providecommand \BibitemShut  [1]{\csname bibitem#1\endcsname}%
\let\auto@bib@innerbib\@empty
\bibitem [{\citenamefont {Kawaguchi}\ and\ \citenamefont
  {Ueda}(2012)}]{Kawaguchi-2012}%
  \BibitemOpen
  \bibfield  {author} {\bibinfo {author} {\bibfnamefont {Y.}~\bibnamefont
  {Kawaguchi}}\ and\ \bibinfo {author} {\bibfnamefont {M.}~\bibnamefont
  {Ueda}},\ }\href
  {https://www.sciencedirect.com/science/article/abs/pii/S0370157312002098?via%3Dihub}
  {\bibfield  {journal} {\bibinfo  {journal} {Physics Reports}\ }\textbf
  {\bibinfo {volume} {520}},\ \bibinfo {pages} {253} (\bibinfo {year}
  {2012})}\BibitemShut {NoStop}%
\bibitem [{\citenamefont {Stamper-Kurn}\ and\ \citenamefont
  {Ueda}(2013)}]{StamperKurnUedaRMP2013}%
  \BibitemOpen
  \bibfield  {author} {\bibinfo {author} {\bibfnamefont {D.~M.}\ \bibnamefont
  {Stamper-Kurn}}\ and\ \bibinfo {author} {\bibfnamefont {M.}~\bibnamefont
  {Ueda}},\ }\href {\doibase 10.1103/RevModPhys.85.1191} {\bibfield  {journal}
  {\bibinfo  {journal} {Rev. Mod. Phys.}\ }\textbf {\bibinfo {volume} {85}},\
  \bibinfo {pages} {1191} (\bibinfo {year} {2013})}\BibitemShut {NoStop}%
\bibitem [{\citenamefont {Ho}(1998)}]{Ho-1998}%
  \BibitemOpen
  \bibfield  {author} {\bibinfo {author} {\bibfnamefont {T.-L.}\ \bibnamefont
  {Ho}},\ }\href
  {https://journals.aps.org/prl/abstract/10.1103/PhysRevLett.81.742} {\bibfield
   {journal} {\bibinfo  {journal} {Phys. Rev. Lett.}\ }\textbf {\bibinfo
  {volume} {81}},\ \bibinfo {pages} {742} (\bibinfo {year} {1998})}\BibitemShut
  {NoStop}%
\bibitem [{\citenamefont {Ohmi}\ and\ \citenamefont
  {Machida}(1998)}]{Ohmi-1998}%
  \BibitemOpen
  \bibfield  {author} {\bibinfo {author} {\bibfnamefont {T.}~\bibnamefont
  {Ohmi}}\ and\ \bibinfo {author} {\bibfnamefont {K.}~\bibnamefont {Machida}},\
  }\href {https://journals.jps.jp/doi/10.1143/JPSJ.67.1822} {\bibfield
  {journal} {\bibinfo  {journal} {Journal of the Physical Society of Japan}\
  }\textbf {\bibinfo {volume} {67}},\ \bibinfo {pages} {1822} (\bibinfo {year}
  {1998})}\BibitemShut {NoStop}%
\bibitem [{\citenamefont {Galitski}\ and\ \citenamefont
  {Spielman}(2013)}]{GalitskiSpielman2015}%
  \BibitemOpen
  \bibfield  {author} {\bibinfo {author} {\bibfnamefont {V.~M.}\ \bibnamefont
  {Galitski}}\ and\ \bibinfo {author} {\bibfnamefont {I.}~\bibnamefont
  {Spielman}},\ }\href {https://www.nature.com/articles/nature11841} {}\bibinfo
  {howpublished} {Nature {\bf 494}, 49} (\bibinfo {year} {2013})\BibitemShut
  {NoStop}%
\bibitem [{\citenamefont {Cooper}\ \emph {et~al.}(2019)\citenamefont {Cooper},
  \citenamefont {Dalibard},\ and\ \citenamefont {Spielman}}]{Cooper-2019}%
  \BibitemOpen
  \bibfield  {author} {\bibinfo {author} {\bibfnamefont {N.~R.}\ \bibnamefont
  {Cooper}}, \bibinfo {author} {\bibfnamefont {J.}~\bibnamefont {Dalibard}}, \
  and\ \bibinfo {author} {\bibfnamefont {I.~B.}\ \bibnamefont {Spielman}},\
  }\href {\doibase 10.1103/RevModPhys.91.015005} {\bibfield  {journal}
  {\bibinfo  {journal} {Rev. Mod. Phys.}\ }\textbf {\bibinfo {volume} {91}},\
  \bibinfo {pages} {015005} (\bibinfo {year} {2019})}\BibitemShut {NoStop}%
\bibitem [{\citenamefont {Wang}\ \emph {et~al.}(2012)\citenamefont {Wang},
  \citenamefont {Yu}, \citenamefont {Fu}, \citenamefont {Miao}, \citenamefont
  {Huang}, \citenamefont {Chai}, \citenamefont {Zhai},\ and\ \citenamefont
  {Zhang}}]{Wang-2012}%
  \BibitemOpen
  \bibfield  {author} {\bibinfo {author} {\bibfnamefont {P.}~\bibnamefont
  {Wang}}, \bibinfo {author} {\bibfnamefont {Z.-Q.}\ \bibnamefont {Yu}},
  \bibinfo {author} {\bibfnamefont {Z.}~\bibnamefont {Fu}}, \bibinfo {author}
  {\bibfnamefont {J.}~\bibnamefont {Miao}}, \bibinfo {author} {\bibfnamefont
  {L.}~\bibnamefont {Huang}}, \bibinfo {author} {\bibfnamefont
  {S.}~\bibnamefont {Chai}}, \bibinfo {author} {\bibfnamefont {H.}~\bibnamefont
  {Zhai}}, \ and\ \bibinfo {author} {\bibfnamefont {J.}~\bibnamefont {Zhang}},\
  }\href {\doibase 10.1103/PhysRevLett.109.095301} {\bibfield  {journal}
  {\bibinfo  {journal} {Phys. Rev. Lett.}\ }\textbf {\bibinfo {volume} {109}},\
  \bibinfo {pages} {095301} (\bibinfo {year} {2012})}\BibitemShut {NoStop}%
\bibitem [{\citenamefont {Lin}\ \emph {et~al.}(2009)\citenamefont {Lin},
  \citenamefont {Compton}, \citenamefont {Jimenez-Garcia}, \citenamefont
  {Porto},\ and\ \citenamefont {Spielman}}]{Lin-2009}%
  \BibitemOpen
  \bibfield  {author} {\bibinfo {author} {\bibfnamefont {Y.-J.}\ \bibnamefont
  {Lin}}, \bibinfo {author} {\bibfnamefont {R.~L.}\ \bibnamefont {Compton}},
  \bibinfo {author} {\bibfnamefont {K.}~\bibnamefont {Jimenez-Garcia}},
  \bibinfo {author} {\bibfnamefont {J.~V.}\ \bibnamefont {Porto}}, \ and\
  \bibinfo {author} {\bibfnamefont {I.~B.}\ \bibnamefont {Spielman}},\ }\href
  {https://www.nature.com/articles/nature08609} {\bibfield  {journal} {\bibinfo
   {journal} {Nature}\ }\textbf {\bibinfo {volume} {462}},\ \bibinfo {pages}
  {628} (\bibinfo {year} {2009})}\BibitemShut {NoStop}%
\bibitem [{\citenamefont {{Lin}}\ \emph {et~al.}(2011)\citenamefont {{Lin}},
  \citenamefont {{Jim{\'e}nez-Garc{\'{\i}}a}},\ and\ \citenamefont
  {{Spielman}}}]{Lin-2011}%
  \BibitemOpen
  \bibfield  {author} {\bibinfo {author} {\bibfnamefont {Y.-J.}\ \bibnamefont
  {{Lin}}}, \bibinfo {author} {\bibfnamefont {K.}~\bibnamefont
  {{Jim{\'e}nez-Garc{\'{\i}}a}}}, \ and\ \bibinfo {author} {\bibfnamefont
  {I.~B.}\ \bibnamefont {{Spielman}}},\ }\href {\doibase 10.1038/nature09887}
  {\bibfield  {journal} {\bibinfo  {journal} {Nature}\ }\textbf {\bibinfo
  {volume} {471}},\ \bibinfo {pages} {83} (\bibinfo {year} {2011})}\BibitemShut
  {NoStop}%
\bibitem [{\citenamefont {Stuhl}\ \emph {et~al.}(2015)\citenamefont {Stuhl},
  \citenamefont {Lu}, \citenamefont {Aycock}, \citenamefont {Genkina},\ and\
  \citenamefont {Spielman}}]{stuhl2015}%
  \BibitemOpen
  \bibfield  {author} {\bibinfo {author} {\bibfnamefont {B.~K.}\ \bibnamefont
  {Stuhl}}, \bibinfo {author} {\bibfnamefont {H.-I.}\ \bibnamefont {Lu}},
  \bibinfo {author} {\bibfnamefont {L.~M.}\ \bibnamefont {Aycock}}, \bibinfo
  {author} {\bibfnamefont {D.}~\bibnamefont {Genkina}}, \ and\ \bibinfo
  {author} {\bibfnamefont {I.~B.}\ \bibnamefont {Spielman}},\ }\href {\doibase
  10.1126/science.aaa8515} {\bibfield  {journal} {\bibinfo  {journal}
  {Science}\ }\textbf {\bibinfo {volume} {349}},\ \bibinfo {pages} {1514}
  (\bibinfo {year} {2015})}\BibitemShut {NoStop}%
\bibitem [{\citenamefont {Campbell}\ \emph {et~al.}(2016)\citenamefont
  {Campbell}, \citenamefont {Price}, \citenamefont {Putra}, \citenamefont
  {Vald{\'e}s-Curiel}, \citenamefont {Trypogeorgos},\ and\ \citenamefont
  {Spielman}}]{Campbell-2016}%
  \BibitemOpen
  \bibfield  {author} {\bibinfo {author} {\bibfnamefont {D.}~\bibnamefont
  {Campbell}}, \bibinfo {author} {\bibfnamefont {R.}~\bibnamefont {Price}},
  \bibinfo {author} {\bibfnamefont {A.}~\bibnamefont {Putra}}, \bibinfo
  {author} {\bibfnamefont {A.}~\bibnamefont {Vald{\'e}s-Curiel}}, \bibinfo
  {author} {\bibfnamefont {D.}~\bibnamefont {Trypogeorgos}}, \ and\ \bibinfo
  {author} {\bibfnamefont {I.}~\bibnamefont {Spielman}},\ }\href
  {https://www.nature.com/articles/ncomms10897?origin=ppub} {\bibfield
  {journal} {\bibinfo  {journal} {Nature communications}\ }\textbf {\bibinfo
  {volume} {7}} (\bibinfo {year} {2016})}\BibitemShut {NoStop}%
\bibitem [{\citenamefont {Vald{\'e}s-Curiel}\ \emph {et~al.}(2017)\citenamefont
  {Vald{\'e}s-Curiel}, \citenamefont {Trypogeorgos}, \citenamefont {Marshall},\
  and\ \citenamefont {Spielman}}]{Valdes-2017}%
  \BibitemOpen
  \bibfield  {author} {\bibinfo {author} {\bibfnamefont {A.}~\bibnamefont
  {Vald{\'e}s-Curiel}}, \bibinfo {author} {\bibfnamefont {D.}~\bibnamefont
  {Trypogeorgos}}, \bibinfo {author} {\bibfnamefont {E.}~\bibnamefont
  {Marshall}}, \ and\ \bibinfo {author} {\bibfnamefont {I.}~\bibnamefont
  {Spielman}},\ }\href
  {https://iopscience.iop.org/article/10.1088/1367-2630/aa6279/pdf} {\bibfield
  {journal} {\bibinfo  {journal} {New Journal of Physics}\ }\textbf {\bibinfo
  {volume} {19}},\ \bibinfo {pages} {033025} (\bibinfo {year}
  {2017})}\BibitemShut {NoStop}%
\bibitem [{\citenamefont {Huang}\ \emph {et~al.}(2016)\citenamefont {Huang},
  \citenamefont {Meng}, \citenamefont {Wang}, \citenamefont {Peng},
  \citenamefont {Zhang}, \citenamefont {Chen}, \citenamefont {Li},
  \citenamefont {Zhou},\ and\ \citenamefont {Zhang}}]{Huang-2016}%
  \BibitemOpen
  \bibfield  {author} {\bibinfo {author} {\bibfnamefont {L.}~\bibnamefont
  {Huang}}, \bibinfo {author} {\bibfnamefont {Z.}~\bibnamefont {Meng}},
  \bibinfo {author} {\bibfnamefont {P.}~\bibnamefont {Wang}}, \bibinfo {author}
  {\bibfnamefont {P.}~\bibnamefont {Peng}}, \bibinfo {author} {\bibfnamefont
  {S.-L.}\ \bibnamefont {Zhang}}, \bibinfo {author} {\bibfnamefont
  {L.}~\bibnamefont {Chen}}, \bibinfo {author} {\bibfnamefont {D.}~\bibnamefont
  {Li}}, \bibinfo {author} {\bibfnamefont {Q.}~\bibnamefont {Zhou}}, \ and\
  \bibinfo {author} {\bibfnamefont {J.}~\bibnamefont {Zhang}},\ }\href
  {https://www.nature.com/articles/nphys3672} {\bibfield  {journal} {\bibinfo
  {journal} {Nature Physics}\ }\textbf {\bibinfo {volume} {12}},\ \bibinfo
  {pages} {540} (\bibinfo {year} {2016})}\BibitemShut {NoStop}%
\bibitem [{\citenamefont {Wu}\ \emph {et~al.}(2016)\citenamefont {Wu},
  \citenamefont {Zhang}, \citenamefont {Sun}, \citenamefont {Xu}, \citenamefont
  {Wang}, \citenamefont {Ji}, \citenamefont {Deng}, \citenamefont {Chen},
  \citenamefont {Liu},\ and\ \citenamefont {Pan}}]{Wu-2016}%
  \BibitemOpen
  \bibfield  {author} {\bibinfo {author} {\bibfnamefont {Z.}~\bibnamefont
  {Wu}}, \bibinfo {author} {\bibfnamefont {L.}~\bibnamefont {Zhang}}, \bibinfo
  {author} {\bibfnamefont {W.}~\bibnamefont {Sun}}, \bibinfo {author}
  {\bibfnamefont {X.-T.}\ \bibnamefont {Xu}}, \bibinfo {author} {\bibfnamefont
  {B.-Z.}\ \bibnamefont {Wang}}, \bibinfo {author} {\bibfnamefont {S.-C.}\
  \bibnamefont {Ji}}, \bibinfo {author} {\bibfnamefont {Y.}~\bibnamefont
  {Deng}}, \bibinfo {author} {\bibfnamefont {S.}~\bibnamefont {Chen}}, \bibinfo
  {author} {\bibfnamefont {X.-J.}\ \bibnamefont {Liu}}, \ and\ \bibinfo
  {author} {\bibfnamefont {J.-W.}\ \bibnamefont {Pan}},\ }\href
  {https://science.sciencemag.org/content/354/6308/83} {\bibfield  {journal}
  {\bibinfo  {journal} {Science}\ }\textbf {\bibinfo {volume} {354}},\ \bibinfo
  {pages} {83} (\bibinfo {year} {2016})}\BibitemShut {NoStop}%
\bibitem [{\citenamefont {Sun}\ \emph {et~al.}(2018)\citenamefont {Sun},
  \citenamefont {Wang}, \citenamefont {Xu}, \citenamefont {Yi}, \citenamefont
  {Zhang}, \citenamefont {Wu}, \citenamefont {Deng}, \citenamefont {Liu},
  \citenamefont {Chen},\ and\ \citenamefont {Pan}}]{Sun-2017}%
  \BibitemOpen
  \bibfield  {author} {\bibinfo {author} {\bibfnamefont {W.}~\bibnamefont
  {Sun}}, \bibinfo {author} {\bibfnamefont {B.-Z.}\ \bibnamefont {Wang}},
  \bibinfo {author} {\bibfnamefont {X.-T.}\ \bibnamefont {Xu}}, \bibinfo
  {author} {\bibfnamefont {C.-R.}\ \bibnamefont {Yi}}, \bibinfo {author}
  {\bibfnamefont {L.}~\bibnamefont {Zhang}}, \bibinfo {author} {\bibfnamefont
  {Z.}~\bibnamefont {Wu}}, \bibinfo {author} {\bibfnamefont {Y.}~\bibnamefont
  {Deng}}, \bibinfo {author} {\bibfnamefont {X.-J.}\ \bibnamefont {Liu}},
  \bibinfo {author} {\bibfnamefont {S.}~\bibnamefont {Chen}}, \ and\ \bibinfo
  {author} {\bibfnamefont {J.-W.}\ \bibnamefont {Pan}},\ }\href
  {https://link.aps.org/doi/10.1103/PhysRevLett.121.150401} {\bibfield
  {journal} {\bibinfo  {journal} {Phys. Rev. Lett.}\ }\textbf {\bibinfo
  {volume} {121}},\ \bibinfo {pages} {150401} (\bibinfo {year}
  {2018})}\BibitemShut {NoStop}%
\bibitem [{\citenamefont {Song}\ \emph {et~al.}(2018)\citenamefont {Song},
  \citenamefont {Zhang}, \citenamefont {He}, \citenamefont {Poon},
  \citenamefont {Hajiyev}, \citenamefont {Zhang}, \citenamefont {Liu},\ and\
  \citenamefont {Jo}}]{Song-2017}%
  \BibitemOpen
  \bibfield  {author} {\bibinfo {author} {\bibfnamefont {B.}~\bibnamefont
  {Song}}, \bibinfo {author} {\bibfnamefont {L.}~\bibnamefont {Zhang}},
  \bibinfo {author} {\bibfnamefont {C.}~\bibnamefont {He}}, \bibinfo {author}
  {\bibfnamefont {T.~F.~J.}\ \bibnamefont {Poon}}, \bibinfo {author}
  {\bibfnamefont {E.}~\bibnamefont {Hajiyev}}, \bibinfo {author} {\bibfnamefont
  {S.}~\bibnamefont {Zhang}}, \bibinfo {author} {\bibfnamefont {X.-J.}\
  \bibnamefont {Liu}}, \ and\ \bibinfo {author} {\bibfnamefont {G.-B.}\
  \bibnamefont {Jo}},\ }\href
  {https://advances.sciencemag.org/content/4/2/eaao4748} {\bibfield  {journal}
  {\bibinfo  {journal} {Science advances}\ }\textbf {\bibinfo {volume} {4}},\
  \bibinfo {pages} {eaao4748} (\bibinfo {year} {2018})}\BibitemShut {NoStop}%
\bibitem [{\citenamefont {Xu}\ \emph {et~al.}(2012)\citenamefont {Xu},
  \citenamefont {Kawaguchi}, \citenamefont {You},\ and\ \citenamefont
  {Ueda}}]{Ueda-12}%
  \BibitemOpen
  \bibfield  {author} {\bibinfo {author} {\bibfnamefont {Z.~F.}\ \bibnamefont
  {Xu}}, \bibinfo {author} {\bibfnamefont {Y.}~\bibnamefont {Kawaguchi}},
  \bibinfo {author} {\bibfnamefont {L.}~\bibnamefont {You}}, \ and\ \bibinfo
  {author} {\bibfnamefont {M.}~\bibnamefont {Ueda}},\ }\href {\doibase
  10.1103/PhysRevA.86.033628} {\bibfield  {journal} {\bibinfo  {journal} {Phys.
  Rev. A}\ }\textbf {\bibinfo {volume} {86}},\ \bibinfo {pages} {033628}
  (\bibinfo {year} {2012})}\BibitemShut {NoStop}%
\bibitem [{\citenamefont {Li}\ \emph {et~al.}(2012)\citenamefont {Li},
  \citenamefont {Pitaevskii},\ and\ \citenamefont {Stringari}}]{Li2012}%
  \BibitemOpen
  \bibfield  {author} {\bibinfo {author} {\bibfnamefont {Y.}~\bibnamefont
  {Li}}, \bibinfo {author} {\bibfnamefont {L.~P.}\ \bibnamefont {Pitaevskii}},
  \ and\ \bibinfo {author} {\bibfnamefont {S.}~\bibnamefont {Stringari}},\
  }\href {http://link.aps.org/doi/10.1103/PhysRevLett.108.225301} {\bibfield
  {journal} {\bibinfo  {journal} {Phys. Rev. Lett.}\ }\textbf {\bibinfo
  {volume} {108}},\ \bibinfo {pages} {225301} (\bibinfo {year}
  {2012})}\BibitemShut {NoStop}%
\bibitem [{\citenamefont {Cole}\ \emph {et~al.}(2012)\citenamefont {Cole},
  \citenamefont {Zhang}, \citenamefont {Paramekanti},\ and\ \citenamefont
  {Trivedi}}]{mott_2d_cole}%
  \BibitemOpen
  \bibfield  {author} {\bibinfo {author} {\bibfnamefont {W.~S.}\ \bibnamefont
  {Cole}}, \bibinfo {author} {\bibfnamefont {S.}~\bibnamefont {Zhang}},
  \bibinfo {author} {\bibfnamefont {A.}~\bibnamefont {Paramekanti}}, \ and\
  \bibinfo {author} {\bibfnamefont {N.}~\bibnamefont {Trivedi}},\ }\href
  {\doibase 10.1103/PhysRevLett.109.085302} {\bibfield  {journal} {\bibinfo
  {journal} {Phys. Rev. Lett.}\ }\textbf {\bibinfo {volume} {109}},\ \bibinfo
  {pages} {085302} (\bibinfo {year} {2012})}\BibitemShut {NoStop}%
\bibitem [{\citenamefont {Li}\ \emph {et~al.}(2013)\citenamefont {Li},
  \citenamefont {Martone}, \citenamefont {Pitaevskii},\ and\ \citenamefont
  {Stringari}}]{Li2013}%
  \BibitemOpen
  \bibfield  {author} {\bibinfo {author} {\bibfnamefont {Y.}~\bibnamefont
  {Li}}, \bibinfo {author} {\bibfnamefont {G.~I.}\ \bibnamefont {Martone}},
  \bibinfo {author} {\bibfnamefont {L.~P.}\ \bibnamefont {Pitaevskii}}, \ and\
  \bibinfo {author} {\bibfnamefont {S.}~\bibnamefont {Stringari}},\ }\href
  {\doibase 10.1103/PhysRevLett.110.235302} {\bibfield  {journal} {\bibinfo
  {journal} {Phys. Rev. Lett.}\ }\textbf {\bibinfo {volume} {110}},\ \bibinfo
  {pages} {235302} (\bibinfo {year} {2013})}\BibitemShut {NoStop}%
\bibitem [{\citenamefont {Hickey}\ and\ \citenamefont
  {Paramekanti}(2014)}]{Hickey2014}%
  \BibitemOpen
  \bibfield  {author} {\bibinfo {author} {\bibfnamefont {C.}~\bibnamefont
  {Hickey}}\ and\ \bibinfo {author} {\bibfnamefont {A.}~\bibnamefont
  {Paramekanti}},\ }\href {\doibase 10.1103/PhysRevLett.113.265302} {\bibfield
  {journal} {\bibinfo  {journal} {Phys. Rev. Lett.}\ }\textbf {\bibinfo
  {volume} {113}},\ \bibinfo {pages} {265302} (\bibinfo {year}
  {2014})}\BibitemShut {NoStop}%
\bibitem [{\citenamefont {Martone}\ \emph {et~al.}(2014)\citenamefont
  {Martone}, \citenamefont {Li},\ and\ \citenamefont
  {Stringari}}]{Martone-2014}%
  \BibitemOpen
  \bibfield  {author} {\bibinfo {author} {\bibfnamefont {G.~I.}\ \bibnamefont
  {Martone}}, \bibinfo {author} {\bibfnamefont {Y.}~\bibnamefont {Li}}, \ and\
  \bibinfo {author} {\bibfnamefont {S.}~\bibnamefont {Stringari}},\ }\href
  {\doibase 10.1103/PhysRevA.90.041604} {\bibfield  {journal} {\bibinfo
  {journal} {Phys. Rev. A}\ }\textbf {\bibinfo {volume} {90}},\ \bibinfo
  {pages} {041604} (\bibinfo {year} {2014})}\BibitemShut {NoStop}%
\bibitem [{\citenamefont {Lan}\ and\ \citenamefont {\"Ohberg}(2014)}]{Lan2014}%
  \BibitemOpen
  \bibfield  {author} {\bibinfo {author} {\bibfnamefont {Z.}~\bibnamefont
  {Lan}}\ and\ \bibinfo {author} {\bibfnamefont {P.}~\bibnamefont {\"Ohberg}},\
  }\href {\doibase 10.1103/PhysRevA.89.023630} {\bibfield  {journal} {\bibinfo
  {journal} {Phys. Rev. A}\ }\textbf {\bibinfo {volume} {89}},\ \bibinfo
  {pages} {023630} (\bibinfo {year} {2014})}\BibitemShut {NoStop}%
\bibitem [{\citenamefont {Pixley}\ \emph {et~al.}(2016)\citenamefont {Pixley},
  \citenamefont {Natu}, \citenamefont {Spielman},\ and\ \citenamefont
  {Das~Sarma}}]{Pixley-2016}%
  \BibitemOpen
  \bibfield  {author} {\bibinfo {author} {\bibfnamefont {J.~H.}\ \bibnamefont
  {Pixley}}, \bibinfo {author} {\bibfnamefont {S.~S.}\ \bibnamefont {Natu}},
  \bibinfo {author} {\bibfnamefont {I.~B.}\ \bibnamefont {Spielman}}, \ and\
  \bibinfo {author} {\bibfnamefont {S.}~\bibnamefont {Das~Sarma}},\ }\href
  {\doibase 10.1103/PhysRevB.93.081101} {\bibfield  {journal} {\bibinfo
  {journal} {Phys. Rev. B}\ }\textbf {\bibinfo {volume} {93}},\ \bibinfo
  {pages} {081101} (\bibinfo {year} {2016})}\BibitemShut {NoStop}%
\bibitem [{\citenamefont {Hurst}\ \emph {et~al.}(2016)\citenamefont {Hurst},
  \citenamefont {Wilson}, \citenamefont {Pixley}, \citenamefont {Spielman},\
  and\ \citenamefont {Natu}}]{Hurst-2016}%
  \BibitemOpen
  \bibfield  {author} {\bibinfo {author} {\bibfnamefont {H.~M.}\ \bibnamefont
  {Hurst}}, \bibinfo {author} {\bibfnamefont {J.~H.}\ \bibnamefont {Wilson}},
  \bibinfo {author} {\bibfnamefont {J.~H.}\ \bibnamefont {Pixley}}, \bibinfo
  {author} {\bibfnamefont {I.~B.}\ \bibnamefont {Spielman}}, \ and\ \bibinfo
  {author} {\bibfnamefont {S.~S.}\ \bibnamefont {Natu}},\ }\href {\doibase
  10.1103/PhysRevA.94.063613} {\bibfield  {journal} {\bibinfo  {journal} {Phys.
  Rev. A}\ }\textbf {\bibinfo {volume} {94}},\ \bibinfo {pages} {063613}
  (\bibinfo {year} {2016})}\BibitemShut {NoStop}%
\bibitem [{\citenamefont {Yan}\ \emph {et~al.}(2017)\citenamefont {Yan},
  \citenamefont {Qian}, \citenamefont {Hui}, \citenamefont {Gong},
  \citenamefont {Zhang},\ and\ \citenamefont {Scarola}}]{YanScarola2017}%
  \BibitemOpen
  \bibfield  {author} {\bibinfo {author} {\bibfnamefont {M.}~\bibnamefont
  {Yan}}, \bibinfo {author} {\bibfnamefont {Y.}~\bibnamefont {Qian}}, \bibinfo
  {author} {\bibfnamefont {H.-Y.}\ \bibnamefont {Hui}}, \bibinfo {author}
  {\bibfnamefont {M.}~\bibnamefont {Gong}}, \bibinfo {author} {\bibfnamefont
  {C.}~\bibnamefont {Zhang}}, \ and\ \bibinfo {author} {\bibfnamefont {V.~W.}\
  \bibnamefont {Scarola}},\ }\href {\doibase 10.1103/PhysRevA.96.053619}
  {\bibfield  {journal} {\bibinfo  {journal} {Phys. Rev. A}\ }\textbf {\bibinfo
  {volume} {96}},\ \bibinfo {pages} {053619} (\bibinfo {year}
  {2017})}\BibitemShut {NoStop}%
\bibitem [{\citenamefont {Nonne}\ \emph {et~al.}(2013)\citenamefont {Nonne},
  \citenamefont {Moliner}, \citenamefont {Capponi}, \citenamefont
  {Lecheminant},\ and\ \citenamefont {Totsuka}}]{Nonne2013symmetry}%
  \BibitemOpen
  \bibfield  {author} {\bibinfo {author} {\bibfnamefont {H.}~\bibnamefont
  {Nonne}}, \bibinfo {author} {\bibfnamefont {M.}~\bibnamefont {Moliner}},
  \bibinfo {author} {\bibfnamefont {S.}~\bibnamefont {Capponi}}, \bibinfo
  {author} {\bibfnamefont {P.}~\bibnamefont {Lecheminant}}, \ and\ \bibinfo
  {author} {\bibfnamefont {K.}~\bibnamefont {Totsuka}},\ }\href
  {https://iopscience.iop.org/article/10.1209/0295-5075/102/37008/meta}
  {\bibfield  {journal} {\bibinfo  {journal} {EPL (Europhysics Letters)}\
  }\textbf {\bibinfo {volume} {102}},\ \bibinfo {pages} {37008} (\bibinfo
  {year} {2013})}\BibitemShut {NoStop}%
\bibitem [{\citenamefont {Hou}\ \emph {et~al.}(2018)\citenamefont {Hou},
  \citenamefont {Hu},\ and\ \citenamefont {Zhang}}]{Hou2018}%
  \BibitemOpen
  \bibfield  {author} {\bibinfo {author} {\bibfnamefont {J.}~\bibnamefont
  {Hou}}, \bibinfo {author} {\bibfnamefont {H.}~\bibnamefont {Hu}}, \ and\
  \bibinfo {author} {\bibfnamefont {C.}~\bibnamefont {Zhang}},\ }\href@noop {}
  {\bibfield  {journal} {\bibinfo  {journal} {arXiv preprint arXiv:1809.04537}\
  } (\bibinfo {year} {2018})}\BibitemShut {NoStop}%
\bibitem [{\citenamefont {Pagano}\ \emph {et~al.}(2014)\citenamefont {Pagano},
  \citenamefont {Mancini}, \citenamefont {Cappellini}, \citenamefont
  {Lombardi}, \citenamefont {Sch{\"a}fer}, \citenamefont {Hu}, \citenamefont
  {Liu}, \citenamefont {Catani}, \citenamefont {Sias}, \citenamefont {Inguscio}
  \emph {et~al.}}]{PaganoFallaini2014}%
  \BibitemOpen
  \bibfield  {author} {\bibinfo {author} {\bibfnamefont {G.}~\bibnamefont
  {Pagano}}, \bibinfo {author} {\bibfnamefont {M.}~\bibnamefont {Mancini}},
  \bibinfo {author} {\bibfnamefont {G.}~\bibnamefont {Cappellini}}, \bibinfo
  {author} {\bibfnamefont {P.}~\bibnamefont {Lombardi}}, \bibinfo {author}
  {\bibfnamefont {F.}~\bibnamefont {Sch{\"a}fer}}, \bibinfo {author}
  {\bibfnamefont {H.}~\bibnamefont {Hu}}, \bibinfo {author} {\bibfnamefont
  {X.-J.}\ \bibnamefont {Liu}}, \bibinfo {author} {\bibfnamefont
  {J.}~\bibnamefont {Catani}}, \bibinfo {author} {\bibfnamefont
  {C.}~\bibnamefont {Sias}}, \bibinfo {author} {\bibfnamefont {M.}~\bibnamefont
  {Inguscio}},  \emph {et~al.},\ }\href
  {https://www.nature.com/articles/nphys2878} {\bibfield  {journal} {\bibinfo
  {journal} {Nature Physics}\ }\textbf {\bibinfo {volume} {10}},\ \bibinfo
  {pages} {198} (\bibinfo {year} {2014})}\BibitemShut {NoStop}%
\bibitem [{\citenamefont {Yang}\ \emph {et~al.}(2018)\citenamefont {Yang},
  \citenamefont {Gri\ifmmode~\check{s}\else \v{s}\fi{}ins}, \citenamefont
  {Chang}, \citenamefont {Zhao}, \citenamefont {Shih}, \citenamefont
  {Giamarchi},\ and\ \citenamefont {Hulet}}]{hulet2018}%
  \BibitemOpen
  \bibfield  {author} {\bibinfo {author} {\bibfnamefont {T.~L.}\ \bibnamefont
  {Yang}}, \bibinfo {author} {\bibfnamefont {P.}~\bibnamefont
  {Gri\ifmmode~\check{s}\else \v{s}\fi{}ins}}, \bibinfo {author} {\bibfnamefont
  {Y.~T.}\ \bibnamefont {Chang}}, \bibinfo {author} {\bibfnamefont {Z.~H.}\
  \bibnamefont {Zhao}}, \bibinfo {author} {\bibfnamefont {C.~Y.}\ \bibnamefont
  {Shih}}, \bibinfo {author} {\bibfnamefont {T.}~\bibnamefont {Giamarchi}}, \
  and\ \bibinfo {author} {\bibfnamefont {R.~G.}\ \bibnamefont {Hulet}},\ }\href
  {\doibase 10.1103/PhysRevLett.121.103001} {\bibfield  {journal} {\bibinfo
  {journal} {Phys. Rev. Lett.}\ }\textbf {\bibinfo {volume} {121}},\ \bibinfo
  {pages} {103001} (\bibinfo {year} {2018})}\BibitemShut {NoStop}%
\bibitem [{\citenamefont {Natu}\ \emph {et~al.}(2015)\citenamefont {Natu},
  \citenamefont {Li},\ and\ \citenamefont {Cole}}]{NatuCole2015}%
  \BibitemOpen
  \bibfield  {author} {\bibinfo {author} {\bibfnamefont {S.~S.}\ \bibnamefont
  {Natu}}, \bibinfo {author} {\bibfnamefont {X.}~\bibnamefont {Li}}, \ and\
  \bibinfo {author} {\bibfnamefont {W.~S.}\ \bibnamefont {Cole}},\ }\href
  {\doibase 10.1103/PhysRevA.91.023608} {\bibfield  {journal} {\bibinfo
  {journal} {Phys. Rev. A}\ }\textbf {\bibinfo {volume} {91}},\ \bibinfo
  {pages} {023608} (\bibinfo {year} {2015})}\BibitemShut {NoStop}%
\bibitem [{\citenamefont {Po}\ \emph {et~al.}(2014)\citenamefont {Po},
  \citenamefont {Chen},\ and\ \citenamefont {Zhou}}]{Po-2014}%
  \BibitemOpen
  \bibfield  {author} {\bibinfo {author} {\bibfnamefont {H.~C.}\ \bibnamefont
  {Po}}, \bibinfo {author} {\bibfnamefont {W.}~\bibnamefont {Chen}}, \ and\
  \bibinfo {author} {\bibfnamefont {Q.}~\bibnamefont {Zhou}},\ }\href
  {http://link.aps.org/doi/10.1103/PhysRevA.90.011602} {\bibfield  {journal}
  {\bibinfo  {journal} {Phys. Rev. A}\ }\textbf {\bibinfo {volume} {90}},\
  \bibinfo {pages} {011602} (\bibinfo {year} {2014})}\BibitemShut {NoStop}%
\bibitem [{\citenamefont {Celi}\ \emph {et~al.}(2014)\citenamefont {Celi},
  \citenamefont {Massignan}, \citenamefont {Ruseckas}, \citenamefont {Goldman},
  \citenamefont {Spielman}, \citenamefont {Juzeli\ifmmode~\bar{u}\else
  \={u}\fi{}nas},\ and\ \citenamefont {Lewenstein}}]{Celi-2014}%
  \BibitemOpen
  \bibfield  {author} {\bibinfo {author} {\bibfnamefont {A.}~\bibnamefont
  {Celi}}, \bibinfo {author} {\bibfnamefont {P.}~\bibnamefont {Massignan}},
  \bibinfo {author} {\bibfnamefont {J.}~\bibnamefont {Ruseckas}}, \bibinfo
  {author} {\bibfnamefont {N.}~\bibnamefont {Goldman}}, \bibinfo {author}
  {\bibfnamefont {I.~B.}\ \bibnamefont {Spielman}}, \bibinfo {author}
  {\bibfnamefont {G.}~\bibnamefont {Juzeli\ifmmode~\bar{u}\else
  \={u}\fi{}nas}}, \ and\ \bibinfo {author} {\bibfnamefont {M.}~\bibnamefont
  {Lewenstein}},\ }\href {\doibase 10.1103/PhysRevLett.112.043001} {\bibfield
  {journal} {\bibinfo  {journal} {Phys. Rev. Lett.}\ }\textbf {\bibinfo
  {volume} {112}},\ \bibinfo {pages} {043001} (\bibinfo {year}
  {2014})}\BibitemShut {NoStop}%
\bibitem [{\citenamefont {Mancini}\ \emph {et~al.}(2015)\citenamefont
  {Mancini}, \citenamefont {Pagano}, \citenamefont {Cappellini}, \citenamefont
  {Livi}, \citenamefont {Rider}, \citenamefont {Catani}, \citenamefont {Sias},
  \citenamefont {Zoller}, \citenamefont {Inguscio}, \citenamefont {Dalmonte}
  \emph {et~al.}}]{mancini2015}%
  \BibitemOpen
  \bibfield  {author} {\bibinfo {author} {\bibfnamefont {M.}~\bibnamefont
  {Mancini}}, \bibinfo {author} {\bibfnamefont {G.}~\bibnamefont {Pagano}},
  \bibinfo {author} {\bibfnamefont {G.}~\bibnamefont {Cappellini}}, \bibinfo
  {author} {\bibfnamefont {L.}~\bibnamefont {Livi}}, \bibinfo {author}
  {\bibfnamefont {M.}~\bibnamefont {Rider}}, \bibinfo {author} {\bibfnamefont
  {J.}~\bibnamefont {Catani}}, \bibinfo {author} {\bibfnamefont
  {C.}~\bibnamefont {Sias}}, \bibinfo {author} {\bibfnamefont {P.}~\bibnamefont
  {Zoller}}, \bibinfo {author} {\bibfnamefont {M.}~\bibnamefont {Inguscio}},
  \bibinfo {author} {\bibfnamefont {M.}~\bibnamefont {Dalmonte}},  \emph
  {et~al.},\ }\href
  {https://science.sciencemag.org/content/349/6255/1510?casa_token=pZXOCvZ20vcAAAAA:XD4X3pyE3LxPob5ZeAghQ68TPys-WG9EoTIbAfB6MkfLze5L-rZV80ECsvqZXD-oxkOc8fOUl1k_YjE}
  {\bibfield  {journal} {\bibinfo  {journal} {Science}\ }\textbf {\bibinfo
  {volume} {349}},\ \bibinfo {pages} {1510} (\bibinfo {year}
  {2015})}\BibitemShut {NoStop}%
\bibitem [{\citenamefont {Zeng}\ \emph {et~al.}(2015)\citenamefont {Zeng},
  \citenamefont {Wang},\ and\ \citenamefont {Zhai}}]{Zeng-2015}%
  \BibitemOpen
  \bibfield  {author} {\bibinfo {author} {\bibfnamefont {T.-S.}\ \bibnamefont
  {Zeng}}, \bibinfo {author} {\bibfnamefont {C.}~\bibnamefont {Wang}}, \ and\
  \bibinfo {author} {\bibfnamefont {H.}~\bibnamefont {Zhai}},\ }\href
  {http://link.aps.org/doi/10.1103/PhysRevLett.115.095302} {\bibfield
  {journal} {\bibinfo  {journal} {Phys. Rev. Lett.}\ }\textbf {\bibinfo
  {volume} {115}},\ \bibinfo {pages} {095302} (\bibinfo {year}
  {2015})}\BibitemShut {NoStop}%
\bibitem [{\citenamefont {Barbarino}\ \emph {et~al.}(2015)\citenamefont
  {Barbarino}, \citenamefont {Taddia}, \citenamefont {Rossini}, \citenamefont
  {Mazza},\ and\ \citenamefont {Fazio}}]{Barbarino-2015}%
  \BibitemOpen
  \bibfield  {author} {\bibinfo {author} {\bibfnamefont {S.}~\bibnamefont
  {Barbarino}}, \bibinfo {author} {\bibfnamefont {L.}~\bibnamefont {Taddia}},
  \bibinfo {author} {\bibfnamefont {D.}~\bibnamefont {Rossini}}, \bibinfo
  {author} {\bibfnamefont {L.}~\bibnamefont {Mazza}}, \ and\ \bibinfo {author}
  {\bibfnamefont {R.}~\bibnamefont {Fazio}},\ }\href
  {https://www.nature.com/articles/ncomms9134?origin=ppub} {\bibfield
  {journal} {\bibinfo  {journal} {Nature communications}\ }\textbf {\bibinfo
  {volume} {6}},\ \bibinfo {pages} {8134} (\bibinfo {year} {2015})}\BibitemShut
  {NoStop}%
\bibitem [{\citenamefont {Orignac}\ and\ \citenamefont
  {Giamarchi}(2001)}]{Orignac-2001}%
  \BibitemOpen
  \bibfield  {author} {\bibinfo {author} {\bibfnamefont {E.}~\bibnamefont
  {Orignac}}\ and\ \bibinfo {author} {\bibfnamefont {T.}~\bibnamefont
  {Giamarchi}},\ }\href {\doibase 10.1103/PhysRevB.64.144515} {\bibfield
  {journal} {\bibinfo  {journal} {Phys. Rev. B}\ }\textbf {\bibinfo {volume}
  {64}},\ \bibinfo {pages} {144515} (\bibinfo {year} {2001})}\BibitemShut
  {NoStop}%
\bibitem [{\citenamefont {Dhar}\ \emph {et~al.}(2012)\citenamefont {Dhar},
  \citenamefont {Maji}, \citenamefont {Mishra}, \citenamefont {Pai},
  \citenamefont {Mukerjee},\ and\ \citenamefont {Paramekanti}}]{Dhar-2012}%
  \BibitemOpen
  \bibfield  {author} {\bibinfo {author} {\bibfnamefont {A.}~\bibnamefont
  {Dhar}}, \bibinfo {author} {\bibfnamefont {M.}~\bibnamefont {Maji}}, \bibinfo
  {author} {\bibfnamefont {T.}~\bibnamefont {Mishra}}, \bibinfo {author}
  {\bibfnamefont {R.~V.}\ \bibnamefont {Pai}}, \bibinfo {author} {\bibfnamefont
  {S.}~\bibnamefont {Mukerjee}}, \ and\ \bibinfo {author} {\bibfnamefont
  {A.}~\bibnamefont {Paramekanti}},\ }\href
  {http://link.aps.org/doi/10.1103/PhysRevA.85.041602} {\bibfield  {journal}
  {\bibinfo  {journal} {Phys. Rev. A}\ }\textbf {\bibinfo {volume} {85}},\
  \bibinfo {pages} {041602} (\bibinfo {year} {2012})}\BibitemShut {NoStop}%
\bibitem [{\citenamefont {Dhar}\ \emph {et~al.}(2013)\citenamefont {Dhar},
  \citenamefont {Mishra}, \citenamefont {Maji}, \citenamefont {Pai},
  \citenamefont {Mukerjee},\ and\ \citenamefont {Paramekanti}}]{Dhar-2013}%
  \BibitemOpen
  \bibfield  {author} {\bibinfo {author} {\bibfnamefont {A.}~\bibnamefont
  {Dhar}}, \bibinfo {author} {\bibfnamefont {T.}~\bibnamefont {Mishra}},
  \bibinfo {author} {\bibfnamefont {M.}~\bibnamefont {Maji}}, \bibinfo {author}
  {\bibfnamefont {R.~V.}\ \bibnamefont {Pai}}, \bibinfo {author} {\bibfnamefont
  {S.}~\bibnamefont {Mukerjee}}, \ and\ \bibinfo {author} {\bibfnamefont
  {A.}~\bibnamefont {Paramekanti}},\ }\href
  {http://link.aps.org/doi/10.1103/PhysRevB.87.174501} {\bibfield  {journal}
  {\bibinfo  {journal} {Phys. Rev. B}\ }\textbf {\bibinfo {volume} {87}},\
  \bibinfo {pages} {174501} (\bibinfo {year} {2013})}\BibitemShut {NoStop}%
\bibitem [{\citenamefont {Petrescu}\ and\ \citenamefont
  {Le~Hur}(2013)}]{Petrescu-2013}%
  \BibitemOpen
  \bibfield  {author} {\bibinfo {author} {\bibfnamefont {A.}~\bibnamefont
  {Petrescu}}\ and\ \bibinfo {author} {\bibfnamefont {K.}~\bibnamefont
  {Le~Hur}},\ }\href {http://link.aps.org/doi/10.1103/PhysRevLett.111.150601}
  {\bibfield  {journal} {\bibinfo  {journal} {Phys. Rev. Lett.}\ }\textbf
  {\bibinfo {volume} {111}},\ \bibinfo {pages} {150601} (\bibinfo {year}
  {2013})}\BibitemShut {NoStop}%
\bibitem [{\citenamefont {Piraud}\ \emph {et~al.}(2015)\citenamefont {Piraud},
  \citenamefont {Heidrich-Meisner}, \citenamefont {McCulloch}, \citenamefont
  {Greschner}, \citenamefont {Vekua},\ and\ \citenamefont
  {Schollw\"ock}}]{Piraud-2015}%
  \BibitemOpen
  \bibfield  {author} {\bibinfo {author} {\bibfnamefont {M.}~\bibnamefont
  {Piraud}}, \bibinfo {author} {\bibfnamefont {F.}~\bibnamefont
  {Heidrich-Meisner}}, \bibinfo {author} {\bibfnamefont {I.~P.}\ \bibnamefont
  {McCulloch}}, \bibinfo {author} {\bibfnamefont {S.}~\bibnamefont
  {Greschner}}, \bibinfo {author} {\bibfnamefont {T.}~\bibnamefont {Vekua}}, \
  and\ \bibinfo {author} {\bibfnamefont {U.}~\bibnamefont {Schollw\"ock}},\
  }\href {http://link.aps.org/doi/10.1103/PhysRevB.91.140406} {\bibfield
  {journal} {\bibinfo  {journal} {Phys. Rev. B}\ }\textbf {\bibinfo {volume}
  {91}},\ \bibinfo {pages} {140406} (\bibinfo {year} {2015})}\BibitemShut
  {NoStop}%
\bibitem [{\citenamefont {Imambekov}\ \emph {et~al.}(2003)\citenamefont
  {Imambekov}, \citenamefont {Lukin},\ and\ \citenamefont
  {Demler}}]{Imambekov2003}%
  \BibitemOpen
  \bibfield  {author} {\bibinfo {author} {\bibfnamefont {A.}~\bibnamefont
  {Imambekov}}, \bibinfo {author} {\bibfnamefont {M.}~\bibnamefont {Lukin}}, \
  and\ \bibinfo {author} {\bibfnamefont {E.}~\bibnamefont {Demler}},\ }\href
  {\doibase 10.1103/PhysRevA.68.063602} {\bibfield  {journal} {\bibinfo
  {journal} {Phys. Rev. A}\ }\textbf {\bibinfo {volume} {68}},\ \bibinfo
  {pages} {063602} (\bibinfo {year} {2003})}\BibitemShut {NoStop}%
\bibitem [{\citenamefont {Rizzi}\ \emph {et~al.}(2005)\citenamefont {Rizzi},
  \citenamefont {Rossini}, \citenamefont {De~Chiara}, \citenamefont
  {Montangero},\ and\ \citenamefont {Fazio}}]{RizziFazio2005}%
  \BibitemOpen
  \bibfield  {author} {\bibinfo {author} {\bibfnamefont {M.}~\bibnamefont
  {Rizzi}}, \bibinfo {author} {\bibfnamefont {D.}~\bibnamefont {Rossini}},
  \bibinfo {author} {\bibfnamefont {G.}~\bibnamefont {De~Chiara}}, \bibinfo
  {author} {\bibfnamefont {S.}~\bibnamefont {Montangero}}, \ and\ \bibinfo
  {author} {\bibfnamefont {R.}~\bibnamefont {Fazio}},\ }\href {\doibase
  10.1103/PhysRevLett.95.240404} {\bibfield  {journal} {\bibinfo  {journal}
  {Phys. Rev. Lett.}\ }\textbf {\bibinfo {volume} {95}},\ \bibinfo {pages}
  {240404} (\bibinfo {year} {2005})}\BibitemShut {NoStop}%
\bibitem [{\citenamefont {Pixley}\ \emph {et~al.}(2017)\citenamefont {Pixley},
  \citenamefont {Cole}, \citenamefont {Spielman}, \citenamefont {Rizzi},\ and\
  \citenamefont {Das~Sarma}}]{PixleyDasSarma2017}%
  \BibitemOpen
  \bibfield  {author} {\bibinfo {author} {\bibfnamefont {J.~H.}\ \bibnamefont
  {Pixley}}, \bibinfo {author} {\bibfnamefont {W.~S.}\ \bibnamefont {Cole}},
  \bibinfo {author} {\bibfnamefont {I.~B.}\ \bibnamefont {Spielman}}, \bibinfo
  {author} {\bibfnamefont {M.}~\bibnamefont {Rizzi}}, \ and\ \bibinfo {author}
  {\bibfnamefont {S.}~\bibnamefont {Das~Sarma}},\ }\href {\doibase
  10.1103/PhysRevA.96.043622} {\bibfield  {journal} {\bibinfo  {journal} {Phys.
  Rev. A}\ }\textbf {\bibinfo {volume} {96}},\ \bibinfo {pages} {043622}
  (\bibinfo {year} {2017})}\BibitemShut {NoStop}%
\bibitem [{\citenamefont {Zhou}\ \emph {et~al.}(2019)\citenamefont {Zhou},
  \citenamefont {Luo}, \citenamefont {Chen}, \citenamefont {Jia},\ and\
  \citenamefont {Zhang}}]{ZhouZhang2019}%
  \BibitemOpen
  \bibfield  {author} {\bibinfo {author} {\bibfnamefont {X.}~\bibnamefont
  {Zhou}}, \bibinfo {author} {\bibfnamefont {X.-W.}\ \bibnamefont {Luo}},
  \bibinfo {author} {\bibfnamefont {G.}~\bibnamefont {Chen}}, \bibinfo {author}
  {\bibfnamefont {S.}~\bibnamefont {Jia}}, \ and\ \bibinfo {author}
  {\bibfnamefont {C.}~\bibnamefont {Zhang}},\ }\href@noop {} {\bibfield
  {journal} {\bibinfo  {journal} {arXiv preprint arXiv:1902.00575}\ } (\bibinfo
  {year} {2019})}\BibitemShut {NoStop}%
\bibitem [{\citenamefont {Cole}\ \emph {et~al.}(2019)\citenamefont {Cole},
  \citenamefont {Lee}, \citenamefont {Mahmud}, \citenamefont {Alavirad},
  \citenamefont {Spielman},\ and\ \citenamefont {Sau}}]{ColeSau2019}%
  \BibitemOpen
  \bibfield  {author} {\bibinfo {author} {\bibfnamefont {W.~S.}\ \bibnamefont
  {Cole}}, \bibinfo {author} {\bibfnamefont {J.}~\bibnamefont {Lee}}, \bibinfo
  {author} {\bibfnamefont {K.~W.}\ \bibnamefont {Mahmud}}, \bibinfo {author}
  {\bibfnamefont {Y.}~\bibnamefont {Alavirad}}, \bibinfo {author}
  {\bibfnamefont {I.~B.}\ \bibnamefont {Spielman}}, \ and\ \bibinfo {author}
  {\bibfnamefont {J.~D.}\ \bibnamefont {Sau}},\ }\href {\doibase
  10.1038/s41598-019-43929-6} {\bibfield  {journal} {\bibinfo  {journal}
  {Scientific Reports}\ }\textbf {\bibinfo {volume} {9}},\ \bibinfo {pages}
  {7471} (\bibinfo {year} {2019})}\BibitemShut {NoStop}%
\bibitem [{\citenamefont {Carusotto}\ and\ \citenamefont
  {Mueller}(2004)}]{CarusottoMueller2004}%
  \BibitemOpen
  \bibfield  {author} {\bibinfo {author} {\bibfnamefont {I.}~\bibnamefont
  {Carusotto}}\ and\ \bibinfo {author} {\bibfnamefont {E.~J.}\ \bibnamefont
  {Mueller}},\ }\href
  {https://iopscience.iop.org/article/10.1088/0953-4075/37/7/058/meta}
  {\bibfield  {journal} {\bibinfo  {journal} {Journal of Physics B: Atomic,
  Molecular and Optical Physics}\ }\textbf {\bibinfo {volume} {37}},\ \bibinfo
  {pages} {S115} (\bibinfo {year} {2004})}\BibitemShut {NoStop}%
\bibitem [{\citenamefont {Mueller}(2004)}]{Mueller2004}%
  \BibitemOpen
  \bibfield  {author} {\bibinfo {author} {\bibfnamefont {E.~J.}\ \bibnamefont
  {Mueller}},\ }\href {https://link.aps.org/doi/10.1103/PhysRevA.69.033606}
  {\bibfield  {journal} {\bibinfo  {journal} {Phys. Rev. A}\ }\textbf {\bibinfo
  {volume} {69}},\ \bibinfo {pages} {033606} (\bibinfo {year}
  {2004})}\BibitemShut {NoStop}%
\bibitem [{ite()}]{itensor}%
  \BibitemOpen
  \href@noop {} {}\bibinfo {note} {ITensor C++ library version 2.0,
  \url{http://itensor.org/}}\BibitemShut {NoStop}%
\bibitem [{\citenamefont {K\"onig}\ and\ \citenamefont
  {Pixley}(2018)}]{KoenigPixley2018}%
  \BibitemOpen
  \bibfield  {author} {\bibinfo {author} {\bibfnamefont {E.~J.}\ \bibnamefont
  {K\"onig}}\ and\ \bibinfo {author} {\bibfnamefont {J.~H.}\ \bibnamefont
  {Pixley}},\ }\href {\doibase 10.1103/PhysRevLett.121.083402} {\bibfield
  {journal} {\bibinfo  {journal} {Phys. Rev. Lett.}\ }\textbf {\bibinfo
  {volume} {121}},\ \bibinfo {pages} {083402} (\bibinfo {year}
  {2018})}\BibitemShut {NoStop}%
\bibitem [{\citenamefont {Powell}\ and\ \citenamefont
  {Sachdev}(2007)}]{PowellSachdev2007}%
  \BibitemOpen
  \bibfield  {author} {\bibinfo {author} {\bibfnamefont {S.}~\bibnamefont
  {Powell}}\ and\ \bibinfo {author} {\bibfnamefont {S.}~\bibnamefont
  {Sachdev}},\ }\href {\doibase 10.1103/PhysRevA.76.033612} {\bibfield
  {journal} {\bibinfo  {journal} {Phys. Rev. A}\ }\textbf {\bibinfo {volume}
  {76}},\ \bibinfo {pages} {033612} (\bibinfo {year} {2007})}\BibitemShut
  {NoStop}%
\bibitem [{\citenamefont {Giamarchi}(2004)}]{GiamarchiBook}%
  \BibitemOpen
  \bibfield  {author} {\bibinfo {author} {\bibfnamefont {T.}~\bibnamefont
  {Giamarchi}},\ }\href {https://books.google.com/books?id=1MwTDAAAQBAJ} {\emph
  {\bibinfo {title} {Quantum Physics in One Dimension}}},\ International Series
  of Monogr\ (\bibinfo  {publisher} {Oxford University Press (Clarendon
  Press)},\ \bibinfo {year} {2004})\BibitemShut {NoStop}%
\bibitem [{\citenamefont {Essler}\ \emph {et~al.}(2009)\citenamefont {Essler},
  \citenamefont {Shlyapnikov},\ and\ \citenamefont
  {Tsvelik}}]{EsslerTsvelik2009}%
  \BibitemOpen
  \bibfield  {author} {\bibinfo {author} {\bibfnamefont {F.}~\bibnamefont
  {Essler}}, \bibinfo {author} {\bibfnamefont {G.}~\bibnamefont {Shlyapnikov}},
  \ and\ \bibinfo {author} {\bibfnamefont {A.}~\bibnamefont {Tsvelik}},\ }\href
  {https://iopscience.iop.org/article/10.1088/1742-5468/2009/02/P02027/meta}
  {\bibfield  {journal} {\bibinfo  {journal} {Journal of Statistical Mechanics:
  Theory and Experiment}\ }\textbf {\bibinfo {volume} {2009}},\ \bibinfo
  {pages} {P02027} (\bibinfo {year} {2009})}\BibitemShut {NoStop}%
\bibitem [{\citenamefont {Fernandes}\ \emph {et~al.}(2019)\citenamefont
  {Fernandes}, \citenamefont {Orth},\ and\ \citenamefont
  {Schmalian}}]{FernandesSchmalian2019}%
  \BibitemOpen
  \bibfield  {author} {\bibinfo {author} {\bibfnamefont {R.~M.}\ \bibnamefont
  {Fernandes}}, \bibinfo {author} {\bibfnamefont {P.~P.}\ \bibnamefont {Orth}},
  \ and\ \bibinfo {author} {\bibfnamefont {J.}~\bibnamefont {Schmalian}},\
  }\href
  {https://www.annualreviews.org/doi/abs/10.1146/annurev-conmatphys-031218-013200?casa_token=Ymy-O12bVAIAAAAA:YLFECRCWlHwnZy43PZkYB1JKSDJFSEJQtdjFKpWmLpgo16cFdWa2BWprUbAvwoco99nPTN7xxh3yBA}
  {\bibfield  {journal} {\bibinfo  {journal} {Annual Review of Condensed Matter
  Physics}\ }\textbf {\bibinfo {volume} {10}},\ \bibinfo {pages} {133}
  (\bibinfo {year} {2019})}\BibitemShut {NoStop}%
\bibitem [{\citenamefont {Arcila-Forero}\ \emph {et~al.}(2016)\citenamefont
  {Arcila-Forero}, \citenamefont {Franco},\ and\ \citenamefont
  {Silva-Valencia}}]{Arcila2016}%
  \BibitemOpen
  \bibfield  {author} {\bibinfo {author} {\bibfnamefont {J.}~\bibnamefont
  {Arcila-Forero}}, \bibinfo {author} {\bibfnamefont {R.}~\bibnamefont
  {Franco}}, \ and\ \bibinfo {author} {\bibfnamefont {J.}~\bibnamefont
  {Silva-Valencia}},\ }\href {\doibase 10.1103/PhysRevA.94.013611} {\bibfield
  {journal} {\bibinfo  {journal} {Phys. Rev. A}\ }\textbf {\bibinfo {volume}
  {94}},\ \bibinfo {pages} {013611} (\bibinfo {year} {2016})}\BibitemShut
  {NoStop}%
\bibitem [{\citenamefont {Sitte}\ \emph {et~al.}(2009)\citenamefont {Sitte},
  \citenamefont {Rosch}, \citenamefont {Meyer}, \citenamefont {Matveev},\ and\
  \citenamefont {Garst}}]{SitteGarst2009}%
  \BibitemOpen
  \bibfield  {author} {\bibinfo {author} {\bibfnamefont {M.}~\bibnamefont
  {Sitte}}, \bibinfo {author} {\bibfnamefont {A.}~\bibnamefont {Rosch}},
  \bibinfo {author} {\bibfnamefont {J.}~\bibnamefont {Meyer}}, \bibinfo
  {author} {\bibfnamefont {K.}~\bibnamefont {Matveev}}, \ and\ \bibinfo
  {author} {\bibfnamefont {M.}~\bibnamefont {Garst}},\ }\href
  {https://journals.aps.org/prl/abstract/10.1103/PhysRevLett.102.176404}
  {\bibfield  {journal} {\bibinfo  {journal} {Physical review letters}\
  }\textbf {\bibinfo {volume} {102}},\ \bibinfo {pages} {176404} (\bibinfo
  {year} {2009})}\BibitemShut {NoStop}%
\bibitem [{\citenamefont {Alberton}\ \emph {et~al.}(2017)\citenamefont
  {Alberton}, \citenamefont {Ruhman}, \citenamefont {Berg},\ and\ \citenamefont
  {Altman}}]{AlbertonAltman2017}%
  \BibitemOpen
  \bibfield  {author} {\bibinfo {author} {\bibfnamefont {O.}~\bibnamefont
  {Alberton}}, \bibinfo {author} {\bibfnamefont {J.}~\bibnamefont {Ruhman}},
  \bibinfo {author} {\bibfnamefont {E.}~\bibnamefont {Berg}}, \ and\ \bibinfo
  {author} {\bibfnamefont {E.}~\bibnamefont {Altman}},\ }\href
  {https://journals.aps.org/prb/abstract/10.1103/PhysRevB.95.075132} {\bibfield
   {journal} {\bibinfo  {journal} {Physical Review B}\ }\textbf {\bibinfo
  {volume} {95}},\ \bibinfo {pages} {075132} (\bibinfo {year}
  {2017})}\BibitemShut {NoStop}%
\bibitem [{\citenamefont {Han}\ \emph {et~al.}(2019)\citenamefont {Han},
  \citenamefont {Lee},\ and\ \citenamefont {Moon}}]{han2019}%
  \BibitemOpen
  \bibfield  {author} {\bibinfo {author} {\bibfnamefont {S.}~\bibnamefont
  {Han}}, \bibinfo {author} {\bibfnamefont {J.}~\bibnamefont {Lee}}, \ and\
  \bibinfo {author} {\bibfnamefont {E.-G.}\ \bibnamefont {Moon}},\ }\href@noop
  {} {\  (\bibinfo {year} {2019})},\ \Eprint
  {http://arxiv.org/abs/arXiv:1911.01435} {arXiv:arXiv:1911.01435
  [cond-mat.str-el]} \BibitemShut {NoStop}%
\bibitem [{\citenamefont {Huijse}\ \emph {et~al.}(2015)\citenamefont {Huijse},
  \citenamefont {Bauer},\ and\ \citenamefont {Berg}}]{HuijseBerg2015}%
  \BibitemOpen
  \bibfield  {author} {\bibinfo {author} {\bibfnamefont {L.}~\bibnamefont
  {Huijse}}, \bibinfo {author} {\bibfnamefont {B.}~\bibnamefont {Bauer}}, \
  and\ \bibinfo {author} {\bibfnamefont {E.}~\bibnamefont {Berg}},\ }\href
  {https://journals.aps.org/prl/abstract/10.1103/PhysRevLett.114.090404}
  {\bibfield  {journal} {\bibinfo  {journal} {Physical review letters}\
  }\textbf {\bibinfo {volume} {114}},\ \bibinfo {pages} {090404} (\bibinfo
  {year} {2015})}\BibitemShut {NoStop}%
\bibitem [{\citenamefont {Ruhman}\ \emph {et~al.}(2015)\citenamefont {Ruhman},
  \citenamefont {Berg},\ and\ \citenamefont {Altman}}]{RuhmanAltman2015}%
  \BibitemOpen
  \bibfield  {author} {\bibinfo {author} {\bibfnamefont {J.}~\bibnamefont
  {Ruhman}}, \bibinfo {author} {\bibfnamefont {E.}~\bibnamefont {Berg}}, \ and\
  \bibinfo {author} {\bibfnamefont {E.}~\bibnamefont {Altman}},\ }\href
  {\doibase 10.1103/PhysRevLett.114.100401} {\bibfield  {journal} {\bibinfo
  {journal} {Phys. Rev. Lett.}\ }\textbf {\bibinfo {volume} {114}},\ \bibinfo
  {pages} {100401} (\bibinfo {year} {2015})}\BibitemShut {NoStop}%
\bibitem [{\citenamefont {Kane}\ \emph {et~al.}(2017)\citenamefont {Kane},
  \citenamefont {Stern},\ and\ \citenamefont {Halperin}}]{KaneStern2017}%
  \BibitemOpen
  \bibfield  {author} {\bibinfo {author} {\bibfnamefont {C.~L.}\ \bibnamefont
  {Kane}}, \bibinfo {author} {\bibfnamefont {A.}~\bibnamefont {Stern}}, \ and\
  \bibinfo {author} {\bibfnamefont {B.~I.}\ \bibnamefont {Halperin}},\ }\href
  {https://journals.aps.org/prx/abstract/10.1103/PhysRevX.7.031009} {\bibfield
  {journal} {\bibinfo  {journal} {Physical Review X}\ }\textbf {\bibinfo
  {volume} {7}},\ \bibinfo {pages} {031009} (\bibinfo {year}
  {2017})}\BibitemShut {NoStop}%
\bibitem [{\citenamefont {Holzhey}\ \emph {et~al.}(1994)\citenamefont
  {Holzhey}, \citenamefont {Larsen},\ and\ \citenamefont
  {Wilczek}}]{HolzheyWilzcek1994}%
  \BibitemOpen
  \bibfield  {author} {\bibinfo {author} {\bibfnamefont {C.}~\bibnamefont
  {Holzhey}}, \bibinfo {author} {\bibfnamefont {F.}~\bibnamefont {Larsen}}, \
  and\ \bibinfo {author} {\bibfnamefont {F.}~\bibnamefont {Wilczek}},\ }\href
  {https://www.sciencedirect.com/science/article/pii/0550321394904022}
  {\bibfield  {journal} {\bibinfo  {journal} {Nuclear Physics B}\ }\textbf
  {\bibinfo {volume} {424}},\ \bibinfo {pages} {443} (\bibinfo {year}
  {1994})}\BibitemShut {NoStop}%
\bibitem [{\citenamefont {Korepin}(2004)}]{Korepin2004}%
  \BibitemOpen
  \bibfield  {author} {\bibinfo {author} {\bibfnamefont {V.~E.}\ \bibnamefont
  {Korepin}},\ }\href {\doibase 10.1103/PhysRevLett.92.096402} {\bibfield
  {journal} {\bibinfo  {journal} {Phys. Rev. Lett.}\ }\textbf {\bibinfo
  {volume} {92}},\ \bibinfo {pages} {096402} (\bibinfo {year}
  {2004})}\BibitemShut {NoStop}%
\bibitem [{\citenamefont {Calabrese}\ and\ \citenamefont
  {Cardy}(2004)}]{CalabreseCardy2004}%
  \BibitemOpen
  \bibfield  {author} {\bibinfo {author} {\bibfnamefont {P.}~\bibnamefont
  {Calabrese}}\ and\ \bibinfo {author} {\bibfnamefont {J.}~\bibnamefont
  {Cardy}},\ }\href {\doibase 10.1088/1742-5468/2004/06/p06002} {\bibfield
  {journal} {\bibinfo  {journal} {Journal of Statistical Mechanics: Theory and
  Experiment}\ }\textbf {\bibinfo {volume} {2004}},\ \bibinfo {pages} {P06002}
  (\bibinfo {year} {2004})}\BibitemShut {NoStop}%
\bibitem [{\citenamefont {Jacob}\ \emph {et~al.}(2012)\citenamefont {Jacob},
  \citenamefont {Shao}, \citenamefont {Corre}, \citenamefont {Zibold},
  \citenamefont {De~Sarlo}, \citenamefont {Mimoun}, \citenamefont {Dalibard},\
  and\ \citenamefont {Gerbier}}]{Jacob-2012}%
  \BibitemOpen
  \bibfield  {author} {\bibinfo {author} {\bibfnamefont {D.}~\bibnamefont
  {Jacob}}, \bibinfo {author} {\bibfnamefont {L.}~\bibnamefont {Shao}},
  \bibinfo {author} {\bibfnamefont {V.}~\bibnamefont {Corre}}, \bibinfo
  {author} {\bibfnamefont {T.}~\bibnamefont {Zibold}}, \bibinfo {author}
  {\bibfnamefont {L.}~\bibnamefont {De~Sarlo}}, \bibinfo {author}
  {\bibfnamefont {E.}~\bibnamefont {Mimoun}}, \bibinfo {author} {\bibfnamefont
  {J.}~\bibnamefont {Dalibard}}, \ and\ \bibinfo {author} {\bibfnamefont
  {F.}~\bibnamefont {Gerbier}},\ }\href {\doibase 10.1103/PhysRevA.86.061601}
  {\bibfield  {journal} {\bibinfo  {journal} {Phys. Rev. A}\ }\textbf {\bibinfo
  {volume} {86}},\ \bibinfo {pages} {061601} (\bibinfo {year}
  {2012})}\BibitemShut {NoStop}%
\bibitem [{\citenamefont {Zibold}\ \emph {et~al.}(2016)\citenamefont {Zibold},
  \citenamefont {Corre}, \citenamefont {Frapolli}, \citenamefont {Invernizzi},
  \citenamefont {Dalibard},\ and\ \citenamefont {Gerbier}}]{Zibold-2016}%
  \BibitemOpen
  \bibfield  {author} {\bibinfo {author} {\bibfnamefont {T.}~\bibnamefont
  {Zibold}}, \bibinfo {author} {\bibfnamefont {V.}~\bibnamefont {Corre}},
  \bibinfo {author} {\bibfnamefont {C.}~\bibnamefont {Frapolli}}, \bibinfo
  {author} {\bibfnamefont {A.}~\bibnamefont {Invernizzi}}, \bibinfo {author}
  {\bibfnamefont {J.}~\bibnamefont {Dalibard}}, \ and\ \bibinfo {author}
  {\bibfnamefont {F.}~\bibnamefont {Gerbier}},\ }\href {\doibase
  10.1103/PhysRevA.93.023614} {\bibfield  {journal} {\bibinfo  {journal} {Phys.
  Rev. A}\ }\textbf {\bibinfo {volume} {93}},\ \bibinfo {pages} {023614}
  (\bibinfo {year} {2016})}\BibitemShut {NoStop}%
\bibitem [{\citenamefont {Farolfi}\ \emph {et~al.}(2019)\citenamefont
  {Farolfi}, \citenamefont {Trypogeorgos}, \citenamefont {Colzi}, \citenamefont
  {Fava}, \citenamefont {Lamporesi},\ and\ \citenamefont
  {Ferrari}}]{FarolfiFerrari2019}%
  \BibitemOpen
  \bibfield  {author} {\bibinfo {author} {\bibfnamefont {A.}~\bibnamefont
  {Farolfi}}, \bibinfo {author} {\bibfnamefont {D.}~\bibnamefont
  {Trypogeorgos}}, \bibinfo {author} {\bibfnamefont {G.}~\bibnamefont {Colzi}},
  \bibinfo {author} {\bibfnamefont {E.}~\bibnamefont {Fava}}, \bibinfo {author}
  {\bibfnamefont {G.}~\bibnamefont {Lamporesi}}, \ and\ \bibinfo {author}
  {\bibfnamefont {G.}~\bibnamefont {Ferrari}},\ }\href
  {https://aip.scitation.org/doi/abs/10.1063/1.5119915} {\bibfield  {journal}
  {\bibinfo  {journal} {Review of Scientific Instruments}\ }\textbf {\bibinfo
  {volume} {90}},\ \bibinfo {pages} {115114} (\bibinfo {year}
  {2019})}\BibitemShut {NoStop}%
\bibitem [{\citenamefont {Bakr}\ \emph {et~al.}(2009)\citenamefont {Bakr},
  \citenamefont {Gillen}, \citenamefont {Peng}, \citenamefont {F{\"o}lling},\
  and\ \citenamefont {Greiner}}]{BakrGreiner2009}%
  \BibitemOpen
  \bibfield  {author} {\bibinfo {author} {\bibfnamefont {W.~S.}\ \bibnamefont
  {Bakr}}, \bibinfo {author} {\bibfnamefont {J.~I.}\ \bibnamefont {Gillen}},
  \bibinfo {author} {\bibfnamefont {A.}~\bibnamefont {Peng}}, \bibinfo {author}
  {\bibfnamefont {S.}~\bibnamefont {F{\"o}lling}}, \ and\ \bibinfo {author}
  {\bibfnamefont {M.}~\bibnamefont {Greiner}},\ }\href
  {https://www.nature.com/articles/nature08482} {\bibfield  {journal} {\bibinfo
   {journal} {Nature}\ }\textbf {\bibinfo {volume} {462}},\ \bibinfo {pages}
  {74} (\bibinfo {year} {2009})}\BibitemShut {NoStop}%
\bibitem [{\citenamefont {Sherson}\ \emph {et~al.}(2010)\citenamefont
  {Sherson}, \citenamefont {Weitenberg}, \citenamefont {Endres}, \citenamefont
  {Cheneau}, \citenamefont {Bloch},\ and\ \citenamefont
  {Kuhr}}]{ShersonBloch2010}%
  \BibitemOpen
  \bibfield  {author} {\bibinfo {author} {\bibfnamefont {J.~F.}\ \bibnamefont
  {Sherson}}, \bibinfo {author} {\bibfnamefont {C.}~\bibnamefont {Weitenberg}},
  \bibinfo {author} {\bibfnamefont {M.}~\bibnamefont {Endres}}, \bibinfo
  {author} {\bibfnamefont {M.}~\bibnamefont {Cheneau}}, \bibinfo {author}
  {\bibfnamefont {I.}~\bibnamefont {Bloch}}, \ and\ \bibinfo {author}
  {\bibfnamefont {S.}~\bibnamefont {Kuhr}},\ }\href
  {https://www.nature.com/articles/nature09378} {\bibfield  {journal} {\bibinfo
   {journal} {Nature}\ }\textbf {\bibinfo {volume} {467}},\ \bibinfo {pages}
  {68} (\bibinfo {year} {2010})}\BibitemShut {NoStop}%
\bibitem [{\citenamefont {Miyake}\ \emph {et~al.}(2011)\citenamefont {Miyake},
  \citenamefont {Siviloglou}, \citenamefont {Puentes}, \citenamefont
  {Pritchard}, \citenamefont {Ketterle},\ and\ \citenamefont
  {Weld}}]{MiyakeWeld2011}%
  \BibitemOpen
  \bibfield  {author} {\bibinfo {author} {\bibfnamefont {H.}~\bibnamefont
  {Miyake}}, \bibinfo {author} {\bibfnamefont {G.~A.}\ \bibnamefont
  {Siviloglou}}, \bibinfo {author} {\bibfnamefont {G.}~\bibnamefont {Puentes}},
  \bibinfo {author} {\bibfnamefont {D.~E.}\ \bibnamefont {Pritchard}}, \bibinfo
  {author} {\bibfnamefont {W.}~\bibnamefont {Ketterle}}, \ and\ \bibinfo
  {author} {\bibfnamefont {D.~M.}\ \bibnamefont {Weld}},\ }\href {\doibase
  10.1103/PhysRevLett.107.175302} {\bibfield  {journal} {\bibinfo  {journal}
  {Phys. Rev. Lett.}\ }\textbf {\bibinfo {volume} {107}},\ \bibinfo {pages}
  {175302} (\bibinfo {year} {2011})}\BibitemShut {NoStop}%
\bibitem [{\citenamefont {Hart}\ \emph {et~al.}(2015)\citenamefont {Hart},
  \citenamefont {Duarte}, \citenamefont {Yang}, \citenamefont {Liu},
  \citenamefont {Paiva}, \citenamefont {Khatami}, \citenamefont {Scalettar},
  \citenamefont {Trivedi}, \citenamefont {Huse},\ and\ \citenamefont
  {Hulet}}]{hartHulet2015}%
  \BibitemOpen
  \bibfield  {author} {\bibinfo {author} {\bibfnamefont {R.~A.}\ \bibnamefont
  {Hart}}, \bibinfo {author} {\bibfnamefont {P.~M.}\ \bibnamefont {Duarte}},
  \bibinfo {author} {\bibfnamefont {T.-L.}\ \bibnamefont {Yang}}, \bibinfo
  {author} {\bibfnamefont {X.}~\bibnamefont {Liu}}, \bibinfo {author}
  {\bibfnamefont {T.}~\bibnamefont {Paiva}}, \bibinfo {author} {\bibfnamefont
  {E.}~\bibnamefont {Khatami}}, \bibinfo {author} {\bibfnamefont {R.~T.}\
  \bibnamefont {Scalettar}}, \bibinfo {author} {\bibfnamefont {N.}~\bibnamefont
  {Trivedi}}, \bibinfo {author} {\bibfnamefont {D.~A.}\ \bibnamefont {Huse}}, \
  and\ \bibinfo {author} {\bibfnamefont {R.~G.}\ \bibnamefont {Hulet}},\ }\href
  {https://www.nature.com/articles/nature14223} {\bibfield  {journal} {\bibinfo
   {journal} {Nature}\ }\textbf {\bibinfo {volume} {519}},\ \bibinfo {pages}
  {211} (\bibinfo {year} {2015})}\BibitemShut {NoStop}%
\bibitem [{\citenamefont {Sachdev}(2011)}]{SachdevBook}%
  \BibitemOpen
  \bibfield  {author} {\bibinfo {author} {\bibfnamefont {S.}~\bibnamefont
  {Sachdev}},\ }\href {https://books.google.com/books?id=F3IkpxwpqSgC} {\emph
  {\bibinfo {title} {Quantum Phase Transitions}}}\ (\bibinfo  {publisher}
  {Cambridge University Press},\ \bibinfo {year} {2011})\BibitemShut {NoStop}%
\end{thebibliography}%

\end{document}